\documentclass[12pt,aps,showpacs]{revtex4}
\usepackage{amsmath,amssymb,amsfonts,epsfig,graphicx,epstopdf}
\newcommand{\beeq}{\begin{equation}}
\newcommand{\eneq}{\end{equation}}
\newcommand{\be}{\begin{eqnarray}}
\newcommand{\ee}{\end{eqnarray}}
\newcommand{\bpic}{\begin{picture}}
\newcommand{\epic}{\end{picture}}
\newcommand{\bs}{\begin{scriptsize}}
\newcommand{\es}{\end{scriptsize}}

\def\la{\raise.16ex\hbox{$\langle$} \, }
\def\ra{\, \raise.16ex\hbox{$\rangle$} }

\def\Box{\kern1pt\vbox{\hrule height 1.2pt\hbox{\vrule width 1.2pt\hskip 3pt
   \vbox{\vskip 6pt}\hskip 3pt\vrule width 0.6pt}\hrule height 0.6pt}\kern1pt}
\def\gtwid{\mathrel{\raise.3ex\hbox{$>$\kern-.75em\lower1ex\hbox{$\sim$}}}}
\def\ltwid{\mathrel{\raise.3ex\hbox{$<$\kern-.75em\lower1ex\hbox{$\sim$}}}}

\begin{document}


\title{Quantum Fluctuations of a Self-interacting Inflaton}

\author{G. Karakaya}\email{gulaykarakaya@itu.edu.tr}

\author{V. K. Onemli}\email{onemli@itu.edu.tr}

\affiliation{Department of Physics, Istanbul Technical
University, Maslak, Istanbul 34469, Turkey}

\begin{abstract}
We present a method to analytically compute the quantum corrected two-point correlation function of a scalar field in leading order at each loop in a homogeneous, isotropic and spatially flat spacetime where the expansion rate is time dependent and express the quantum corrected power spectrum $\Delta^2(k)$ as a time derivative of the coincident correlation function evaluated at time $t_k$ of the first horizon crossing of a mode with comoving wave number $k$. To facilitate the method, we consider the simplest version of inflation driven by a massive, minimally coupled inflaton endowing a quartic self-interaction---with positive or negative self-coupling. We compute the quantum corrected two-point correlation function, power spectrum, spectral index $n(k)$ and the running of the spectral index $\alpha(k)$ for the inflaton fluctuations at one-loop order. Numerical estimates of the $n(k)$ and $\alpha(k)$ and the cosmological measurements are in agreement, within reasonable ranges of values for the physical parameters in the model.
\end{abstract}
\pacs{98.80.Cq, 04.62.+v}
\maketitle \vskip 0.1in \vspace{.1cm}

\section{Introduction}
\label{sec:intro}
A method to compute the quantum corrected two-point correlation function of a self-interacting spectator scalar field, in leading order at each loop, during de Sitter inflation---where the expansion rate $H$ is constant---was introduced in Refs. \cite{vacuum,GKVO}. The computation incorporated Starobinsky's stochastic approach \cite{Star,StarYok} and the techniques of quantum field theory. The quantum corrected power spectrum was attained~\cite{GKVO} as a time derivative of the coincidence limit of the computed correlation function evaluated at the time $t_k$ of first horizon crossing.
The method was employed to study a massless \cite{vacuum} and massive \cite{GKVO} minimally coupled scalars with a quartic self-interaction.

In this paper, we generalize the method to a more general Friedman-Robinson-Walker spacetime where the expansion rate $H$ is time dependent. A massive, minimally coupled, self-interacting scalar (inflaton) field with potential $V(\varphi)\!=\!\frac{m^2}{2}\varphi^2\!\pm\!\frac{\lambda}{4!}\varphi^4$, where $0\!<\!\lambda\!\ll\!1$, drives the inflation in the model we consider, and hence, the expansion rate is time dependent. The plus sign in the quartic self-interaction implies a slightly steeper feature in the potential whereas the minus sign implies a slightly flattened feature. The alternative sign choices in the potential yield only a trivial sign difference between the results of any $\mathcal{O}(\lambda)$ computation. Note that a spectator scalar with this potential exhibits \cite{OW1,OW2,BOW,KO,KOW1,O1} enhanced quantum effects---in the massless limit---during de Sitter inflation. The effects are reduced \cite{GKVO} as the mass increases.

Physical origin of the quantum fluctuations is the Heisenberg's uncertainty principle which manifests itself as the particle-antiparticle pair production and annihilation out of and into the vacuum. The observed scalar perturbations on the cosmic microwave background (CMB) anisotropy are believed to be the amplified imprints of quantum fluctuations of an  inflaton which seeded the large scale structure in the universe. The issue of fluctuations in inflationary spacetimes and their observable implications has been of more than passing interest \cite{sfl} in cosmology. One naturally inquires, if there are correlations between the theoretical predictions for primordial inflaton fluctuations and the measurements of CMB temperature anisotropy, galaxy power spectra and gravitational lensing survey correlation functions. Indeed, all measurements agree that the number of the observed entity of a particular size in the corresponding distribution, in each case of the inquiries, increases as the size of the entity increases.

During inflation the physical size grows more rapidly than the horizon size. Therefore, for each scale of cosmological interest with comoving wave number $k$ there exists a certain comoving time $t\!=\!t_k$, as pointed out earlier, at which the scale exits the horizon. The Fourier mode $k$ of the fluctuation field is, therefore, a subhorizon [ultraviolet (UV)] mode when $t\!\!<\!t_k$ but becomes a superhorizon [infrared (IR)] mode when $t\!>\!t_k$. Many subhorizon fluctuation modes are shifted continuously to the superhorizon scales during inflation.

As soon as a particular mode exits the horizon, the crest of the mode loses causal contact with the trough and therefore cannot propagate. Hence, the super-horizon mode ''freezes in'' after the horizon exit. It does not behave like a wave at all. After the end of inflationary phase, during the phases of radiation domination and matter domination, the horizon size grows more rapidly than the scale factor $a(t)$---hence, than the physical size. Thus, as the physical horizon size successively reaches the physical wavelengths of the super-horizon modes, the comoving scales that exited the horizon during inflation start reentering the horizon with the same amplitudes as they were when they exited---except for a small factor which depends on the ratio of pressure to energy density in the universe when they exit and reenter---and the crests and troughs of the corresponding modes regain causal contact and propagate. The modes that exit the horizon latest reenter earliest and play the key role in seeding the large scale structure formation in the early universe.

Starobinsky realized \cite{Star,StarYok} that the UV modes are irrelevant for the late time behavior of fields in the scalar potential models and amputated them. The free field expansion of a field contains arbitrarily large wavenumbers (UV modes) and therefore the expectation values of coincident products of the fields can arbor UV divergences. All of these features are absent in the Starobinsky's realization of the free field which is constructed by taking the IR limit of the mode function and retaining only the IR modes at a particular time. Using the IR truncated field in computations is an approximation which gets only the UV-finite, secular parts of the full result. In scalar potential models, Starobinsky's formalism gives {\rm exactly} the same leading terms that the Schwinger-Keldish (in-in) formalism gives at each perturbative order. For example, the coincidence limit of two-point correlation function of a massless minimally coupled scalar with a quartic self-interaction in de Sitter spacetime, computed applying Schwinger-Keldish formalism \cite{BOW,KO} and applying Starobinsky's formalism \cite{KO} yield exactly the same leading terms at one- and two-loop order. Other examples showing the exact agreement include: scalars interacting with fermions \cite{W3,W4}, scalar quantum electrodynamics \cite{PTW1,PTW2} and scalars with derivative interactions \cite{W1,TWstgrav,MW}.

Several aspects of IR dynamics in scalar potential models have been studied with various approaches that include, employing stochastic spectral expansion \cite{MRST}, extending the stochastic formalism \cite{TT}, applying complementary and principal series analysis \cite{AMPP1,AMPP2}, computing effective actions \cite{MR,R1,DB1,CW1} and potentials \cite{JSS}, implementing Fokker-Planck equation and $\delta N$ formalism \cite{AFNVW,PVAW}, employing $1/N$ expansion \cite{GS1}, adopting reduced density matrix method \cite{DB2} and applying renormalization group analysis \cite{GS2}. We follow Starobinsky's approach \cite{Star,StarYok} and use the standard techniques of quantum field theory, in this paper.

The two-point correlation function of the inflaton fluctuation field we compute is a measure of the relationship between the random fluctuations at two events $(t, \vec{x})$ and $(t', \vec{x}\,')$. Thus, it provides information about how
independent the fluctuation $\delta\varphi(t, \vec{x})$ is from the fluctuation $\delta\varphi(t', \vec{x}\,')$. The spatial two-point correlation function, therefore, measures the emergence probability of a fluctuation $\delta\varphi(t, \vec{x})$ at a separation $\Delta x\!\!=\!\!\|\vec{x}\!\!-\!\!\vec{x}\,'\|$ from the fluctuation $\delta\varphi(t, \vec{x}\,')$, at a given time $t$. The fluctuation spectrum contains contributions from all ranges of fluctuations of different sizes (modes). The Fourier transform of the spatial two-point correlation function can be used to define the time dependent power spectrum $\Delta^2(t, k)$, which quantifies information about the probability of how many random quantum fluctuation of each size ought to contribute to the spectrum at a given time $t$. The usual, time-independent power spectrum \cite{Dod} of the fluctuations $\lim_{t\gg t_k}\Delta^2_{\delta\varphi}(t, k)\!\equiv\!\Delta^2_{\delta\varphi}(k)$ can be obtained \cite{GKVO} from the coincident correlation function of the IR truncated fluctuation field.

The $\Delta^2_{\delta\varphi}(k)$ of the primordial inflaton fluctuations can be measured from the power spectrum of the CMB which encodes size distribution of the temperature fluctuations, i.e., how many hot and cold spots there are of each angular size in the CMB. Hence, the measured CMB data is the ultimate arbiter of the inflationary models. The models, for example, generically predict a slightly negative (red) tilt in the power spectrum, i.e., a deviation between $-0.08$ and $-0.02$ from that of the Harrison-Zeldovich scale invariant (constant amplitude) spectrum with zero tilt generated during a de Sitter inflation. (Red-tilt implies that the amplitudes of inflaton fluctuations grow toward the larger scales.) The exact value of the tilt depends on the details of the particular inflationary scenario and can be used to discriminate between the models. We compute the quantum corrected tilt and its running in our model up to $\mathcal{O}(\lambda^2)$. Numerical estimates obtained from the results are in accordance with the observations.

The remainder of this paper is organized as follows. In Sec.~\ref{sec:model} we present the model, obtain the field equations, solve the expansion rate, scale factor, slow roll parameter and the background field in terms of the initial values. In Sec.~\ref{sec:InfFluct} we obtain the mode function for the free fluctuation field, give the free field expansion and express the full field in terms of the free field and the Green's function. We compute the tree-order and one-loop two point correlators for the fluctuations of the IR truncated inflaton in our model, up to $\mathcal{O}(\lambda^2)$, in Sec.~\ref{sec:twopointcorrelator}. The coincidence limit of the quantum corrected two-point correlation function is given in Sec.~\ref{sec:coincorr}. In Sec.~\ref{sec:power}, using the coincident correlation function, we compute the quantum corrected power spectrum, spectral index and running of the spectral index for the inflaton fluctuations; estimate the tilt and its running numerically and compare them with the observational values. The conclusions are summarized in Sec.~\ref{sec:conclusions}. The Appendixes comprise the computational details of the tree-order and one-loop correlators.

\section{The model}
\label{sec:model}

We consider a massive, minimally coupled self-interacting scalar field driven inflation. The renormalized Lagrangian density of the model
\begin{equation}
\mathcal{L} \!=\!\left[\frac{R}{16\pi G}\!-\!\frac{1\!\!+\!\delta Z}{2}\partial_{\mu} \varphi \, \partial_{\nu}
\varphi\, g^{\mu\nu}\!-\!\! V(\varphi)\right]\!\sqrt{-g} \; ,\label{lagden}
\end{equation}
where the $R$ and $g$ denote the Ricci scalar and the determinant of the metric $g_{\mu\nu}$, respectively. [We adopt the convention where
a Greek index $\mu\!=\!0,1,2, \dots,(D\!-\!\!1)$, hence the components of four-position vector $x$ are $x^\mu\!=\!(x^0\!,\vec x)$ with comoving time $x^0\!\equiv\!t$ and $\partial_\mu\!=\!(\partial_0,\vec\nabla)$.] The $\delta Z$ stand for the field strength renormalization counterterm. The renormalized potential\beeq\hspace{-0.9cm}
V(\varphi)\!=\!\frac{1\!\!+\!\delta Z}{2}m^2\varphi^2\!+\!V_{\rm pert}(\varphi) \;,\label{V}
\eneq
where $m$ is the renormalized mass. The part of the potential that we treat perturbatively,
\beeq
\hspace{-0.7cm} V_{\rm pert}(\varphi)\!\equiv\!\frac{\delta m^2}{2}
\varphi^2\!\pm\!\frac{\lambda\!+\!\delta\lambda}{4!}\varphi^4\;,\label{Vpert}
\eneq
where $\lambda$ is the coupling strength and $\delta m^2$ and $\delta\lambda$ stand for the mass renormalization and the coupling strength renormalization counterterms, respectively. Counterterms are just constants and their numerical coefficients are identical in any geometry. Thus, as in flat spacetime and de Sitter geometry, the orders of $\lambda$ of the counterterms are $\delta Z\!\!\sim\!\mathcal{O}(\lambda^2)$, $\delta m^2\!\!\sim\!\mathcal{O}(\lambda)$ and $\delta\lambda\!\!\sim\!\mathcal{O}(\lambda^2)$.

The metric tensor $g_{\mu\nu}$ is chosen as that of a $D$-dimensional spatially flat Friedmann-Robertson-Walker spacetime with no metric perturbations. Therefore, the invariant
line element \beeq \hspace{-0.7cm} ds^2\!\!=\! g_{\mu\nu} dx^{\mu}
dx^{\nu}\!\!=\!-dt^2\!\!+\!a^2(t) d\vec{x} \cdot d\vec{x}
\; .\label{lineelement}\eneq
Note that, just as the tensor power spectrum is suppressed by a factor of the slow-roll parameter
\beeq
\hspace{-0.7cm}\epsilon(t)\!\equiv\!-\frac{\dot{H}(t)}{H^2(t)}\ll 1 \; , \label{epsilon}
\eneq
where an overdot denotes derivative with respect to comoving time $t$ and $H(t)\equiv\dot{a}/a$ stands for the Hubble parameter, the metric perturbations are suppressed by $\sqrt{\epsilon}$.

The inflaton in this background obeys the field equation,
\be
\hspace{0cm}\ddot{\varphi}(x)\!+\!(D\!-\!1) H(t) \dot{\varphi}(x)\!-\!\!
\left[\frac{\nabla^2}{a^2}-\!m^2\right]\!\!\varphi(x)\!=\!-\frac{\frac{\partial V_{\rm pert}(\varphi)}{\partial\varphi}(x)}{1\!\!+\!\delta Z}\; ,\label{feq}
\ee
where the derivative of the perturbative potential
\beeq
\hspace{-0.4cm}\frac{\partial V_{\rm pert}(\varphi)}{\partial\varphi}\!=\!\delta m^2
\varphi\!\pm\!\frac{1}{6}(\lambda\!+\!\delta\lambda) \varphi^3\; .
\eneq
We consider the inflaton field $\varphi(x)$ as a sum of an unperturbed (averaged) background field $\bar\varphi(t)$ that depends only on time and a small perturbation $\delta\varphi(x)$ on this background field, \beeq\hspace{-0.2cm}\varphi(x)\!=\!\bar\varphi(t)\!+\!\delta\varphi(x)\; .\label{decomp}\eneq
Thus Eq.~(\ref{feq}) yields the following equations for the homogenous part $\bar\varphi(t)$ and for the fluctuations $\delta\varphi(t, \vec{x})$ around it,
\be
&&\hspace{-0.4cm}\ddot{\bar\varphi}(t)\!+\!(D\!-\!1)H(t)\dot{\bar\varphi}(t)\!+\!m^2\bar\varphi(t)\!=\!-\frac{1}{1\!\!+\!\delta Z}\!\left\{\delta m^2
\bar\varphi(t)\!\pm\!\frac{\lambda\!+\!\delta\lambda}{6} \bar\varphi^3(t)\right\}\; ,\label{barvarphieq}\\
&&\hspace{0.1cm}\delta\ddot{\varphi}(x)\!+\!(D\!-\!1)H(t)\,\delta\dot{\varphi}(x)\!-\!\!
\left[\frac{\nabla^2}{a^2}-\!m^2\right]\!\!\delta\varphi(x)\!=\!-\frac{\frac{\partial V_{\rm pert}(\delta\varphi)}{\partial\delta\varphi}(x)}{1\!\!+\!\delta Z}\; ,\label{deltavarphifull}
\ee
where
\beeq
\frac{\partial V_{\rm pert}(\delta\varphi)}{\partial\delta\varphi}(x)\!\equiv\!\delta m^2
\delta\varphi(x)\!\pm\!\frac{\lambda\!+\!\delta\lambda}{6}\Bigl[3 \bar\varphi^2(t)\,\delta\varphi(x)\!+\!3\bar\varphi(t)\,(\delta\varphi(x))^2\!\!+\!(\delta\varphi(x))^3\Bigr]\; .\label{Vpertdeltavarphi}
\eneq

The scale factor $a(t)$ and the background field $\bar\varphi(t)$ are related by two non-trivial Einstein's field equations---corresponding to purely temporal and purely spatial components---as
\be
&&\hspace{-0.0cm}(D\!-\!2)\frac{D\!-\!1}{2}H^2\!=\!8\pi G\Bigl[\frac{{\dot{\bar\varphi}}^2(t)}{2}\!+\!V(\bar\varphi)\Bigr]\label{A}\;,\\
&&\hspace{-1.2cm}-(D\!-\!2)\frac{D\!-\!1}{2}H^2\Bigl[1\!-\!\frac{2\epsilon(t)}{D\!-\!1}\Bigr]\!=\!8\pi G\Big[\frac{{\dot{\bar\varphi}}^2(t)}{2}\!-\!V(\bar\varphi)\Big]\;.\label{B}
\ee
Equations~(\ref{A}) and (\ref{B}) can be inverted to eliminate the background inflaton field $\bar\varphi(t)$,
\be
&&\hspace{0cm}\dot{\bar\varphi}^2(t)\!=\!\frac{(D\!-\!2)H^2(t)}{8 \pi G} \, \epsilon(t) \;
,\label{dotvarphi}\\
&&\hspace{-0.75cm}V(\bar\varphi)\!=\!\frac{(D\!-\!2)H^2(t)}{16 \pi G} \Bigl[
D\!-\!1\!-\!\epsilon(t)\Bigr] \, .\label{potent}
\ee
Using Eqs.~(\ref{V}) and (\ref{Vpert}) in Eq.~(\ref{potent}), at $\mathcal{O}(\lambda)$, we find
\beeq
\hspace{-0.1cm}H^2\!\!=\!\frac{8\pi G}{(D\!-\!2)(D\!-\!1\!-\!\epsilon)}\Bigl[\left(m^2\!\!+\!\delta m^2\right)\bar{\varphi}^2\!\pm\!\frac{\lambda}{12}\bar{\varphi}^4\!+\!\mathcal{O}(\lambda^2)\Bigr]\;.\label{}
\eneq
The mass counterterm $\delta m^2$, which is of $\mathcal{O}(\lambda)$, is multiplied by $\bar{\varphi}^2$. Hence, it is suppressed by a factor of $\bar{\varphi}^2$ compared to the quartic self-interaction term which is also of $\mathcal{O}(\lambda)$. Because $4\lesssim \bar\varphi\lesssim 10^6$ in Planckian units, the former cannot contribute as significant as the latter. Therefore, we have
\beeq
H\!\!=\!\sqrt{\frac{8\pi G}{(D\!-\!2)(D\!-\!1)}}m^2\frac{\bar{\varphi}}{m}\Bigl[1\!\!\pm\!\frac{\lambda}{24}
\Bigl(\frac{\bar\varphi}{m}\Bigr)^2\!\!\!+\!\mathcal{O}(\epsilon)\!+\!\mathcal{O}(\lambda^2)\Bigr]\;.\label{H}
\eneq
Recall that the quartic self-interaction term $\frac{\lambda}{4!}{\bar{\varphi}}^4$ is treated perturbatively next to the the mass term $\frac{m^2}{2}{\bar{\varphi}}^2$. Hence, $1\gg\frac{\lambda}{12}
\frac{{\bar{\varphi}}^2}{m^2}$ in Eq.~(\ref{H}).

To get the expansion rate $H$ from Eq.~(\ref{H}), as a function of time and initial values $H_{i0}$ and $\epsilon_{i0}$, we first need to solve field equation~(\ref{barvarphieq}) for $\bar\varphi(t)$.  In the slow roll regime where $|\ddot{\bar\varphi}|\ll H|\dot{\bar\varphi}|$, and at $\mathcal{O}(\lambda)$, Eq.~(\ref{barvarphieq}) implies
\beeq
(D\!-\!1)H\dot{\bar\varphi}\!+\!m^2\!\bar\varphi\!\simeq\!\mp\frac{\lambda}{6}\bar\varphi^3\; .\label{barphislowroll}
\eneq
We neglected the other $\mathcal{O}(\lambda)$ term---$\delta m^2\bar\varphi$---suppressed by a factor of $\bar\varphi^2$ on the right side of Eq.~(\ref{barvarphieq}), as we did in obtaining Eq.~(\ref{H}). Employing Eq.~(\ref{H}) in Eq.~(\ref{barphislowroll}) yields
\beeq
\dot{\bar\varphi}\!\simeq\!-\frac{m}{\sqrt{8\pi G}}\sqrt{\frac{D\!-\!2}{D\!-\!1}}\left[1\!\!\pm\!\frac{\lambda}{8}\Bigl(\frac{\bar\varphi}{m}\Bigr)^2\!\!\!+\!\mathcal{O}(\epsilon)\!+\!\mathcal{O}(\lambda^2)\right] \; .\label{dotfi}
\eneq
The solution of Eq.~(\ref{dotfi}), for the plus sign choice in the $\mathcal{O}(\lambda)$-term, can be given as\beeq
\bar\varphi(t)\!\simeq\!\sqrt{\frac{8}{\lambda}}\,m\tan\left(\!-\sqrt{\frac{\lambda}{8}\frac{D\!-\!2}{(D\!-\!1)8\pi G}}\;t\!+\!\arctan\Bigl(\!{\sqrt{\frac{\lambda}{8}}\frac{\bar\varphi_i}{m}}\Bigr)\!\right)\;.\label{brvph}
\eneq
For the minus sign choice in Eq.~(\ref{dotfi}), taking $\lambda\!\rightarrow\!-\!\lambda$ and analytically continuing the tangent function in Eq.~(\ref{brvph}), yields the background field solution as
\beeq
\bar\varphi(t)\!\simeq\!\sqrt{\frac{8}{\lambda}}\,m\tanh\left(\!-\sqrt{\frac{\lambda}{8}\frac{D\!-\!2}{(D\!-\!1)8\pi G}}\;t\!+\!{\rm{arctanh}}\Bigl(\!{\sqrt{\frac{\lambda}{8}}\frac{\bar\varphi_i}{m}}\Bigr)\!\right)\;.\label{brvphhyp}
\eneq
Using the addition formulae for the $\tan(x)$ and $\tanh(x)$ functions in Eqs.~(\ref{brvph}) and (\ref{brvphhyp}) respectively and then expanding the results in series, up to $\mathcal{O}(\lambda^2)$, leads to
\be
\hspace{0.8cm}\bar\varphi(t)\!\simeq\!\bar\varphi_i\left\{\!1\!-\!\sqrt{\frac{D\!-\!2}{(D\!-\!1)8\pi G}}\!\left[\frac{m}{\bar\varphi_i}\!\pm\!\frac{\lambda}{8}\frac{\bar\varphi_i}{m}\right]\!t\!\pm\!\frac{\lambda}{8}\frac{D\!-\!2}{(D\!-\!1)8\pi G}\,t^2\!\right\}\;,\label{barfi}
\ee
where $\bar\varphi_i$ is the initial value of $\bar\varphi(t)$ at $t\!=\!t_i\!=\!0$. Note that $\bar\varphi_i=\bar\varphi_{i0}$ where the subscript $0$ denotes the non-interacting ($\lambda\!\rightarrow\!0$) result. Note also that, in the noninteracting regime, one recovers\beeq
\hspace{0.1cm}\lim_{\lambda\rightarrow 0}\bar\varphi(t)\simeq-m\sqrt{\frac{D\!-\!2}{(D\!-\!1)8\pi G}}\,t\!+\!\bar\varphi_i\;,
\eneq
the well known solution in the simplest version of the chaotic inflation driven by a mass term.

The fluctuation field $\delta\varphi(x)$ in Eq.~(\ref{decomp}), on the other hand, is obtained by solving Eq.~(\ref{deltavarphifull}) as\beeq
\hspace{0.6cm}\delta\varphi(x)\!=\!\delta\varphi_0(x)\!-\!\!\!\int_0^t\!\!dt'a^{D-1}(t')\!\int\! d^{D-1}x' G_{\rm ret}(x ; x')\frac{\frac{\partial V_{\rm pert}(\delta\varphi)}{\partial\delta\varphi}(x')}{1\!\!+\!\delta Z}\; ,\label{fuldeltavarphi}
\eneq
where $\frac{\partial V_{\rm pert}(\delta\varphi)}{\partial\delta\varphi}$ is given in Eq.~(\ref{Vpertdeltavarphi}) and the retarded Green's function is computed in Eq.~(\ref{IRgreen}). We use solution~(\ref{fuldeltavarphi}) to compute the quantum corrected two-point correlation function in Sec.~\ref{sec:twopointcorrelator}. In the rest of this section, throughout Secs.~\ref{subsec:H}-C, we use homogeneous background solution~(\ref{barfi}) to express the expansion rate $H(t)$, scale factor $a(t)$ and slow-roll parameter $\epsilon(t)$ in terms of the initial values $H_{i0}$ and $\epsilon_{i0}$.

\subsection{The expansion rate $H(t)$ in terms of initial values $H_{i0}$ and $\epsilon_{i0}$}
\label{subsec:H}
Employing Eq.~(\ref{barfi}) in Eq.~(\ref{H}) we obtain the Hubble parameter
\beeq
H(t)\!=\!H_0(t)\!\pm\!\lambda H_\lambda(t)\!+\!\mathcal{O}(\lambda^2)\; ,\label{Hzerolambda}
\eneq
as
\beeq
H(t)\!=\! H_i\!+\!\dot{H}_it\!+\!\frac{1}{2!}\ddot{H}_it^2\!+\!\frac{1}{3!}\dddot{H}_it^3\!+\!\mathcal{O}(\lambda^2)\; .\label{Htaylor}
\eneq
The initial values of the Hubble parameter $H_i$ and its time derivatives $\dot{H}_i$, $\ddot{H}_i$ and $\dddot{H}_i$ in Eq.~(\ref{Htaylor}) can be written in terms of the tree-order initial values
\be
&&\hspace{0cm}H_{i0}\!=\!\sqrt{\frac{8\pi G}{(D\!-\!2)(D\!-\!1)}}\,m\bar{\varphi}_i\; ,\label{initialH}\\
&&\hspace{1.3cm}\dot{H}_{i0}\!=\!-\frac{m^2}{D\!-\!1}\; ,\label{initialdotH}
\ee
as
\be
&&\hspace{0cm}H_i\!=\!H_{i0}\!\!\left[1\!\!\pm\!\frac{\lambda}{4!}\Bigl(\frac{{\bar{\varphi}_i}}{m}\Bigr)^2\right]
\!\!+\!\mathcal{O}(\lambda^2)\!\equiv\!H_{i0}\!\pm\!\lambda H_{i\lambda}\!\!+\!\mathcal{O}(\lambda^2) \; ,\label{Hi}
\ee
\be
&&\hspace{0cm}\dot{H}_i\!=\!\dot{H}_{i0}\!\!\left[1\!\!\pm\!\frac{\lambda}{4}
\Bigl(\frac{{\bar{\varphi}_i}}{m}\Bigr)^2\right]\!\!+\!\mathcal{O}(\lambda^2)\!\equiv\!\dot{H}_{i0}\!\pm\!\lambda \dot{H}_{i\lambda}\!\!+\!\mathcal{O}(\lambda^2)\; ,\label{dotHi}\\
&&\hspace{0cm}\ddot{H}_i\!\!=\!\pm H_{i0}\frac{\lambda}{2}\frac{D\!-\!2}{(D\!-\!1)8\pi G}\!+\!\mathcal{O}(\lambda^2)\!\equiv\!\ddot{H}_{i0}\!\pm\!\lambda \ddot{H}_{i\lambda}\!\!+\!\mathcal{O}(\lambda^2)\; ,\label{ddotHi}\\
&&\hspace{0cm}\dddot{H}_i\!\!=\!\pm\dot{H}_{i0}\frac{\lambda}{4}\frac{D\!-\!2}{(D\!-\!1)8\pi G}\!+\!\mathcal{O}(\lambda^2)\!\equiv\!\dddot{H}_{i0}\!\pm\!\lambda \dddot{H}_{i\lambda}\!\!+\!\mathcal{O}(\lambda^2)\; .\label{dddotHi}
\ee
The ratio ${\bar{\varphi}_i}/{m}$ in Eqs.~(\ref{Hi}) and (\ref{dotHi}) can be expressed in terms of the initial values of the Hubble parameter ${H}_{i0}$ and the slow roll parameter
\beeq
\epsilon_{i0}\!=\!-\frac{\dot{H}_{i0}}{H^2_{i0}}\!=\!\frac{m^2}{(D\!-\!1)H^2_{i0}}\; ,\label{epsi0}
\eneq
where we used Eq.~(\ref{initialdotH}). Equations~(\ref{initialH}) and (\ref{epsi0}) yield
\beeq
\bar{\varphi}_i\!=\!\bar{\varphi}_{i0}\!=\!\sqrt{\xi}\epsilon^{-1}_{i0}m\!=\!\sqrt{\frac{D\!-\!\!2}{8\pi G \epsilon_{i0}}}\; .\label{phibolum}
\eneq
Here, we define a dimensionless number
\beeq
\xi\!\equiv\!\frac{D\!-\!2}{(D\!-\!1)8\pi G H^2_{i0}}\; . \label{GAM}
\eneq
Employing Eq.~(\ref{phibolum}) in Eqs.~(\ref{Hi}) and (\ref{dotHi}) yields the ratios of $H_{i\lambda}$ and $H_{i0}$ and of $\dot{H}_{i\lambda}$ and $\dot{H}_{i0}$, in terms $\xi$ and $\epsilon_{i0}$, as
\beeq
\frac{H_{i\lambda}}{H_{i0}}\!=\!\frac{1}{6}\frac{\dot{H}_{i\lambda}}{\dot{H}_{i0}}\!=\!\frac{1}{4!}\xi\epsilon^{-2}_{i0}
\;.\label{ratioHinitial}
\eneq
Equations (\ref{ddotHi}) and (\ref{dddotHi}) imply that the initial values $\ddot{H}_{i0}\!=\!0$, $\dddot{H}_{i0}\!=\!0$, $\ddot{H}_{i\lambda}\!\neq\!0$ and $\dddot{H}_{i\lambda}\!\neq\!0$.
In fact, the ratios\beeq
\frac{\dot{H}_{i\lambda}}{H_{i\lambda}}\!=\!-6\epsilon_{i0}H_{i0}\; ,\;\;\frac{\ddot{H}_{i\lambda}}{H_{i\lambda}}\!=\!12\epsilon^2_{i0}H^2_{i0}\;\; {\rm and}\;\;\frac{\dddot{H}_{i\lambda}}{H_{i\lambda}}\!=\!-6\epsilon^3_{i0}H^3_{i0}\; .\label{coffHlambda}
\eneq
Equations (\ref{Hzerolambda})-(\ref{Htaylor}) and (\ref{ratioHinitial})-(\ref{coffHlambda}) imply that
\be
&&\hspace{3.3cm}H_0(t)\!=\!H_{i0}\Bigl[1\!\!+\!\frac{\dot{H}_{i0}}{{H}_{i0}}t\Bigr]
\!\!=\!H_{i0}\Bigl[1\!-\!\epsilon_{i0}{H}_{i0}t\Bigr]\;,\label{Hz0}\\
&&\hspace{-1.2cm}H_\lambda(t)\!=\!H_{i\lambda}\Bigl[\!1\!\!+\!\frac{\dot{H}_{i\lambda}}{H_{i\lambda}}t\!\!+\!\frac{1}{2}\frac{\ddot{H}_{i\lambda}}{H_{i\lambda}}t^2
\!\!+\!\frac{1}{3!}\frac{\dddot{H}_{i\lambda}}{H_{i\lambda}}t^3\Bigr]
\!\!=\!H_{i0}\frac{\xi}{4!}\epsilon^{-2}_{i0}
\Bigl[\!1\!\!-\!6\epsilon_{i0}{H}_{i0}t\!\!+\!6\!\left(\epsilon_{i0}{H}_{i0}t\right)^2
\!\!-\!\left(\epsilon_{i0}{H}_{i0}t\right)^3\Bigr].\label{Hlam}
\ee
Thus, combining Eqs.~(\ref{Hz0})-(\ref{Hlam}) in Eq.~(\ref{Hzerolambda}) yields the expansion rate as
\beeq
H(t)\!=\!H_{i0}\!\left[1\!\!-\!\!\epsilon_{i0}H_{i0}t\!\!\pm\!\frac{\lambda\xi}{4!}
\Bigl[\epsilon^{-2}_{i0}\!\!-\!6\,\epsilon^{-1}_{i0}H_{i0}t\!+\!6\left(H_{i0}t\right)^2
\!-\!\epsilon_{i0}\left(H_{i0}t\right)^3\Bigr]\!\!+\!\mathcal{O}(\lambda^2)\right]\; .\label{Hwhole}
\eneq
Among the $\mathcal{O}(\lambda)$ terms, the first two ones dominate as long as $0\!<\!\!\epsilon_{i0}\!\!<\!\!\frac{1}{2(H_{i0}t)}[\sqrt{{5}/{3}}\!-\!1]$. This means that for a range of $0\!\!<\!\!\epsilon_{i0}\!\!<\!\!0.0025$ they remain as leading terms during the first $60$ e-foldings. For a larger $\epsilon_{i0}$ the third term becomes of order the sum of the first two terms during the last epochs of the inflation. The forth term (the cubic one), however, is suppressed during inflation for an $\epsilon_{i0}\!<\!6/(H_{i0}t)$. Thus for $\epsilon_{i0}\!\ll\!0.1$---satisfied in the slow-roll regime---the forth term may be neglected during the first $60$ e-foldings. Thus, to express time $t$ in terms of the scale factor $a$, which provides the essential change of variable to evaluate the integrals of the one-loop correlator in Sec.~\ref{subsect:1loopcorr}, we neglect the fourth term as a {\it zeroth-order approximation} but we do {\it include} it in our {\it first iteration}. Thus, integrating Eq.~(\ref{Hwhole}) we find the cubic polynomial
\beeq
\left(H_{i0}t\right)^3\!\!\mp\!\frac{6}{\lambda\xi}\epsilon_{i0}
\Bigl[1\!\!\pm\!\frac{\lambda\xi}{4}\epsilon^{-2}_{i0}\Bigr]\!\!\left(H_{i0}t\right)^2\!\!\pm\!\frac{12}{\lambda\xi}
\Bigl[1\!\!\pm\!\frac{\lambda\xi}{4!}\epsilon^{-2}_{i0}\Bigr]\!\!\left(H_{i0}t\right)
\!\mp\!\frac{12}{\lambda\xi}\ln(a)\cong0\;,\label{cubicpol}
\eneq
which has three distinct real roots; see Appendix~\ref{App:comovt}. We define a new variable
\beeq
q\!\equiv\!\sqrt{1\!\!-\!\!2\epsilon_{i0}\!\ln(a)}\; ,\label{Q}
\eneq
and express the relevant root of polynomial~(\ref{cubicpol}) in terms of $q$ in Eq.~(\ref{comovtimecube}) which implies that the comoving time
\beeq
t\!\cong\!\epsilon^{-1}_{i0}\frac{1\!\!-\!q}{H_{i0}}\!\!
\left[1\!\!\pm\!\frac{\lambda\xi}{4!}\epsilon^{-2}_{i0}\Bigl[1\!\!-\!\!2q\Bigr]
\!\!+\!\mathcal{O}(\lambda^2)\right]\; .\label{time}
\eneq
Using Eq.~(\ref{time}) in Eq.~(\ref{Hwhole})---without neglecting the fourth term---yields the expansion rate~as
\be
\hspace{0.7cm}H(t)\!=\!H_{i0}\!\left[q\!\pm\!\frac{\lambda}{4!}\xi\epsilon^{-2}_{i0}
\Bigl[q^3\!\!+\!q^2\!\!-\!1\Bigr]\!\!+\!\mathcal{O}(\lambda^2)\right]
\; .\label{HQ}
\ee
Hence, Eq.~(\ref{HQ}) implies
\beeq
H_0(t)\!=\!H_{i0}q_0(t) \;,\label{HzeroQ}
\eneq
where \beeq
q_0\!=\!\sqrt{1\!\!-\!\!2\epsilon_{i0}\!\ln(a_0)}\; .\label{qsifir}
\eneq
Therefore,
\beeq
\dot{H}_0(t)\!=\!H_{i0}\dot{q}_0(t)  \;.\label{dotHzeroQ}
\eneq
Note that the time derivative of Eq.~(\ref{Q}) yields
\beeq
\dot{q}(t)\!=\!-\epsilon_{i0}\frac{H(t)}{q(t)}\; ,\label{dotQ}
\eneq
which implies
\beeq
\dot{q}_0(t)\!=\!-\epsilon_{i0}\frac{H_0(t)}{q_0(t)}\!=\!-\epsilon_{i0}H_{i0}\!\!=\!{\rm const.} \; .\label{dotQnew}
\eneq
Using Eq.~(\ref{dotQnew}) in Eq.~(\ref{dotHzeroQ}) yields\beeq
\dot{H}_0(t)\!=\!-\epsilon_{i0}H^2_{i0}\!\!=\!\dot{H}_{i0}\!\!=\!{\rm const.} \; ,\label{dotHzero}
\eneq
in agreement with Eq.~(\ref{epsi0}). Expansion rate $H(t)$, by definition, is the time derivative of $\ln(a(t))$. Therefore, in the next section, the scale factor $a(t)$ is obtained integrating the $H(t)$ and then exponentiating the result.

\subsection{The scale factor $a(t)$ in terms of initial values $H_{i0}$ and $\epsilon_{i0}$}
\label{subsec:a}
The logarithm of scale factor \beeq
\ln(a(t))\!=\!\ln(a_0(t))\!\pm\!\!\lambda\ln(a(t))_\lambda\!\!+\!\mathcal{O}(\lambda^2)\;,\label{loga}
\eneq
is obtained by integrating expansion rate~(\ref{HQ}) via making change of variable~(\ref{time}) which implies
\beeq
dt\!=\!-\frac{dq}{H_{i0}}\epsilon^{-1}_{i0}\!\!\left[1\!\!\pm\!\frac{\lambda}{8}\xi\epsilon^{-2}_{i0}
\Bigl[1\!-\!\frac{4}{3}q\Bigr]\!\!+\!\mathcal{O}(\lambda^2)\right]\; .
\eneq
The result of the integral yields
\beeq
\ln(a_0(t))\!=\!\epsilon^{-1}_{i0}\frac{\mathcal{E}_0}{2}\; ,\label{lna0t}
\eneq
where we define
\beeq
\mathcal{E}_0(t)\!\equiv\!1\!-\!q^2_0(t)\!=\!2\epsilon_{i0}\ln(a_0(t))\; ,
\eneq
and
\beeq
\ln(a(t))_\lambda\!\!=\!-\frac{
\xi}{4!}\epsilon^{-3}_{i0}\frac{\left[1\!\!-\!q_0\right]^4}{4}\;.\label{lna0l}
\eneq
Exponentiating Eq.~(\ref{loga}), using Eqs.~(\ref{lna0t}) and (\ref{lna0l}), gives the scale factor as
\beeq
a(t)\!=\!a_0(t)\exp\!\left(\!\mp\frac{\lambda}{4!}\frac{
\xi}{4}\epsilon^{-3}_{i0}\!\left[1\!\!-\!q_0\right]^4\!\!+\!\mathcal{O}(\lambda^2)\!\right)
\!\!=\!a_0(t)\!\left[1\!\!\mp\!\frac{\lambda}{4!}\frac{
\xi}{4}\epsilon^{-3}_{i0}\!\left[1\!\!-\!q_0\right]^4\!\!+\!\mathcal{O}(\lambda^2)\right]\; ,\label{sclfct}
\eneq
where
\beeq
a_0(t)\!=\!\exp\Bigl(\!H_{i0}t_0\Bigl[1\!-\!\frac{\epsilon_{i0}}{2}H_{i0}t_0\Bigr]\Bigr)
\!=\exp\Bigl(\epsilon^{-1}_{i0}\frac{\mathcal{E}_0}{2}\Bigr)\; .\label{a0t}
\eneq
To get Eq.~(\ref{a0t}), we used Eq.~(\ref{Hz0}) in the first equality and Eq.~(\ref{lna0t}) in the second.

Scale factor~(\ref{sclfct}) reaches its maximum value at time
\beeq
t_{\rm m}\!\cong\!(\epsilon_{i0}H_{i0})^{-1}\; ,
\eneq
where $\dot{a}\!=\!0$.  Inflation, on the other hand, continues as long as $\ddot{a}\!>\!0$. This implies that inflation ends at
\beeq
t_{\rm e}\!\cong\!t_{\rm m}[1\!-\!\sqrt{\epsilon_{i0}}\,]\!=\!t_{\rm m}\!\!-\!(\sqrt{\epsilon_{i0}}H_{i0})^{-1}\; .\label{tend}
\eneq
The e-folding condition $a(t_{\rm e})\!\gtrsim\!e^{60}$ and the fact that $q\!>\!0$ constrain the $\epsilon_{i0}$. They imply that $\epsilon_{i0}\!<\!0.0083$. There is, however, a more stringent constraint on the $\epsilon_{i0}$ as we shall obtain in Sec.~\ref{sec:power}. For the spectral index not to be singular, a slow-roll parameter with an initial value $\epsilon_{i0}\!<\!0.0041$ is required. Hence, $m\!<\!0.111H_{i0}$ in $D\!=\!4$ dimensions.

Note that Eqs.~(\ref{loga}), (\ref{lna0t}) and (\ref{lna0l}) also yield
\beeq
\hspace{0.4cm} q^2\!=\!1\!\!-\!\!2\epsilon_{i0}\ln(a)\!=\!q^2_0\!\left[1\!\!\pm\!\frac{\lambda}{4!}\frac{
\xi}{2}\epsilon^{-2}_{i0}q^{-2}_0\left[1\!\!-\!q_0\right]^4\!\!\!+\!\mathcal{O}(\lambda^2)\right]\;.\label{qsqrd}
\eneq
Hence, using Eq.~(\ref{qsqrd}), Eq.~(\ref{HQ}) can be recast as
\be
\hspace{0.9cm}H(t)\!=\!H_{i0}\!\left[q_0\!\pm\!\frac{\lambda}{4!}\xi\epsilon^{-2}_{i0}
\Bigl[q^3_0\!\!+\!q^2_0\!\!-\!1\!\!+\!q^{-1}_0\frac{\left[1\!\!-\!q_0\right]^4}{4}\Bigr]\!\!+\!\mathcal{O}(\lambda^2)\right]
\; ,
\ee
which implies that Hubble parameter~(\ref{Hzerolambda}) at $\mathcal{O}(\lambda)$ is\beeq
H_\lambda(t)\!=\!H_{i0}\frac{\xi}{4!}\epsilon^{-2}_{i0}
\Bigl[q^3_0\!\!+\!q^2_0\!\!-\!1\!\!+\!q^{-1}_0\frac{\left[1\!\!-\!q_0\right]^4}{4}\Bigr]\; .\label{HLq}
\eneq
In the next section, we similarly obtain the slow roll parameter, which characterizes the flatness of the potential that derives inflation.

\subsection{The slow-roll parameter $\epsilon(t)$ in terms of initial values $H_{i0}$ and $\epsilon_{i0}$}
\label{subsec:eps}

Slow-roll parameter~(\ref{epsilon}) can also be expressed in terms of the initial values $\epsilon_{i0}$ and $H_{i0}$. Note that change of variable~(\ref{Q}) and Eq.~(\ref{dotQ}) imply
\beeq
\frac{d}{dt}\!=\!-\epsilon_{i0}q^{-1}H\frac{d}{dq}\; .\label{chnvarder}
\eneq
Hence, using Eq.~(\ref{chnvarder}) in Eq.~(\ref{HQ}) we obtain
\beeq\epsilon\!\!=\!\!-\frac{\dot{H}}{H^2}\!=\!\frac{\epsilon_{i0}}{qH}\frac{dH}{dq}
\!=\!\epsilon_{i0}\!\left[q^{-2}\!\!\pm\!\!\frac{\lambda}{4!}\xi\epsilon^{-2}_{i0}q^{-3}\Bigl[2q^3\!\!+\!q^2\!\!+\!\!1\Bigr]\right]
\!\!+\!\mathcal{O}(\lambda^2)\!\equiv\!\epsilon_0\!\pm\!\lambda\epsilon_\lambda\!\!+\!\mathcal{O}(\lambda^2) \;.\label{EPS}\eneq
Therefore,
\beeq
\epsilon_0\!=\!-\frac{\dot{H}_0}{H_0^2}\!=\!\epsilon_{i0}q^{-2}_0\!=\!\frac{m^2q^{-2}_0}{(D\!-\!1)H^2_{i0}}
\!=\!\frac{m^2}{(D\!-\!1)H^2_0}\;,\label{e0}
\eneq
where we used Eqs.~(\ref{epsi0}) and (\ref{HzeroQ}), and
\beeq
\epsilon_\lambda\!=\!-\frac{\dot{H}_\lambda}{H_0^2}\!-\!2\epsilon_0\frac{H_\lambda}{H_0}
\!=\!\frac{\xi}{4!}\epsilon^{-1}_{i0}q^{-3}_0\Bigl[2q^3_0\!+\!q^{2}_0\!+\!1\!-\!q^{-1}_0
\frac{\left[1\!\!-\!q_0\right]^4}{2}\Bigr]\; ,\label{el}
\eneq
where we used Eqs.~(\ref{qsqrd}) and (\ref{EPS}). This completes the discussion of geometry in the interacting theory. Next, we obtain the solution of the background field $\bar\varphi(t)$ up to $\mathcal{O}(\lambda^2)$.

\subsection{The background inflaton field $\bar\varphi(t)$ in terms of initial values $H_{i0}$ and $\epsilon_{i0}$}
\label{subsec:backgrndinfl}

The background inflaton field $\bar\varphi(t)$ is obtained in Eq.~(\ref{barfi}). Employing Eqs.~(\ref{phibolum}) and (\ref{GAM}) in Eq.~(\ref{barfi}) yields
\be
\bar\varphi(t)\!\simeq\!\bar\varphi_{i0}\!\!\left[1\!-\!\epsilon_{i0}H_{i0}t
\!\mp\!\frac{\lambda}{8}\xi\Bigl[\epsilon^{-1}_{i0}\!\left(H_{i0}t\right)\!-\!\left(H_{i0}t\right)^2\Bigr]
\!\!+\!\mathcal{O}(\lambda^2)\right]\; .\label{barPH}
\ee
Using Eq.~(\ref{time}) in Eq.~(\ref{barPH}) gives
\be
\bar\varphi(t)\!\simeq\!\bar\varphi_{i0}\!\!\left[q\!\mp\!\frac{\lambda}{4!}\xi\epsilon^{-2}_{i0}\Bigl[\!1\!\!-\!q^2\Bigr]
\!\!+\!\mathcal{O}(\lambda^2)\right]\!\!=\!\bar\varphi_{i0}\!\!\left[q_0\!\mp\!\frac{\lambda}{4!}\xi\epsilon^{-2}_{i0}\Bigl[
\mathcal{E}_{0}\!-\!q_0^{-1}\frac{\left[1\!\!-\!q_0\right]^4}{4}\Bigr]\!\!+\!\mathcal{O}(\lambda^2)\right]\; .\label{barbarvar}
\ee
Recall that the initial value $\bar\varphi_{i0}$ is given terms of $\epsilon_{i0}$ in Eq.~(\ref{phibolum}).

Results we obtained throughout Secs.~\ref{subsec:H}-D are needed to study the fluctuations of the self-interacting inflaton to which the remainder of this paper is devoted.

\section{Inflaton Fluctuations}
\label{sec:InfFluct}

Quantum nature of the universe imply production of virtual pair of particles out of vacuum and causes fluctuations in the field. These quantum fluctuations are enhanced during inflation. As a particular fluctuation mode exits the horizon, it becomes acausal and ''freezes-in.'' After the end of inflation as the horizon extends and includes these modes they become causal and provide the origin of large scale structure in the universe. To compute the effects of a quartic self-interaction on the correlations (Secs.~\ref{sec:twopointcorrelator}-\ref{sec:coincorr}) and power spectrum (Sec.~\ref{sec:power}) of inflaton fluctuations, we first obtain the mode expansion for the fluctuation field of the inflaton at tree-order in Sec.~\ref{subsect:freethry} and then use it to get the $\mathcal{O}(\lambda)$ correction to the fluctuation field at one-loop order in Sec.~\ref{subsect:interactingtheory}.

\subsection{Fluctuations of the tree-order inflaton field}
\label{subsect:freethry}
Fluctuation field $\delta\varphi(x)$ of the full inflaton satisfies Eq.~(\ref{deltavarphifull}) whose solution is given in Eq.~(\ref{fuldeltavarphi}) in terms of the fluctuation field $\delta\varphi_0(x)$ of the tree-order inflaton and the retarded Green's function $G_{\rm ret}(x;x')$.
The mode expansion for the fluctuation field of the tree-order inflaton can be given as
\beeq
\delta\varphi_0(x)\!=\!\!\int\!\!\frac{d^{D-1}k}{(2\pi)^{D-1}}
\Bigl\{\delta\Phi_0(x,\vec{k}) A(\vec{k})\!+\!
\delta\Phi_0^*(x,\vec{k}) A^*(\vec{k})
\Bigr\} \; ,\label{phizero} \eneq
where the mode function of the fluctuation field
\beeq
\delta{\Phi}_0(x,\vec{k})\!\equiv\!u_0(t,k) e^{i \vec{k} \cdot \vec{x}}\; ,\label{fmodefl}\eneq
is the spatial plane wave solution of the linearized effective field equation, evaluated at the full solution of the background effective field equation. The tree-order amplitude function $u_0(t,k)$ of the plane waves that are  superposed in Eq.~(\ref{phizero}), therefore, obeys\beeq
\ddot{u}_0(t,k)\!+\!(D\!-\!1) H(t) \dot{u}_0(t,k)\!+\!\!
\left[\frac{k^2}{a^2}\!+\!m^2\right]\!u_0(t,k)\!=\!0 \; .\label{Fouriertransfieldeq}
\eneq
The mode coordinates $A(\vec{k})$ and $A^*(\vec{k})$ in Eq.~(\ref{phizero}) are functions of the spatial Fourier transforms of the $\delta\varphi_0(x)$ and its first time derivative evaluated on the initial value surface. Canonical quantization promotes the mode coordinates to the annihilation and creation operators $\hat{A}(\vec{k})$ and $\hat{A}^\dagger(\vec{k})$ satisfying the usual commutation relations
\beeq
\left[\hat{A}(\vec{k}),
\hat{A}(\vec{k}\,')\right]\!=\!0\;,\left[\hat{A}^\dagger(\vec{k}),
\hat{A}^\dagger(\vec{k}\,')\right]\!=\!0\;,\left[\hat{A}(\vec{k}),
\hat{A}^\dagger(\vec{k}\,')\right]\!=\!(2\pi)^{D-1}\delta^{D-1}(\vec{k}\!-\!\vec{k}\,')\;
,\label{anncre}
\eneq
and imposes the Wronskian normalization condition on the amplitude function,
\beeq
u_0(t,k)\,\dot{u}_0^*(t,k)\!-\!u_0^*(t,k)\,\dot{u}_0(t,k)\!=\!\frac{i}{a^{D-1}}\; .\label{Wronskiannonzero}
\eneq
Rescaling the amplitude function as $a^{\frac{D-1}{2}}(t)\,u_0(t,k)\!\!\equiv\!\!v_0(t,k)$ reduces differential equation~(\ref{Fouriertransfieldeq}) to the form of the harmonic oscillator equation with a time dependent frequency
\be
\ddot{v}_0(t,k)\!+\!\!
\left[\frac{k^2}{a^2(t)}\!+\!m^2\!-\!\frac{D\!-\!\!1}{2}\dot{H}(t)\!-\!\Big(\frac{D\!-\!\!1}{2}H(t)\Big)^2\right]
\!v_0(t,k)\!=\!0 \; .\label{eqmotPhi}
\ee
To solve Eq.~(\ref{eqmotPhi}), following the approach by Finelli, Marozzi, Vacca, and Venturi given in Ref.~\cite{FinMarVacVen}, we make the ansatz \beeq
v_0(t,k)\!\equiv\!\zeta^\mu Z_\nu(\chi\zeta)\qquad {\rm with}\qquad \zeta\!=\!\frac{k}{Ha}, \; \mu\!=\!\mu(H), \;\nu\!=\!\nu(H), \; {\rm and} \;\chi\!=\!\chi(H).
\eneq
and express the time derivative as derivatives with respect to $\zeta$ and $H$,
\be
\frac{\partial}{\partial t}\!=\!-H\!\!\left[(1\!-\!\epsilon)\,\zeta\frac{\partial}{\partial \zeta}\!+\!\epsilon H\!\frac{\partial}{\partial H}\right]\; .\label{timeder}
\ee
Employing Eq.~(\ref{timeder}) in Eq.~(\ref{eqmotPhi}), neglecting the terms of $\mathcal{O}(\ddot{H})$ and $\mathcal{O}(\epsilon^2)$, yields
\beeq
\zeta^2\frac{\partial^2Z_\nu}{\partial \zeta^2}\!+\!\zeta\frac{\partial Z_\nu}{\partial \zeta}\!+\!\left(\chi^2\zeta^2\!-\!\nu^2\right)\!Z_\nu\!=\!0\; ,\label{bessel}
\eneq
with parameters \beeq
\mu\!=\!-\frac{\epsilon}{2}\!+\!\mathcal{O}(\epsilon^2)\; ,\;\;\chi\!=\!1\!\!+\!\epsilon\!\!+\!\mathcal{O}(\epsilon^2)\; ,
\eneq and
\beeq \nu^2\!=\!\Big(\frac{D\!-\!\!1}{2}\Big)^2\!\!\!-\!\frac{m^2}{H^2}\!+\!\frac{(D\!-\!\!2)(D\!-\!\!1)}{2}
\,\epsilon\!+\!\mathcal{O}(\epsilon^2)\; .\label{primitnu}
\eneq
Using Eqs.~(\ref{epsi0}) and~(\ref{HQ}) implies
\beeq
\frac{m^2}{H^2}\!=\!(D\!-\!1)\,\epsilon\!+\!\mathcal{O}(\lambda\epsilon_\lambda)\; ,
\eneq
and hence
\beeq
\nu^2\!=\!\Big(\frac{D\!-\!\!1}{2}\Big)^2\!\left[1\!\!+\!2\,\frac{D\!-\!4}{D\!-\!1}\,\epsilon\!+\!\mathcal{O}(\epsilon^2)\right]\; .\label{nu}
\eneq
The solution of Bessel's equation~(\ref{bessel}) is given in terms of Hankel's function of the first kind and its complex conjugate \beeq
Z_\nu(\chi\zeta)\!=\!C_1 H^{(1)}_\nu(\chi\zeta)\!+\!C_2 H^{(2)}_\nu(\chi\zeta)\; .
\eneq
Thus, the solution of Eq.~(\ref{Fouriertransfieldeq}) can be given as\beeq
u_0(t,k)\!=\!a^{-\frac{D-1}{2}}\zeta^\mu Z_\nu(\chi\zeta)\!=\!a^{-\frac{D-1}{2}}
\!\Big(\!\frac{k}{Ha}\Big)^{-\frac{\epsilon}{2}}\Big[C_1 H^{(1)}_\nu\!\Big(\!\frac{(1\!\!+\!\epsilon)k}{Ha}\Big)\!\!+\!C_2 H^{(2)}_\nu\!\Big(\!\frac{(1\!\!+\!\epsilon)k}{Ha}\Big)\Big]\;.\label{modesoln}
\eneq
Wronskian normalization~(\ref{Wronskiannonzero}) constrains the constants $C_1$ and $C_2$,
\beeq
|C_1|^2\!-\!|C_2|^2\!=\!\frac{\pi}{4H}(1\!\!+\!\epsilon)\,\Big(\!\frac{k}{Ha}\Big)^{\epsilon}
\Bigl[1\!\!+\!\mathcal{O}(\epsilon^2)\Bigr]\; .
\eneq
A solution which corresponds to Bunch-Davies mode function in de Sitter spacetime can be chosen, up to $\mathcal{O}(\epsilon^2)$, as\beeq
C_1\!\!=i\!\left[\frac{\pi}{4H}(1\!\!+\!\epsilon)\Big(\!\frac{k}{Ha}\Big)^{\epsilon}\right]^{\frac{1}{2}}\qquad {\rm and} \qquad C_2\!\!=\!0\;.\label{c1c2}
\eneq
Inserting Eq.~(\ref{c1c2}) into Eq.~(\ref{modesoln}) yields the amplitude of mode function~(\ref{fmodefl}) up to $\mathcal{O}(\epsilon^2)$,
\beeq
u_0(t, k)\!=\!ia^{-\frac{D-1}{2}}\!\left[\frac{\pi}{4H}(1\!\!+\!\epsilon)\right]^{\frac{1}{2}}\!\! H^{(1)}_\nu\!\Big(\!\frac{(1\!\!+\!\epsilon)k}{Ha}\Big)\; .\label{modefncfluct}
\eneq
Recall that, at comoving time $t\!=\!t_k$, the mode with wave number
\beeq
k\!\equiv\!\frac{H(t_k)a(t_k)}{1\!\!+\!\epsilon(t_k)}\; ,\label{TK}
\eneq
exits the horizon, becomes a superhorizon (IR) mode and ''freezes in.'' At a given time $t$, we consider superhorizon modes \beeq k\!<\!\!\frac{Ha}{1\!\!+\!\epsilon}\; ,\eneq
for which the argument of the Hankel function $\frac{(1+\epsilon)k}{Ha}\!<\!1$. Thus,
using the power series expansion of the Hankel function we obtain the IR limit of the amplitude function in leading order
\beeq
u_0(t, k)\longrightarrow\frac{2^{\nu-1}\Gamma(\nu)}{\sqrt{\pi}a^{\frac{\delta}{2}}} \Bigl(\!\frac{H}{1\!\!+\!\epsilon}\Bigr)^{\nu-\frac{1}{2}}k^{-\nu}\!\!\left[1\!\!+\!
\mathcal{O}\!\left(\!\Bigl(\frac{(1\!\!+\!\epsilon)k}{2Ha}\Bigr)^2\right)\!\right]\!
\Bigl[1\!\!+\!\mathcal{O}(\epsilon^2)\Bigr]\; ,\label{modeIR}
\eneq
where we define the parameter $\delta\!\equiv\!D\!\!-\!\!1\!\!-\!\!2\nu$.
Employing Eq.~(\ref{modeIR}) in Eq.~(\ref{phizero}) we obtain the IR truncated fluctuation field of the tree-order inflaton as
\beeq
\delta\bar{\varphi_0}(t, \vec{x})\!=\!\frac{2^{\nu-1}\Gamma(\nu)}{\sqrt{\pi}\,a^{\frac{\delta}{2}}} \Bigl(\!\frac{H}{1\!\!+\!\epsilon}\Bigr)^{\nu-\frac{1}{2}}\!\!\!\int\!\!\frac{d^{D-1}k}{(2\pi)^{D-1}}
\frac{\Theta\left(\frac{Ha}{1+\epsilon}\!-\!k\right)}{k^\nu}
\Bigl[e^{i \vec{k} \cdot \vec{x}} \hat{A}(\vec{k})\!+\!
e^{-i\vec{k} \cdot \vec{x}} \hat{A}^{\dagger}(\vec{k})
\Bigr] \; .\label{varphizero} \eneq
The ratio $\frac{H(t)\, a(t)}{1+\epsilon(t)}$ in the argument of the $\Theta$-function  is monotonically increasing function of time until \beeq
t_{\rm f}\!\cong\!t_{\rm m}
[1\!-\!3^{{1}/{4}}\sqrt{\epsilon_{i0}}]\; ,
\eneq
which is slightly less than $t_e$ defined in Eq.~(\ref{tend}). Hence, we consider inflation during $0\!<\!t\!<\!t_{\rm f}$ in the model we consider. Mode expansion~(\ref{varphizero}) is used to obtain the full fluctuation field in the next section.

\subsection{Fluctuations of the full inflaton field}
\label{subsect:interactingtheory}

Fluctuations $\delta{\varphi}(x)$ of a self-interacting inflaton field is given in Eq.~(\ref{fuldeltavarphi}) in terms of the fluctuations $\delta\varphi_0(x)$ of the tree-order field and the retarded Green's function $G(x;x')$. The $\delta\varphi_0(x)$ is given via Eqs.~(\ref{phizero}), (\ref{fmodefl}) and (\ref{modefncfluct}). [The IR limit $\delta\bar{\varphi}_0(x)$ of the free field is obtained in Eqs.~(\ref{modeIR}) and (\ref{varphizero}).] To get the full fluctuations $\delta{\varphi}(x)$ and its IR limit, the last ingredient we need is the Green's function which can be given in terms of the commutator function of the $\delta\varphi_0(x)$ as\beeq
G(x;x')\!=\!i\Theta(t\!-\!t')\left[\delta\varphi_0(x), \delta\varphi_0(x')\right]\; .\label{commutatorgreen}
\eneq
The commutator function
\beeq
\hspace{-0.7cm}\left[\delta\varphi_0(x), \delta\varphi_0(x')\right]\!=\!\!\int\!\!\frac{d^{D-1}k}{(2\pi)^{D-1}}\Bigl\{ u(t,k) u^*(t'\!,k)
\!-\!u^*(t,k) u(t'\!,k) \Bigr\}e^{i \vec{k} \cdot (\vec{x}-\vec{x}'\!)}\;,
\eneq
where the terms inside the curly brackets
\be
&&\hspace{1.5cm}u(t,k) u^*(t'\!,k)
\!-\!u^*(t,k) u(t'\!,k)\!=\!i\frac{\pi}{2}[\,aa']^{-\frac{D-1}{2}}\!\!
\left[\!\frac{(1\!\!+\!\epsilon)(1\!\!+\!\epsilon')}{HH'}\right]^\frac{1}{2}\nonumber\\
&&\hspace{-1.2cm}\times\Bigg\{\!\Big[\!\cot\!\left(\nu\pi\right)\!-\!\cot\!\left(\nu'\pi\right)\!\Big]
{J}_{\nu}(\mathfrak{z})\,{J}_{\nu'}(\mathfrak{z}')\!+\!\csc\!\left(\nu'\pi\right){J}_{\nu}(\mathfrak{z})\,{J}_{-\nu'}(\mathfrak{z}')
\!-\!\csc\!\left(\nu\pi\right){J}_{-\nu}(\mathfrak{z})\,{J}_{-\nu'}(\mathfrak{z}')\Bigg\}  \; .\label{commutfunc}
\ee
Here, we define $\mathfrak{z}\!\equiv\!\chi\zeta\!=\!\frac{(1\!+\!\epsilon)k}{Ha}$ and $\mathfrak{z}'\!\equiv\!\chi'\zeta'\!=\!\frac{(1\!+\!\epsilon')k}{H'a'}$. During slow roll inflation $\nu\!=\!\frac{D-1}{2}\!+\!\mathcal{O}(\epsilon)\!=\!\nu'$ in $D$-dimensions. [In $D\!=\!4$, on the other hand, we have $\nu\!=\!\frac{3}{2}\!+\!\mathcal{O}(\epsilon^2)\!=\!\nu'$, see Eq.~(\ref{nu}).] Therefore, the IR limit of Green's function~(\ref{commutatorgreen}), in leading order, is\be
\hspace{-0.3cm}G(x;x')\!\rightarrow\!\frac{\Theta(t\!\!-\!t')}{D\!-\!1}\!\left[\frac{H}{1\!\!+\!\epsilon}\right]^{\frac{D}{2}-1}
\!\!\left[\frac{H'}{1\!\!+\!\epsilon'}\right]^{\frac{D}{2}-1}\!\!
\Bigg\{\!\!\left[\frac{1\!\!+\!\epsilon'}{H'a'}\right]^{D-1}
\!\!\!-\!\!\left[\frac{1\!\!+\!\epsilon}{Ha}\right]^{D-1}\!\Bigg\}\delta^{D-1}(\vec{x}\!-\!\vec{x}')\; .\label{IRgreen}
\ee
The fluctuation field of the full inflaton, in the IR limit, can be obtained mingling Eqs.~(\ref{Vpertdeltavarphi}), (\ref{fuldeltavarphi}) and (\ref{IRgreen}): When the source term in Eq.~(\ref{deltavarphifull}) weighted by the Green's function is integrated throughout the spacetime from the initial time $t_i=0$ to a late time $t$ yields the $\delta\varphi$. In leading order, therefore, the latter term in the curly brackets in Eq.~(\ref{IRgreen}) can be neglected next to the former which dominates throughout the range of integration. Hence, \be
&&\delta\varphi(x)\!\simeq\!\delta\varphi_0(x)\!-\!\frac{1}{D\!-\!1}\!\left[\frac{H}{1\!\!+\!\epsilon}\right]^{\frac{D}{2}-1}\!\!\!\!\int_0^t\!\!dt'
\Theta(t\!-\!t')\!\left[\frac{1\!\!+\!\epsilon'}{H'}\right]^\frac{D}{2}
\frac{\frac{\partial V_{\rm pert}(\delta\varphi)}{\partial\delta\varphi}(t',\vec{x})}{1\!\!+\!\delta Z}\; .\label{IRdeltavar}
\ee
In the above integrand, the counterterms in $\frac{1}{1\!+\!\delta Z}\frac{\partial V_{\rm pert}(\delta\varphi)}{\partial\delta\varphi}$ can be neglected. This can be seen by comparing the orders of $\lambda$ of the counterterms and the powers of the fields $\bar\varphi(t)$ and $\delta\varphi(x)$ involved in various terms in Eq.~(\ref{Vpertdeltavarphi}). The terms proportional to the $\delta\lambda$ and $\delta Z$ can be neglected because they are of $\mathcal{O}(\lambda^2)$. [$\delta m^2\!\!\sim\!\mathcal{O}(\lambda)$.] Because $\bar\varphi\!\!\gg\!\!|\delta\varphi|$, the term quadratic in background field is the dominant term which yields the leading contribution.  Recall that $4\!\!\lesssim\!\!\bar\varphi\!\!<\!\!10^6$ in Planckian units. Thus, the fact that $|\delta\varphi|\!\ll\!\bar\varphi$ does not necessarily imply that $|\delta\varphi|$ is small by itself. Hence, we keep the term quadratic in $\delta\varphi$---that is also linear in the background field---and the term cubic in $\delta\varphi$, but neglect the term linear in $\delta\varphi$ next to the dominant term. Therefore, the fluctuation field of the full inflaton in our model is\be
&&\hspace{-0.4cm}\delta\varphi(x)\!\simeq\!\delta\varphi_0(x)
\!\mp\!\frac{\lambda}{6(D\!-\!\!1)}\!\left[\frac{H_0}{1\!\!+\!\epsilon_0}\right]^{\frac{D}{2}-1}\!\!\!\!\int_0^t\!\!dt'
\Theta(t\!-\!t')\!\!
\left[\frac{1\!\!+\!\epsilon'_0}{H'_0}\right]^\frac{D}{2}\!\!
\Bigg\{3\bar\varphi_0^2(t')\delta\varphi_0(t'\!, \vec{x})\nonumber\\
&&\hspace{2.7cm}+3\bar\varphi_0(t')\Bigl[\delta\varphi_0(t'\!, \vec{x})\Bigr]^2
\!\!\!\!+\!\Bigl[\delta\varphi_0(t'\!,\vec{x})\Bigr]^3\Bigg\}\!\!+\!\mathcal{O}(\lambda^2)\;.\label{fullfieldexp}
\ee
We employ Eq.~(\ref{fullfieldexp}) to compute quantum corrected two-point correlation function of the inflaton fluctuations
at tree- and one-loop order in the next section.

\section{Two-point correlation function of the fluctuations}
\label{sec:twopointcorrelator}

The two-point correlation function of the IR truncated fluctuations of the full field for two distinct events
\be
\hspace{-0.8cm}\langle\Omega|
\delta\bar{\varphi}(x)\delta\bar{\varphi}(x')|\Omega\rangle\!\!\!&=&\!\!\!\langle\Omega|
\delta\bar{\varphi}(x)\delta\bar{\varphi}(x')|\Omega\rangle_{\rm tree}\!\!+\!\!\langle\Omega|
\delta\bar{\varphi}(x)\delta\bar{\varphi}(x')|\Omega\rangle_{\rm 1-loop}\!\!+\!\mathcal{O}(\lambda^2) ,\label{tpcf}
\ee
with $t'\!\leq\!t$ and $\vec{x}\,'\!\!\neq\!\vec{x}$, can be obtained for the inflaton with a quartic self-interaction using Eq.~(\ref{fullfieldexp}). It yields, at tree-order
\beeq
\langle\Omega|\delta\bar{\varphi}(x)\delta\bar{\varphi}(x')|\Omega\rangle_{\rm tree}\!=\!\langle\Omega|
\delta\bar{\varphi}_0(x)\delta\bar{\varphi}_0(x')|\Omega\rangle\; ,
\eneq
and at one-loop order
\be
&&\hspace{4.7cm}\langle\Omega|
\delta\bar{\varphi}(x)\delta\bar{\varphi}(x')|\Omega\rangle_{\rm 1-loop}\nonumber\\
&&\hspace{-1cm}=\mp\frac{\lambda}{6(D\!-\!\!1)}\!
\Bigg\{\!\!\left[\frac{H'_0}{1\!\!+\!\epsilon'_0}\right]^{\frac{D}{2}-1}\!\!\!\int_0^{t'}\!\!\!d\tilde{t}\!
\left[\frac{1\!\!+\!\tilde{\epsilon}_0}{\tilde{H}_0}\right]^\frac{D}{2}\!\!\!\langle\Omega|
\delta\bar{\varphi}_0(x)
\Bigl[3\bar\varphi_0^2(\tilde{t})\delta\bar{\varphi}_0(\tilde{t},\vec{x}\,')\!+\![\delta\bar{\varphi}_0(\tilde{t},\vec{x}\,')]^3\Bigr]
|\Omega\rangle\nonumber\\
&&\hspace{0cm}+\!\left[\frac{H_0}{1\!\!+\!\epsilon_0}\right]^{\frac{D}{2}-1}\!\!\!\!
\int_0^{t}\!\!dt''\Bigl[\frac{1\!\!+\!\epsilon_0''}{H_0''}\Bigr]^\frac{D}{2}\!\langle\Omega|
\Bigl[3\bar\varphi_0^2(t'')\delta\bar{\varphi}_0(t''\!,\vec{x})\!+\![\delta\bar{\varphi}_0(t''\!,\vec{x})]^3\Bigr]
\delta\bar{\varphi}_0(x')|\Omega\rangle\Bigg\}\; ,\label{1loopcorr}
\ee
which are computed in Secs.~\ref{subsect:treecorr}-C.

\subsection{Tree-order correlator}
\label{subsect:treecorr}

The tree-order two-point correlation function of the IR truncated inflaton fluctuations is obtained
using Eqs.~(\ref{anncre}) and (\ref{varphizero}) as
\be
&&\hspace{3cm}\langle\Omega|
\delta\bar{\varphi}(x)\delta\bar{\varphi}(x')|\Omega\rangle_{\rm tree}\!=\!\langle\Omega|\delta\bar{\varphi}_0(x) \delta\bar{\varphi}_0(x')|\Omega\rangle\nonumber\\
&&\hspace{-1.3cm}=\!\!\frac{1}{4\pi} \Bigl(\!\frac{H}{1\!\!+\!\epsilon}\Bigr)^{\nu-\frac{1}{2}}\Bigl(\!\frac{H'}{1\!\!+\!\epsilon'}\Bigr)^{\nu'-\frac{1}{2}}
\frac{2^{\nu}\Gamma(\nu)}{a^{\frac{\delta}{2}}}\frac{2^{\nu'}\Gamma(\nu\,')}{a'^{\frac{\delta'}{2}}}
\!\!\int\!\!\frac{d^{D-1}k}{(2\pi)^{D-1}}\Theta\Bigl(\!\frac{Ha}{1\!\!+\!\epsilon}\!-\!k\Bigr)
\Theta\Bigl(\!\frac{H'a'}{1\!\!+\!\epsilon'}\!-\!k\Bigr)
\frac{e^{i \vec{k} \cdot (\vec{x}-\vec{x}')}}{k^{\nu+\nu'}}
 \; .\label{treeint}\ee
For $0\!\leq\!t'\!\leq\!t$, we have $\frac{H'a'}{1+\epsilon'}\!\leq\!\frac{Ha}{1+\epsilon}$. Thus, defining $\Delta x\!\!\equiv\!\|\Delta\vec{x}\|\!\!=\!\|\vec{x}\!-\!\vec{x}\,'\|$ and $\theta$ as the angle between the vectors $\vec{k}$ and $\Delta\vec{x}$, the integral on the right side of Eq.~(\ref{treeint}) yields
\beeq
\int\!\!\frac{d\Omega_{D-1}}{(2\pi)^{D-1}}\!\!\int_0^\infty\!\!\!dk\,
\Theta\Bigl(\!\frac{H'a'}{1\!\!+\!\epsilon'}\!-\!k\Bigr)\,\frac{e^{ik \Delta x \cos{(\theta)}}}{k^{1-\frac{\delta+\delta'}{2}}}\!=\!\frac{\Gamma\!\!\left(\frac{D}{2}\right)}
{\pi^{\frac{D}{2}}\Gamma(D\!-\!\!1)}(\Delta x^2)^{-\frac{\delta+\delta'}{4}}\!\!\int_{\alpha_i}^{\alpha'}\!\!d y\frac{\sin(y)}{y^{2-\frac{\delta+\delta'}{2}}}\; ,\label{intKandang}
\eneq
where we performed angular integrations, made a change of variable $y\!\equiv\!k\Delta x$ and defined\beeq \alpha\!\equiv\!\frac{Ha\Delta x}{1\!\!+\!\epsilon}\;,\label{defnalp}
\eneq
which implies $\alpha'\!=\!\frac{H'a'\Delta x}{1+\epsilon'}$ and $\alpha_i\!=\!\frac{H_i\Delta x}{1+\epsilon_i}$. The remaining integral can be evaluated in terms of the exponential integral function, defined in Eq.~(\ref{expintdefn}), as
\be
\int\!\!d y\frac{\sin(y)}{y^{2-\frac{\delta+\delta'}{2}}}\!=\!-\frac{i}{2}y^{\frac{\delta+\delta'}{2}-1}
\Bigl[E_{2-\frac{\delta+\delta'}{2}}\left(iy\right)
\!-\!E_{2-\frac{\delta+\delta'}{2}}\left(-iy\right)\Bigr]\, .\label{intEXPint}
\ee
Note that the result is real for a purely real parameter $y$. Combining Eqs.~(\ref{intKandang})-(\ref{intEXPint}) in Eq.~(\ref{treeint}) we find the tree-order correlator as
\be
&&\hspace{-0.3cm}\langle\Omega|\delta\bar\varphi_0(x) \delta\bar\varphi_0(x')|\Omega\rangle\!=\!
\frac{-i\Gamma\!\!\left(\frac{D}{2}\right)}{8\pi^{\frac{D}{2}+1}\Gamma(D\!-\!\!1)} \Bigl(\!\frac{H}{1\!\!+\!\epsilon}\Bigr)^{\frac{D}{2}-1}\!\Bigl(\!\frac{H'}{1\!\!+\!\epsilon'}\Bigr)^{\frac{D}{2}-1}
\frac{2^{\nu}\,\Gamma(\nu)}{\alpha^{\frac{\delta}{2}}}\frac{2^{\nu'}\Gamma(\nu\,')}
{\alpha'^{\frac{\delta'}{2}}}\nonumber\\
&&\hspace{-1.1cm}\times\!\Bigg\{\!\alpha'^{\frac{\delta+\delta\,'}{2}-1}\Bigl[E_{2-\frac{\delta+\delta\,'}{2}}(i\alpha')
\!-\!E_{2-\frac{\delta+\delta\,'}{2}}(-i\alpha')\Bigr]
\!\!-\!\alpha_i^{\frac{\delta+\delta\,'}{2}-1}\Bigl[E_{2-\frac{\delta+\delta\,'}{2}}(i\alpha_i)
\!-\!E_{2-\frac{\delta+\delta\,'}{2}}(-i\alpha_i)\Bigr]\!\Bigg\}\; .
\label{corrtreeord}\ee
One can also express the tree-order correlator in terms of the incomplete gamma function $\Gamma(\beta, z)$, defined in Eq.~(\ref{incompgammadefn}). The exponential integral function is related to the incomplete gamma function,
\beeq
E_\beta(z)\!=\!z^{\beta-1}\Gamma(1\!\!-\!\beta, z)\; .
\eneq
Therefore, tree-order correlator~(\ref{corrtreeord}) is recast as
\be
&&\hspace{-1.2cm}\langle\Omega|\delta\bar\varphi_0(x) \delta\bar\varphi_0(x')|\Omega\rangle\!=\!
\frac{\Gamma\!\!\left(\frac{D}{2}\right)}{8\pi^{\frac{D}{2}+1}\Gamma(D\!-\!\!1)} \Bigl(\!\frac{H}{1\!\!+\!\epsilon}\Bigr)^{\frac{D}{2}-1}\!\Bigl(\!\frac{H'}{1\!\!+\!\epsilon'}\Bigr)^{\frac{D}{2}-1}
\frac{2^{\nu}\,\Gamma(\nu)}{\alpha^{\frac{\delta}{2}}}\frac{2^{\nu'}\Gamma(\nu\,')}
{\alpha'^{\frac{\delta'}{2}}}(-1)^{-\frac{\delta+\delta'}{4}}\nonumber\\
&&\hspace{-1.25cm}\times\!\Bigg\{\!\Gamma\!\Bigl(\!\frac{\delta\!\!+\!\delta\,'}{2}\!-\!\!1,i\alpha'\Bigr)
\!\!-\!\Gamma\!\Bigl(\!\frac{\delta\!\!+\!\delta\,'}{2}\!-\!\!1,i\alpha_i\Bigr)
\!\!+\!(-1)^{-\frac{\delta+\delta'}{2}}\Bigl[\Gamma\!\Bigl(\!\frac{\delta\!\!+\!\delta\,'}{2}\!-\!\!1,\!-i\alpha'\Bigr)
\!\!-\!\!\Gamma\!\Bigl(\!\frac{\delta\!\!+\!\delta\,'}{2}\!-\!\!1,\!-i\alpha_i\Bigr)\!\Bigr]\!\Bigg\} .
\label{corrtreeordincomp}\ee
This form is useful in obtaining a series representation of the correlator using
expansion~(\ref{gammaaspowerseries}) of the incomplete gamma function.

A simpler form of tree-order correlator~(\ref{corrtreeord}) or (\ref{corrtreeordincomp}) is obtained recalling the fact that Eq.~(\ref{nu}) implies $\nu\!\!=\!(D\!-\!1)/2\!+\!\mathcal{O}(\epsilon)$ in $D$-dimensions and $\nu\!\!=\!3/2\!+\!\mathcal{O}(\epsilon^2)$ for $D\!\!=\!4$. The tree-order correlator for $\nu\!\!=\!(D\!-\!1)/2$ is given in Eq.~(\ref{Dcor}). Its $D\!\rightarrow\!4$ limit, up to $\mathcal{O}(\epsilon^2)$, is
\beeq
\hspace{-0.8cm}\langle\Omega|\delta\bar\varphi_0(x) \delta\bar\varphi_0(x')|\Omega\rangle\!=\!\frac{HH'}{4\pi^2(1\!\!+\!\epsilon)(1\!\!+\!\epsilon')}\Bigl[{\rm ci}(\alpha')\!-\!\frac{\sin(\alpha')}{\alpha'}\!-\!{\rm ci}(\alpha_i)\!+\!\frac{\sin(\alpha_i)}{\alpha_i}\Bigr]\; ,\label{D4cor}
\eneq
where the ${\rm ci}(z)$ is the cosine integral function defined in Eq.~(\ref{cosintdefn}). Using identity~(\ref{cieksisin}) in Eq.~(\ref{D4cor}) a power series expansion of the tree-order correlator is obtained
\be
\hspace{-0cm}\langle\Omega|\delta\bar\varphi_0(x) \delta\bar\varphi_0(x')|\Omega\rangle\!=\!\frac{HH'}{4\pi^2(1\!\!+\!\epsilon)(1\!\!+\!\epsilon')}
\Bigl[\ln\Bigl(\!\frac{H'a'}{H_i}\!\Bigr)\!-\!\ln\Bigl(\!\frac{1\!\!+\!\epsilon'}{1\!\!+\!\epsilon_i}\!\Bigr)
\!\!+\!\!\sum_{n=1}^\infty\!\frac{(-1)^n\!\!\left[{\alpha'}^{2n}\!\!-\!\alpha_i^{2n}\right]}{2n(2n\!\!+\!1)!}\Bigr] \;.\label{D4corpower}
\ee

The tree-order correlator is evaluated at the full solution of the background effective field equation, as in the case of tree-order mode function~(\ref{modefncfluct}). The Hubble parameter, scale factor and slow-roll parameter in Eqs.~(\ref{D4cor}) and (\ref{D4corpower}) have $\mathcal{O}(\lambda)$ corrections which can be given in terms of $q_0$ [Eq.~(\ref{qsifir})] and the initial values $H_{i0}$ and $\epsilon_{i0}$. The ratio \beeq
\frac{H}{1\!\!+\!\epsilon}\!=\!\frac{H_0}{1\!\!+\!\epsilon_0}\!
\left[1\!\pm\!\lambda\Bigl[\frac{H_\lambda}{H_0}\!-\!\frac{\epsilon_\lambda}{1\!\!+\!\epsilon_0}\Bigr]
\!\!+\!\mathcal{O}(\lambda^2)\right]\, ,\label{Hep1}
\eneq
where $H_0$, $H_\lambda$, $\epsilon_0$ and $\epsilon_\lambda$ are given in Eqs.~(\ref{HzeroQ}), (\ref{HLq}), (\ref{e0}) and (\ref{el}), respectively. They yield the various terms in Eq.~(\ref{Hep1}) as
\be
&&\hspace{-0.2cm}\frac{H_0}{1\!\!+\!\epsilon_0}\!=\!\frac{H_{i0}q_0^3}{q_0^2\!\!+\!\epsilon_{i0}}\; ,\;
\frac{H_\lambda}{H_0}\!=\!\frac{\xi}{4!}\epsilon^{-2}_{i0}\!\left[q_0^2\!\!+\!q_0\!\!-\!q^{-1}_0
\!\!+\!q^{-2}_0\frac{\left[1\!-\!q_0\right]^4}{4}\right]\;,\nonumber\\
&&\hspace{0.7cm}\frac{\epsilon_\lambda}{1\!\!+\!\epsilon_0}
\!=\!\frac{\xi}{4!}\frac{\epsilon^{-1}_{i0}}{q_0^2\!\!+\!\epsilon_{i0}}\!\left[2q_0^2\!+\!q_0\!\!+\!q^{-1}_0
\!\!-\!q^{-2}_0\frac{\left[1\!-\!q_0\right]^4}{2}\right]
\;. \label{Hep2}
\ee
Their initial values are obtained by recalling that $q_{i0}\!\!=\!\!1$. Note also that $\frac{H'}{1+\epsilon'}$ in Eqs.~(\ref{D4cor})-(\ref{D4corpower}) can be inferred from Eqs.~(\ref{Hep1})-(\ref{Hep2}).

The logarithm of the scale factor is given in Eq.~(\ref{loga}). Therefore, the first logarithmic term in Eq.~(\ref{D4corpower}),
\be
&&\hspace{1.5cm}\ln\Bigl(\!\frac{H'a'}{H_i}\!\Bigr)\!=\!\ln\Bigl(\!\frac{H_0'a'_0}{H_{i0}}\!\Bigr)
\!\pm\!\lambda\!\left[\frac{H'_\lambda}{H'_0}\!-\!\frac{H_{i\lambda}}{H_{i0}}\!+\!\ln(a')_\lambda\right]
\!\!+\!\mathcal{O}(\lambda^2)\\
&&\hspace{-0.5cm}=\ln(q'_0)\!+\!\epsilon^{-1}_{i0}\frac{\mathcal{E}'_{0}}{2}\!\pm\!\frac{\lambda}{4!}
\xi\epsilon^{-2}_{i0}\!\left[q_0'^2\!\!+\!q_0'\!\!-\!1\!\!-\!q_0'^{-1}
\!\!-\!\Bigl[\epsilon^{-1}_{i0}\!\!-\!q_0'^{-2}\Bigr]\frac{\left[1\!\!-\!q_0'\right]^4}{4}\right]
\!\!+\!\mathcal{O}(\lambda^2)\; ,\label{lnaHlnQ}
\ee
and the second logarithmic term in Eq.~(\ref{D4corpower}),
\be
&&\hspace{2.2cm}\ln\Bigl(\!\frac{1\!\!+\!\epsilon'}{1\!\!+\!\epsilon_i}\!\Bigr)
\!\!=\!\ln\Bigl(\!\frac{1\!\!+\!\epsilon'_0}{1\!\!+\!\epsilon_{i0}}\!\Bigr)
\!\pm\!\lambda\!\left[\frac{\epsilon'_\lambda}{1\!\!+\!\epsilon'_0}
\!-\!\frac{\epsilon_{i\lambda}}{1\!\!+\!\epsilon_{i0}}\right]\!\!+\!\mathcal{O}(\lambda^2)\\
&&\hspace{-1.1cm}=\!\ln\Bigl(\frac{q_0'^2\!\!+\!\epsilon_{i0}}{q_0'^2\!\left(1\!\!+\!\epsilon_{i0}\right)}\Bigr)
\!\pm\!\frac{\lambda}{4!}\xi\epsilon^{-1}_{i0}\!
\left[\!\frac{1}{q_0'^2\!\!+\!\epsilon_{i0}}\Bigl[2q_0'^2\!\!+\!q_0'\!\!+\!q_0'^{-1}
\!\!-\!q_0'^{-2}\frac{\left[1\!\!-\!q_0'\right]^4}{2}\Bigr]
\!\!-\!\frac{4}{1\!\!+\!\epsilon_{i0}}\right]\!\!+\!\mathcal{O}(\lambda^2)\; .\label{lnratioeps}
\ee
The $\alpha'$ [Eq.~(\ref{defnalp})] in Eqs.~(\ref{D4cor})-(\ref{D4corpower}), on the other hand, can be given as
\beeq
\alpha'\!\!=\!\alpha'_0\!\pm\!\lambda\alpha'_\lambda\!\!+\!\mathcal{O}(\lambda^2)\; ,
\eneq
where
\be
\hspace{-0.2cm}\alpha'_0\!=\!\frac{a'_0H'_0\Delta x}{1\!\!+\!\epsilon'_0}\!=\!\frac{q_0'^3H_{i0}\Delta x}
{q_0'^2\!\!+\!\epsilon_{i0}}\exp\Bigl(\epsilon^{-1}_{i0}\frac{\mathcal{E}'_{0}}{2}\Bigr)\; ,\label{alpzrprm}
\ee
and
\be
&&\hspace{3.8cm}\alpha'_\lambda\!\!=\!\alpha'_0\Bigl[\frac{H'_\lambda}{H'_0}
\!-\!\frac{\epsilon'_\lambda}{1\!\!+\!\epsilon'_0}\!+\!\ln(a')_\lambda\Bigr]\nonumber\\
&&\hspace{-1cm}=\!\alpha'_0\frac{\xi}{4!}\epsilon^{-2}_{i0}\!\left[
q_0'^2\!\!+\!q_0'\!\!-\!q_0'^{-1}\!\!\!-\!\!\Bigl[\epsilon^{-1}_{i0}\!\!\!-\!\!q_0'^{-2}\Bigr]\!
\frac{\left[1\!\!-\!q_0'\right]^4}{4}\!-\!\frac{\epsilon_{i0}}{q_0'^2\!\!+\!\epsilon_{i0}}\Bigl[\!2q_0'^2
\!\!+\!q_0'\!\!+\!q_0'^{-1}
\!\!\!-\!q_0'^{-2}\frac{\left[1\!\!-\!q_0'\right]^4}{2}\Bigr]\!\right] \;.\label{alphprm2}
\ee
The powers of $\alpha'$ in Eq.~(\ref{D4corpower}), to wit,  $\alpha'^{2n}\!\!=\!\alpha'^{2n}_0\Bigl[1\!\pm\!2n\lambda\frac{\alpha'_\lambda}{\alpha'_0}\Bigr]
\!\!+\!\mathcal{O}(\lambda^2)$ and its initial value $\alpha'^{2n}_i$ can be inferred from Eqs.~(\ref{alpzrprm}) and (\ref{alphprm2}). Hence, tree-order correlator~(\ref{D4corpower}) evaluated at the full solution of the background effective field equation can be given as
\be
&&\hspace{2cm}\langle\Omega|\delta\bar\varphi_0(x) \delta\bar\varphi_0(x')|\Omega\rangle\nonumber\\
&&\hspace{-1.2cm}\equiv\!\langle\Omega|\delta\bar\varphi_0(x) \delta\bar\varphi_0(x')|\Omega\rangle_{\bf 0}\!\pm\!\lambda\langle\Omega|\delta\bar\varphi_0(x) \delta\bar\varphi_0(x')|\Omega\rangle_{\bf\lambda}\!\!+\!\mathcal{O}(\lambda^2) .
\ee
The tree-order correlator in the noninteracting limit
\be
&&\hspace{-1.2cm}\langle\Omega|\delta\bar\varphi_0(x) \delta\bar\varphi_0(x')|\Omega\rangle_{\bf 0}
\!\!=\!\!\frac{H_0H_0'}{4\pi^2(1\!\!+\!\epsilon_0\!)(1\!\!+\!\epsilon_0'\!)}
\Bigl[\ln\Bigl(\!\frac{H_0'a_0'}{H_{i0}}\!\Bigr)\!\!-\!\ln\Bigl(\!\frac{1\!\!+\!\epsilon'_0}{1\!\!+\!\epsilon_{i0}}\!\Bigr)
\!\!+\!\!\sum_{n=1}^\infty\!\!\frac{(-1)^n\!\left[\alpha_0'^{2n}\!\!-\!\alpha_{i0}^{2n}\right]}
{2n(2n\!\!+\!\!1)!}\Bigr]\; .\label{treecorrzeroorder0}
\ee
Employing Eqs.~(\ref{HzeroQ}) and (\ref{e0}) in Eq.~(\ref{treecorrzeroorder0}) yields
\beeq
\langle\Omega|\delta\bar\varphi_0(x) \delta\bar\varphi_0(x')|\Omega\rangle_{\bf 0}
\!\!=\!\!\frac{H^2_{i0}}{4\pi^2}f_{00}(t, t'\!, \Delta x)\; ,\label{treecorrzeroorder1}
\eneq
where
\be
&&\hspace{-1cm}f_{00}\!\equiv\!\frac{(q_0q_0')^3}
{(q^2_0\!\!+\!\epsilon_{i0})(q'^2_0\!\!+\!\epsilon_{i0})}\!\left[\epsilon^{-1}_{i0}\frac{\mathcal{E}'_{0}}{2}
\!-\!\ln\Bigl(\frac{q_0'^2\!\!+\!\epsilon_{i0}}{q_0'^3\!\left(1\!\!+\!\epsilon_{i0}\right)}\Bigr)
\!\!+\!\!\sum_{n=1}^\infty\!\!\frac{(-1)^n\!\left[\alpha_0'^{2n}\!\!-\!\alpha_{i0}^{2n}\right]}
{2n(2n\!\!+\!\!1)!}\right]\; .
\label{treecorrzeroorder2}
\ee
The $\alpha_0'$, given in Eq.~(\ref{alpzrprm}), implies $\alpha_{i0}\!=\!\frac{H_{i0}\Delta x}{1\!+\epsilon_{i0}}$. Choosing the physical distance $a(x')\Delta x$ a constant fraction $K$ of Hubble length so that the comoving separation\beeq
\Delta x\!=\!\frac{K}{H_0(t')a_0(t')}\; ,\label{comovsep}
\eneq
in Eq.~(\ref{treecorrzeroorder2}) yields $f_{00}(t, t'\!, K)$.\begin{figure}
\includegraphics[width=11.5cm,height=6.5cm]{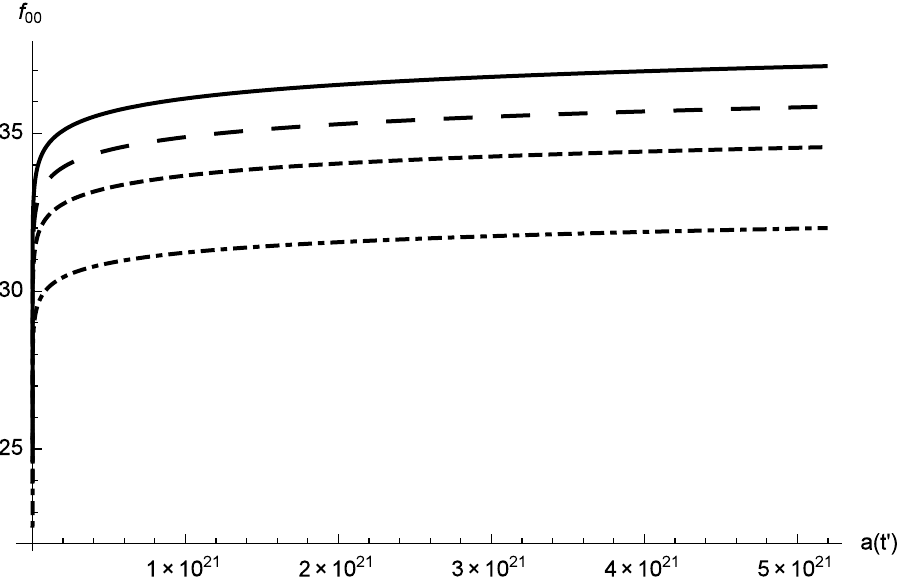}
\caption{Plots of the function $f_{00}(t,t', K)$, defined in Eq.~(\ref{treecorrzeroorder2}) with $\Delta x\!=\!K/H_0(t')a_0(t')$, versus $a_0(t')$ for  different values of $\epsilon_{i0}\!=\!m^2/3H^2_{i0}$. The scale factor $a_0(t)$ and the fraction $K$ are chosen to be $e^{50}$ and $1/2$, respectively. $a_0(t')$ runs from $1$ to $a_0(t)$. The solid, large-dashed, dashed and dot-dashed plots are for $\epsilon_{i0}\!=\!0.0025, 0.00275, 0.003$ and $0.0035$, respectively.}
\label{fig:flucttreecorr}
\end{figure} The plots of the function $f_{00}(t, t'\!, K)$ versus $a(t')$, for $K\!\!=\!\!1/2$ and four different initial values of the slow-roll parameter $\epsilon_{i0}\!=\!\frac{m^2}{3H^2_{i0}}$, are given in Fig.~\ref{fig:flucttreecorr}. They show that correlation~(\ref{treecorrzeroorder1}), for a given $\epsilon_{i0}$, grows and asymptotes to an almost constant value at late times during inflation. The growth is sensitive to the value of $\epsilon_{i0}$. Even a slight increase in the $\epsilon_{i0}$ causes a noticeable suppression in the growth. This is because the production rate and lifetime of virtual inflatons---the origin of fluctuations---are suppressed as the mass increases.

\begin{figure}
\includegraphics[width=11.5cm,height=6.5cm]{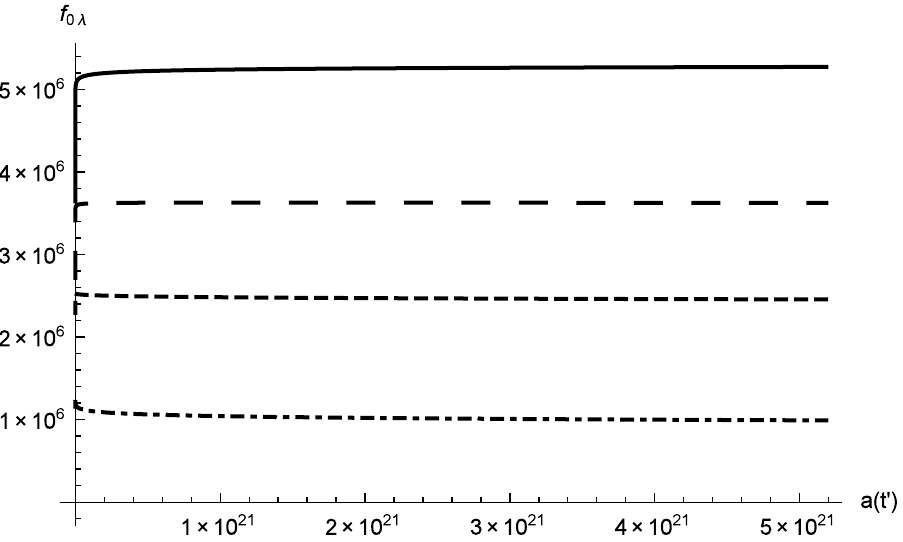}
\caption{Plots of the function $f_{0\lambda}(t,t', K)$, defined in Eq.~(\ref{f0lambda}). The ranges of parameters $a_0(t)$, $a_0(t')$, $K$ and $\epsilon_{i0}$ are chosen same as in Fig.~\ref{fig:flucttreecorr}.}
\label{fig:flucttreelambdacorr}
\end{figure}

As pointed out earlier, evaluation of the tree-order correlator at the full solution of the background effective field equation yields an $\mathcal{O}(\lambda)$ correction,
\be
&&\hspace{-0.1cm}\langle\Omega|\delta\bar\varphi_0(x) \delta\bar\varphi_0(x')|\Omega\rangle_{\bf\lambda}
\!\!=\!\!\left[\frac{H_\lambda}{H_0}\!-\!\frac{\epsilon_\lambda}{1\!\!+\!\epsilon_0}
\!+\!\frac{H'_\lambda}{H'_0}
\!-\!\frac{\epsilon'_\lambda}{1\!\!+\!\epsilon'_0}\right]\!\langle\Omega|\delta\bar\varphi_0(x) \delta\bar\varphi_0(x')|\Omega\rangle_{\bf 0}\nonumber\\
&&\hspace{-1.4cm}
+\frac{H_0H_0'}{4\pi^2(1\!\!+\!\epsilon_0)(1\!\!+\!\epsilon_0')}
\Bigg\{\!\!\!\left[\frac{H'_{\lambda}}{H'_0}\!-\!\frac{\epsilon'_{\lambda}}{1\!\!+\!\epsilon'_0}
\!+\!\ln(a')_\lambda\right]\!\!\sum_{n=0}^\infty\!\frac{(-1)^n\alpha_0'^{2n}}
{(2n\!\!+\!\!1)!}\!-\!\!\left[\frac{H_{i\lambda}}{H_{i0}}\!-\!\frac{\epsilon_{i\lambda}}{1\!\!+\!\epsilon_{i0}}\right]
\!\!\sum_{n=0}^\infty\!\frac{(-1)^n\alpha_{i0}^{2n}}{(2n\!\!+\!\!1)!}\!\Bigg\}.\label{corrordl}
\ee
Employing Eq.~(\ref{Hep2}) in Eq.~(\ref{corrordl}) yields
\be
\langle\Omega|\delta\bar\varphi_0(x) \delta\bar\varphi_0(x')|\Omega\rangle_{\bf\lambda}
\!\!=\!\!\frac{H^2_{i0}}{4\pi^2}\frac{\xi}{4!}f_{0\lambda}(t, t'\!, \Delta x)\; ,\label{fzerolambda}
\ee
where $\xi/4!\!\sim\!700$ in a GUT scale inflation for which $H_{i0}\!\sim\!10^{-3}M_{Pl}$ and the dimensionless function
\be
&&\hspace{-0.1cm}f_{0\lambda}
\!=\!\epsilon^{-2}_{i0}\Bigg\{\!\Bigg\{\!\Bigg[q_0^2\!\!+\!q_0\!\!-\!q_0^{-1}\!\!
\!+\!q_0^{-2}\frac{[1\!\!-\!q_0]^4}{4}
\!-\!\frac{\epsilon_{i0}}{q_0^2\!\!+\!\epsilon_{i0}}\!\left(\!\!2q_0^2\!\!+\!q_0\!\!+\!q_0^{-1}\!\!
\!-\!q_0^{-2}\frac{[1\!\!-\!q_0]^4}{2}\!\right)\!\!\Bigg]\!\!\!+\!\!\Bigg[q_0\!\!\rightarrow\!q'_0\Bigg]\!\Bigg\}
f_{00}\nonumber\\
&&\hspace{-0.4cm}+\frac{(q_0q_0')^3}
{(q^2_0\!\!+\!\epsilon_{i0})(q'^2_0\!\!+\!\epsilon_{i0})}
\Bigg\{\!\Bigg[q'^2_0\!\!+\!q'_0\!\!-\!q_0'^{-1}\!\!-\!\!\Bigl[\epsilon^{-1}_{i0}\!\!-\!q_0'^{-2}\Bigr]
\!\frac{[1\!\!-\!q'_0]^4}{4}\!-\!\frac{\epsilon_{i0}}{q'^2_0\!\!+\!\epsilon_{i0}}
\Bigl[ 2q_0'^2\!\!+\!q'_0\!\!+\!q_0'^{-1}\!-\!q_0'^{-2}\frac{[1\!\!-\!q'_0]^4}{2}\Bigr]\!\Bigg]\nonumber\\
&&\hspace{3.7cm}\times\!\!\sum_{n=0}^\infty
\!\frac{(-1)^n\alpha_0'^{2n}}{(2n\!\!+\!\!1)!}\!-\!\Bigl[1\!\!-\!\frac{4\epsilon_{i0}}
{1\!\!+\!\epsilon_{i0}}\Bigr]\!\sum_{n=0}^\infty\!
\frac{(-1)^n\alpha_{i0}^{2n}}{(2n\!\!+\!\!1)!}\!\Bigg\}\Bigg\}\;.\label{f0lambda}
\ee
For comoving separation~(\ref{comovsep}) and $K\!\!=\!\!1/2$, plots of the function $f_{0\lambda}(t, t'\!, K)$ versus $a(t')$ are given in Fig.~\ref{fig:flucttreelambdacorr} considering the values of $\epsilon_{i0}$ chosen in Fig~\ref{fig:flucttreecorr}.
The plots show that $\mathcal{O}(\lambda)$ correction~(\ref{fzerolambda}), for a given $\epsilon_{i0}$, grows at early times and asymptotes to a constant at late times during inflation. The asymptotic value is substantially suppressed as $\epsilon_{i0}$ increases even slightly. In Secs.~\ref{subsubsubsect:equalspacecorr}-2 we compute the temporal and coincident tree-order correlators, respectively.

\subsubsection{Temporal tree-order correlator}
\label{subsubsubsect:equalspacecorr}
The temporal tree-order correlator is the equal space $(\vec{x}\,'\!\!\rightarrow\!\!\vec{x})$ limit of correlator~(\ref{corrtreeordincomp}),
\be
&&\hspace{4cm}\langle\Omega|\delta\bar\varphi_0(t,\vec{x}) \delta\bar\varphi_0(t'\!,\vec{x})|\Omega\rangle\nonumber\\
&&\hspace{-1cm}=\!\frac{\Gamma\!\!\left(\frac{D}{2}\right)}
{2\pi^{\frac{D}{2}+1}\Gamma(D\!-\!\!1)} \Bigl(\!\frac{H}{1\!\!+\!\epsilon}\Bigr)^{\frac{D}{2}-1}\!\Bigl(\!\frac{H'}{1\!\!+\!\epsilon'}\Bigr)^{\frac{D}{2}-1}
\frac{2^{\nu}\Gamma(\nu)}{(\frac{Ha}{1+\epsilon})^{\frac{\delta}{2}}}
\,\frac{2^{\nu'}\Gamma(\nu')}{(\frac{H'a'}{1+\epsilon'})^{\frac{\delta'}{2}}}
\Bigl[\frac{(\frac{H'a'}{1+\epsilon'})^{\frac{\delta+\delta'}{2}}
\!\!-\!(\frac{H_i}{1+\epsilon_i})^{\frac{\delta+\delta'}{2}}}
{\delta\!+\!\delta'}\Bigr]\;.
\label{corrtreeeqspace}\ee
Recall that in $D\!\rightarrow\!4$ limit we have $\nu\!\!=\!\nu'\!\!\rightarrow\!3/2\!+\!\mathcal{O}(\epsilon^2)$ and $\delta\!\!=\!\delta'\!\!\rightarrow\!0\!+\!\mathcal{O}(\epsilon^2)$. At this order,\be
\!\!\langle\Omega|\delta\bar\varphi_0(t,\vec{x}) \delta\bar\varphi_0(t'\!,\vec{x})|\Omega\rangle\!\rightarrow\!
\frac{HH'}{4\pi^2(1\!\!+\!\epsilon)(1\!\!+\!\epsilon')}\Bigl[
\ln\Bigl(\!\frac{H'a'}{H_i}\Bigr)\!-\!\ln\Bigl(\!\frac{1\!\!+\!\epsilon'}{1\!\!+\!\epsilon_i}\Bigr)\Bigr]\; .\label{corrtreqspD4}
\ee
Expanding the correlator as\be
&&\hspace{2.5cm}\langle\Omega|\delta\bar\varphi_0(t,\vec{x}) \delta\bar\varphi_0(t'\!,\vec{x})|\Omega\rangle\nonumber\\
&&\hspace{-1cm}\equiv\!\langle\Omega|\delta\bar\varphi_0(t,\vec{x}) \delta\bar\varphi_0(t'\!,\vec{x})|\Omega\rangle_{\bf 0}\!\pm\!\lambda\langle\Omega|\delta\bar\varphi_0(t,\vec{x}) \delta\bar\varphi_0(t'\!,\vec{x})|\Omega\rangle_{\bf\lambda}\!\!+\!\mathcal{O}(\lambda^2) .\label{coreqsp1}
\ee
implies
\be
&&\hspace{-0.9cm}\langle\Omega|\delta\bar\varphi_0(t,\vec{x}) \delta\bar\varphi_0(t'\!,\vec{x})|\Omega\rangle_{\bf 0}
\!\!=\!\!\frac{H_0H_0'}{4\pi^2(1\!\!+\!\epsilon_0\!)(1\!\!+\!\epsilon_0'\!)}
\Bigl[\ln\Bigl(\!\frac{H_0'a_0'}{H_{i0}}\!\Bigr)\!\!-\!\ln\Bigl(\!\frac{1\!\!+\!\epsilon'_0}{1\!\!+\!\epsilon_{i0}}
\!\Bigr)\Bigr]\nonumber\\
&&\hspace{0.4cm}=\!\frac{H^2_{i0}}{4\pi^2}\frac{(q_0q_0')^3}
{(q^2_0\!\!+\!\epsilon_{i0})(q'^2_0\!\!+\!\epsilon_{i0})}\!\left[\epsilon^{-1}_{i0}\frac{\mathcal{E}'_0}{2}
\!-\!\ln\Bigl(\frac{q_0'^2\!\!+\!\epsilon_{i0}}{q_0'^3\!\left(1\!\!+\!\epsilon_{i0}\right)}\Bigr)\right] ,\label{coreqsp2}
\ee
and
\be
&&\hspace{0.5cm}\langle\Omega|\delta\bar\varphi_0(t,\vec{x})\delta\bar\varphi_0(t'\!,\vec{x})|\Omega\rangle_{\bf\lambda}
\!\!=\!\left[\frac{H_\lambda}{H_0}\!-\!\frac{\epsilon_\lambda}{1\!\!+\!\epsilon_0}
\!+\!\frac{H'_\lambda}{H'_0}
\!-\!\frac{\epsilon'_\lambda}{1\!\!+\!\epsilon'_0}\right]\!\langle\Omega|\delta\bar\varphi_0(t,\vec{x}) \delta\bar\varphi_0(t'\!,\vec{x})|\Omega\rangle_{\bf 0}\nonumber\\
&&\hspace{-0.4cm}
+\frac{H_0H_0'}{4\pi^2(1\!\!+\!\epsilon_0)(1\!\!+\!\epsilon_0')}
\Bigg\{\!\!\!\left[\frac{H'_{\lambda}}{H'_0}\!-\!\!\frac{\epsilon'_{\lambda}}{1\!\!+\!\epsilon'_0}
\!+\!\ln(a')_\lambda\right]\!\!-\!\!\left[\frac{H_{i\lambda}}{H_{i0}}\!-\!\!\frac{\epsilon_{i\lambda}}
{1\!\!+\!\epsilon_{i0}}\right]\!\!\!\Bigg\}\!\!=\!\frac{\xi}{4!}\epsilon^{-2}_{i0}
\Bigg\{\!\!\Bigg\{\!\Bigg[q_0^2\!\!+\!q_0\!\!-\!q_0^{-1}\!\!
\!+\!q_0^{-2}\frac{[1\!\!-\!q_0]^4}{4}\nonumber\\
&&\hspace{-0.4cm}
-\frac{\epsilon_{i0}}{q_0^2\!\!+\!\epsilon_{i0}}\!\left(\!\!2q_0^2\!\!+\!q_0\!\!+\!q_0^{-1}\!\!
\!-\!q_0^{-2}\frac{[1\!\!-\!q_0]^4}{2}\!\right)\!\!\Bigg]
\!\!\!+\!\!\Bigg[\!q_0\!\!\rightarrow\!q'_0\!\Bigg]\!\!\Bigg\}\!\langle\Omega|\delta\bar\varphi_0(t,\!\vec{x}) \delta\bar\varphi_0(t'\!,\!\vec{x})|\Omega\rangle_{\bf 0}
\!\!+\!\!\frac{H^2_{i0}}{4\pi^2}\frac{(q_0q_0')^3}
{(q^2_0\!\!+\!\epsilon_{i0})(q'^2_0\!\!+\!\epsilon_{i0})}\nonumber\\
&&\hspace{-0.4cm}\times\!\Bigg\{\!\Bigg[q'^2_0\!\!+\!q'_0\!\!-\!q_0'^{-1}\!\!\!-\!\!\Bigl[\epsilon^{-1}_{i0}\!\!\!-\!q_0'^{-2}\Bigr]
\!\frac{[1\!\!-\!q'_0]^4}{4}\!-\!\frac{\epsilon_{i0}}{q'^2_0\!\!+\!\epsilon_{i0}}
\!\!\left[ 2q_0'^2\!\!+\!q'_0\!\!+\!q_0'^{-1}\!\!\!-\!q_0'^{-2}\frac{[1\!\!-\!q'_0]^4}{2}\right]\!\!\Bigg]
\!\!-\!\!\Bigl[\!1\!\!-\!\frac{4\epsilon_{i0}}
{1\!\!+\!\epsilon_{i0}}\Bigr]\!\!\Bigg\}\!\!\Bigg\}.
\ee
In Sec.~\ref{subsubsect:equalspacetimecorr} we take the equal time limits of correlators~(\ref{corrtreeeqspace}) and (\ref{corrtreqspD4}) to get the coincident tree-order correlator respectively in $D$, and in four dimensions which is used to compute the one-loop correlator in Sec.~\ref{subsect:1loopcorr}.

\subsubsection{Coincident tree-order correlator}
\label{subsubsect:equalspacetimecorr}
Equal time  $t'\!\!\rightarrow t$ limit of tree-order correlator~(\ref{corrtreeeqspace}) is
\be
\hspace{0.8cm}\langle\Omega|\delta\bar\varphi^2_0(x)|\Omega\rangle\!=\!\frac{
\Gamma\!\!\left(\frac{D}{2}\right)}
{4\pi^{\frac{D}{2}+1}\Gamma(D\!-\!\!1)} \Bigl(\!\frac{H}{1\!\!+\!\epsilon}\Bigr)^{D-2}\,
\frac{2^{2\nu}\,\Gamma^2(\nu)}{\delta}\Bigl[1\!-\!\Bigl(\frac{\alpha_i}{\alpha}\Bigr)^\delta\Bigr]\;.
\label{corrtreeeqspacetime}\ee
In $D\!=\!4$, the equal time limit of Eq.~(\ref{corrtreqspD4}) yields the coincident correlator as
\be
\hspace{0.4cm}\langle\Omega|\delta\bar\varphi^2_0(x)|\Omega\rangle\!\rightarrow\!
\frac{H^2}{4\pi^2(1\!\!+\!\epsilon)^2}\Bigl[
\ln\Bigl(\!\frac{Ha}{H_i}\Bigr)\!-\!\ln\Bigl(\!\frac{1\!\!+\!\epsilon}{1\!\!+\!\epsilon_i}\Bigr)\Bigr]\; .\label{cointreebit}
\ee
Perturbative expansion of the parameters $H$, $a$ and $\epsilon$ implies\be
&&\hspace{-0.2cm}\langle\Omega|\delta\bar\varphi^2_0(x)|\Omega\rangle\!\equiv\!
\langle\Omega|\delta\bar\varphi^2_0(x)|\Omega\rangle_{\bf 0}\!\pm\!\lambda\langle\Omega|\delta\bar\varphi^2_0(x)|\Omega\rangle_{\bf\lambda}\!\!+\!\mathcal{O}(\lambda^2) \;.\label{coreqspcoin}
\ee
The $\mathcal{O}(\lambda^0)$ term, i.e., the coincident correlator in the noninteracting theory
\be
&&\hspace{-1.35cm}\langle\Omega|\delta\bar\varphi^2_0(x)|\Omega\rangle_{\bf 0}\!\!=\!\!
\frac{H^2_0}{4\pi^2(1\!\!+\!\epsilon_0\!)^2}\!\!
\left[\ln\Bigl(\!\frac{H_0a_0}{H_{i0}}\!\Bigr)\!\!-\!\ln\Bigl(\!\frac{1\!\!+\!\epsilon_0}{1\!\!+\!\epsilon_{i0}}
\!\Bigr)\!\right]\!\!\!=\!\!\frac{H^2_{i0}}{4\pi^2}\frac{q^6_0}
{\left[q^2_0\!\!+\!\epsilon_{i0}\right]^2}\!\!\left[\epsilon^{-1}_{i0}\frac{\mathcal{E}_0}{2}\!-\!
\ln\Bigl(\!\frac{q_0^2\!\!+\!\epsilon_{i0}}{q_0^3\!\left(1\!\!+\!\epsilon_{i0}\right)}\!\Bigr)\!\right]\!,
\label{0tree0}
\ee
where the logarithm\beeq
\ln\Bigl(\!\frac{q_0^2\!\!+\!\epsilon_{i0}}{q_0^3\!\left(1\!\!+\!\epsilon_{i0}\right)}\!\Bigr)\!\!=\!-\!\ln(q_0)
\!+\!\epsilon_{i0}\mathcal{E}_0q^{-2}_0
\!\!+\!\mathcal{O}(\epsilon^2_{i0})\; .
\eneq

The $\mathcal{O}(\lambda)$ correction in Eq.~(\ref{coreqspcoin}) induced as a reaction to the change in the expansion rate---hence in the scale factor and the slow-roll parameter---due to the self-interactions of the inflaton is
\be
&&\hspace{-0.4cm}\langle\Omega|\delta\bar\varphi^2_0(x)|\Omega\rangle_{\bf\lambda}
\!\!=\!\!\left[2\langle\Omega|\delta\bar\varphi^2_0(x)|\Omega\rangle_{\bf 0}\!+\!\frac{H^2_0}{4\pi^2(1\!\!+\!\epsilon_0\!)^2}\!\right]\!\!\left[\!\frac{H_\lambda}{H_0}
\!-\!\frac{\epsilon_\lambda}{1\!\!+\!\epsilon_0}\!
\right]
\!\!\!+\!\frac{H^2_0}{4\pi^2(1\!\!+\!\epsilon_0\!)^2}\!\!\left[
\ln(a)_\lambda\!-\!\frac{H_{i\lambda}}{H_{i0}}\!+\!\frac{\epsilon_{i\lambda}}
{\!1\!\!+\!\epsilon_{i0}}\right]\nonumber\\
&&\hspace{-0.3cm}=\!\frac{H^2_{i0}}{4\pi^2}\frac{q_0^6}{(q_0^2\!\!+\!\epsilon_{i0})^2}\frac{\xi}{4!}\epsilon^{-2}_{i0}
\Bigg\{\!\!\!\left[\epsilon^{-1}_{i0}\mathcal{E}_0\!\!+\!1\!
-\!\!2\ln\Bigl(\frac{q_0^2\!\!+\!\epsilon_{i0}}{q_0^3\!\left(1\!\!+\!\epsilon_{i0}\right)}\Bigr)\right]
\!\!\left[q_0^2\!\!+\!q_0\!\!-\!q_0^{-1}\!\!+\!q_0^{-2}\frac{[1\!\!-\!q_0]^4}{4}\!-\!\frac{\epsilon_{i0}}
{q_0^2\!\!+\!\epsilon_{i0}}\Bigl[2q_0^2\!\!+\!q_0\right.\nonumber\\
&&\hspace{3.7cm}\left.+q_0^{-1}
\!\!-\!q_0^{-2}\frac{[1\!\!-\!q_0]^4}{2}
\Bigr]\right]\!\!-\!\epsilon^{-1}_{i0}
\frac{[1\!\!-\!q_0]^4}{4}
\!-\!1\!\!+\!\frac{4\epsilon_{i0}}{1\!\!+\!\epsilon_{i0}}\Bigg\}\;.
\ee
Expanding the terms inside the curly brackets in powers of $\epsilon_{i0}$ and $\mathcal{E}_0$ yields\be
&&\hspace{0.cm}\langle\Omega|\delta\bar\varphi^2_0(x)|\Omega\rangle_{\bf\lambda}
\!=\!-\frac{H_{i0}^2}{576\pi^4}\frac{\pi}{GH^2_{i0}}\frac{q_0^4}{\left[q_0^2\!\!+\!\epsilon_{i0}\right]^2}\Bigg\{\!\epsilon^{-3}_{i0}\Bigg[\!1\!\!-\!q_0
\!-\!\frac{7}{2}\mathcal{E}_0\Bigl[\!1\!\!-\!\frac{5}{7}q_0\Bigr]\!\!+\!\frac{25}{8}\mathcal{E}^2_0
\Bigl[\!1\!\!-\!\frac{4}{25}q_0\Bigr]\!\!-\!\frac{3}{4}\mathcal{E}^3_0\Bigg]
\nonumber\\
&&\hspace{-0.3cm}-\epsilon^{-2}_{i0}q_0^{-2}\Bigg[\!1\!\!-\!q_0\!
\!+\!\!\frac{3}{2}\Bigl[1\!\!-\!\frac{2}{3}q_0\Bigr]\!\ln(q^2_0)
\!-\!\!\frac{3}{2}\mathcal{E}_0\Bigl[1\!\!+\!\frac{4}{3}q_0
\!+\!\!\frac{7}{3}\Bigl[1\!\!-\!\frac{2}{7}q_0\Bigr]\!\ln(q^2_0)\Bigr]
\!\!+\!\frac{17}{8}\mathcal{E}^2_0\Bigl[1\!\!+\!\frac{12}{17}q_0
\!\!+\!\!\frac{21}{17}\ln(q^2_0)\Bigr]\nonumber\\
&&\hspace{-0.5cm}-\frac{11}{8}\mathcal{E}^3_0\Bigl[\!1\!\!+\!\frac{5}{11}\!\ln(q^2_0)\Bigr]\!\Bigg]
\!\!\!-\!\epsilon^{-1}_{i0}q_0^{-4}3\Bigg[\!1\!\!-\!q_0\!\!+\!\frac{1\!\!-\!3q_0}{3}\ln(q^2_0)
\!-\!\!\frac{11}{3}\mathcal{E}_0
\Bigl[\!1\!\!-\!\frac{19}{22}q_0\!
\!+\!\frac{1}{11}\Bigl[\!1\!\!-\!\frac{9}{2}q_0\Bigr]\!\ln(q^2_0)\Bigr]\!\!+\!\frac{49}{12}\mathcal{E}^2_0\nonumber\\
&&\hspace{-0.35cm}\times\!\Bigl[\!1\!\!-\!\frac{20}{49}q_0\!
\!-\!\frac{3}{49}\Bigl[\!1\!\!+\!2q_0\Bigr]\!\ln(q^2_0)\Bigr]\!\!-\!\frac{23}{12}\mathcal{E}^3_0
\Bigl[\!1\!\!-\!\frac{3}{23}\ln(q^2_0)\Bigr]\!\!+\!\!\frac{5}{12}\mathcal{E}^4_0\Bigg]
\!\!\!+\!q_0^{-6}3\Bigg[\!1\!\!-\!q_0\!\!+\!\frac{1\!\!-\!3q_0}{3}\ln(q^2_0)\!-\!\frac{11}{3}\mathcal{E}_0\nonumber\\
&&\hspace{-0.35cm}\times\!
\Bigl[\!1\!\!-\!\frac{7}{22}q_0\!
\!+\!\frac{1}{11}\Bigl[\!1\!\!-\!\frac{9}{2}q_0\Bigr]\!\ln(q^2_0)\Bigr]
\!\!+\!\frac{71}{12}\mathcal{E}^2_0\Bigl[\!1\!\!+\!\frac{12}{71}q_0\!
\!-\!\frac{3}{71}\Bigl[\!1\!\!+\!2q_0\Bigr]\!\ln(q^2_0)\Bigr]\!\!-\!\frac{67}{12}\mathcal{E}^3_0
\Bigl[\!1\!\!+\!\frac{8}{67}q_0\!\!-\!\frac{3}{67}\ln(q^2_0)\Bigr]\nonumber\\
&&\hspace{6cm}+\frac{59}{24}\mathcal{E}^4_0\!
\!-\!\frac{5}{24}\mathcal{E}^5_0\Bigg]\!\!+\!\mathcal{O}(\epsilon_{i0})\Bigg\}\;.\label{coincorrtree}
\ee
Self-interactions of the inflaton field yield quantum corrections to the two-point correlation function. The one-loop correction at $\mathcal{O}(\lambda)$ is computed in the next section.

\subsection{One-loop correlator}
\label{subsect:1loopcorr}
Computation of the one-loop contribution~(\ref{1loopcorr}) to two-point correlation~(\ref{tpcf}) involves evaluations of four VEVs. Two of the VEVs which are quadratic both in the background and fluctuation fields are evaluated in Sec.~\ref{subsubsect:backgrndsqr}. The remaining VEVs which are quartic in the fluctuation field are evaluated in Sec.~\ref{subsubsect:flctqrt}.

\subsubsection{The VEVs quadratic in the background and fluctuation fields}
\label{subsubsect:backgrndsqr}

One-loop correlator~(\ref{1loopcorr}) has two VEVs that are quadratic in the background and fluctuation fields---in addition to the two VEVs that are quartic in the fluctuation field. In $D\!=\!4$ dimensions, they are
\be
&&\hspace{0.2cm}\mathcal{V}_{\bar\varphi^2_0\delta\bar{\varphi}^2_0}(q_0, q'_0, \Delta x)\!\equiv\!\mp\frac{\lambda}{6}\Bigg\{\!\!\frac{H_0'}{1\!\!+\!\epsilon_0'}\langle\Omega|
\delta\bar{\varphi}_0(x)\!\!\int_0^{t'_0}\!\!\!d\tilde{t}_0\,\Bigl[\frac{1\!\!+\!\tilde\epsilon_0}{\tilde{H}_0}\Bigr]^2
\!\!\bar\varphi_0^2(\tilde{t}_0)\,\delta\bar{\varphi}_0(\tilde{t}_0,\vec{x}\,')
|\Omega\rangle_{\bf 0}\nonumber\\
&&\hspace{1.9cm}+\frac{H_0}{1\!\!+\!\epsilon_0}\langle\Omega|\!\!
\int_0^{t_0}\!\!dt''_0\Bigl[\frac{1\!\!+\!\epsilon_0''}{H_0''}\Bigr]^2\!\!\bar\varphi^2_0(t''_0)\,
\delta\bar{\varphi}_0(t''_0,\vec{x})\,
\delta\bar{\varphi}_0(x')|\Omega\rangle_{\bf 0}\!\Bigg\}\; .\label{corrpropphi2}
\ee
Evaluations of the integrals in Eq.~(\ref{corrpropphi2}) are outlined in Appendix~\ref{app:oneloop-quad}. Combining Eqs.~(\ref{1stint}), (\ref{2ndint}), (\ref{3rdint}) and (\ref{4thint}) in Eq.~(\ref{1stvevintg}) yields the first integral in Eq.~(\ref{corrpropphi2}) as
\be
&&\hspace{2.3cm}\frac{H'_0}{1\!\!+\!\epsilon'_0}\!\int_0^{t'_0}\!\!\!d\tilde{t}_0
\Bigl[\frac{1\!\!+\!\tilde\epsilon_0}{\tilde{H}_0}\Bigr]^2
\!\!\bar\varphi_0^2(\tilde{t}_0)\langle\Omega|
\delta\bar{\varphi}_0(t_0,\vec{x})\,\delta\bar{\varphi}_0(\tilde{t}_0,\vec{x}\,')
|\Omega\rangle\nonumber\\
&&\hspace{1.1cm}=\!\frac{1}{32\pi^3G}\frac{q_0^3\,q_0'^3}{(q_0^2\!\!+\!\epsilon_{i0})(q_0'^2\!\!+\!\epsilon_{i0})}
\Bigg\{\!\epsilon^{-3}_{i0}\frac{\mathcal{E}'^2_0}{4}
\!-\!\epsilon^{-2}_{i0}\Bigl[\!\left[1\!\!+\!q'^2_0\right]\!\!+\!(2\!-\!\gamma)
\,\mathcal{E}'_0\Bigr]\!\ln(q'_0)\nonumber\\
&&\hspace{-0.2cm}-\epsilon^{-1}_{i0}\!\Bigl[\ln^2(q'_0)\!-\!(2\!-\!\gamma)\ln(q'^2_0)\!-\!1\!\!+\!q'^2_0\Bigr]
\!\!-\!\ln(q'^2_0)
\!-\!\frac{q'^{-2}_0}{2}\!+\!\frac{q'^2_0}{2}
\!-\!\Bigl[\epsilon^{-2}_{i0}\mathcal{E}'_0\!-\!\epsilon^{-1}_{i0}\ln(q'^2_0)\Bigr]\nonumber\\
&&\hspace{2.7cm}\times\!\Bigl[{\rm ci}(\alpha_i)\!-\!\frac{\sin(\alpha_i)}{\alpha_i}\!-\!\ln(\alpha_i)\Bigr]\!\!+\!\frac{\mathbf{A}(q'_0,\Delta x)}{2}
\!+\!\mathcal{O}(\epsilon_{i0})  \Bigg\} ,\label{1loop1stint}
\ee
where
\be
&&\hspace{-0.1cm}\mathbf{A}(q'_0,\Delta x)\!\equiv\!\!\sum_{n=1}^\infty\!\frac{(-1)^n(H_{i0}\Delta x)^{2n}}{2n(2n\!\!+\!\!1)!} 2e^{\frac{n}{\epsilon_{i0}}}\Bigg\{\!
\epsilon^{-2}_{i0}\Bigl[q'^{2n+2}_0\!E_{-n}\!\Bigl(\!\frac{nq'^2_0}{\epsilon_{i0}}\Bigr)
\!\!-\!\!E_{-n}\!\Bigl(\!\frac{n}{\epsilon_{i0}}\Bigr)\!\Bigr]\nonumber\\
&&\hspace{-0.7cm}-(2n\!\!-\!\!1)\!\left[\epsilon^{-1}_{i0}\!\Bigl[q'^{2n}_0E_{1-n}\!\Bigl(\!\frac{nq'^2_0}{\epsilon_{i0}}\Bigr)
\!\!-\!\!E_{1-n}\!\Bigl(\!\frac{n}{\epsilon_{i0}}\Bigr)\!\Bigr]\!\!-\!\!n\Bigl[q'^{2n-2}_0E_{2-n}\!\Bigl(\!\frac{nq'^2_0}{\epsilon_{i0}}\Bigr)
\!\!-\!\!E_{2-n}\!\Bigl(\!\frac{n}{\epsilon_{i0}}\Bigr)\!\Bigr]\right]\!\!\Bigg\}.
\ee

The second integral in Eq.~(\ref{corrpropphi2}) is obtained mingling Eq.~(\ref{1loop1stint}) and
Eqs.~(\ref{1l1p})-(\ref{1l2p}). The result is
\be
&&\hspace{2.9cm}\frac{H_0}{1\!\!+\!\epsilon_0}\!
\int_0^{t_0}\!\!dt_0''\Bigl[\frac{1\!\!+\!\epsilon_0''}{H_0''}\Bigr]^2\!\!\bar\varphi^2_0(t_0'')
\langle\Omega|\delta\bar{\varphi}_0(t_0'',\vec{x})\,
\delta\bar{\varphi}_0(t'_0,\vec{x}\,')|\Omega\rangle\nonumber\\
&&\hspace{-0.4cm}=\!\frac{1}{32\pi^3G}\frac{q_0^3\,q_0'^3}{(q_0^2\!\!+\!\epsilon_{i0})(q_0'^2\!\!+\!\epsilon_{i0})}
\Bigg\{\!\epsilon^{-3}_{i0}\frac{\mathcal{E}'^2_0}{4}
\!-\!\epsilon^{-2}_{i0}\Bigl[1\!\!+\!q'^2_0\Bigr]\!\ln(q'_0)\!-\!\epsilon^{-1}_{i0}
\Bigl[\ln^2(q'_0)\!-\!\mathcal{E}'_0\Bigr]\!\!+\!\!\Bigl[\epsilon^{-2}_{i0}\mathcal{E}'_0
\!\!-\!\epsilon^{-1}_{i0}\!\ln(q'^2_0)\Bigr]\nonumber\\
&&\hspace{-0.4cm}\times\!\Bigl[\ln(\alpha_{i0})\!-\!\!2\!+\!\gamma\Bigr]\!\!-\!\!\Bigl[\epsilon^{-2}_{i0}
\mathcal{E}_0\!-\!\epsilon^{-1}_{i0}\!\ln(q'^2_0)\Bigr]\!\Bigl[{\rm ci}(\alpha_{i0})\!-\!\frac{\sin(\alpha_{i0})}{\alpha_{i0}}\Bigr]
\!\!-\!\!\Bigg[\epsilon_{i0}^{-2}\Bigl[q^2_0\!\!-\!q'^2_0\Bigr]
\!\!+\!\epsilon_{i0}^{-1}\Bigl[\ln(q^2_0)\!-\!\ln(q'^2_0)\Bigr]\!\Bigg]\nonumber\\
&&\hspace{2.5cm}\times\!\Bigl[{\rm ci}(\alpha'_0)\!-\!\frac{\sin(\alpha'_0)}{\alpha'_0}\Bigr]\!\!-\!\ln(q'^2_0)
\!+\!\frac{q'^2_0}{2}\!-\!\frac{q'^{-2}_0}{2}\!+\!\frac{\mathbf{A}(q'_0,\Delta x)}{2}\!+\!\mathcal{O}(\epsilon_{i0})\!\Bigg\}\;.
\label{1loop2ndint}
\ee

\begin{figure}
\includegraphics[width=11.5cm,height=6.5cm]{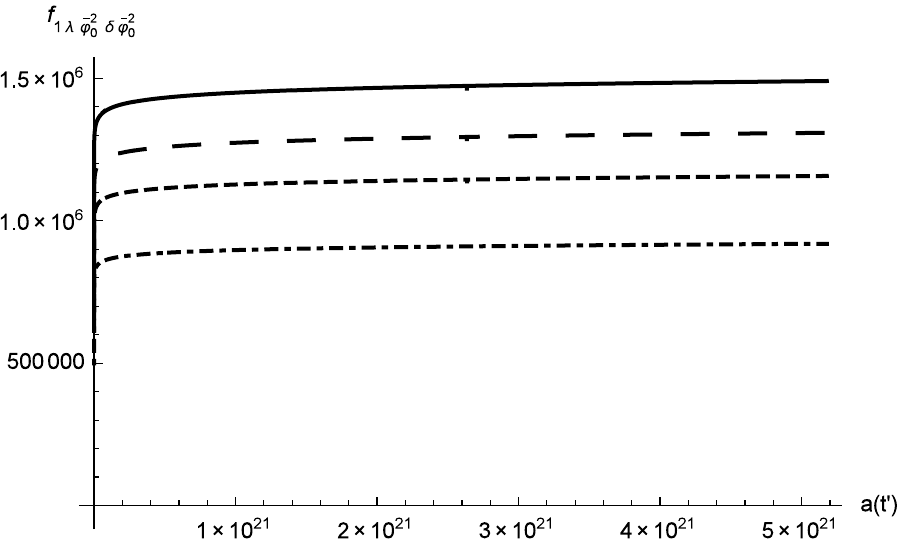}
\caption{Plots of the function $f_{1\lambda\bar\varphi^2_0\delta\bar{\varphi}^2_0}(t, t'\!, K)$, defined in Eq.~(\ref{totalproptophiekare}). The ranges of parameters $a_0(t)$, $a_0(t')$, $K$ and $\epsilon_{i0}$ are chosen same as in Fig.~\ref{fig:flucttreecorr}.}
\label{fig:fluctoneloopquad}
\end{figure}

The sum of the VEVs that are quadratic in the background and fluctuation fields at $\mathcal{O}(\lambda)$ is obtained combining Eqs. (\ref{1loop1stint}) and (\ref{1loop2ndint}) in Eq.~(\ref{corrpropphi2}),
\beeq
\hspace{0.7cm}\mathcal{V}_{\bar\varphi^2_0\delta\bar{\varphi}^2_0}
\!\!\equiv\!\!\mp\lambda\frac{H^2_{i0}}{192\pi^4}\frac{\pi}{GH^2_{i0}}f_{1\lambda\bar\varphi^2_0\delta\bar{\varphi}^2_0}(t, t'\!, \Delta x)\; ,\label{Vphisqr}
\eneq
where
\be
&&\hspace{-0.45cm}f_{1\lambda\bar\varphi^2_0\delta\bar{\varphi}^2_0}
\!\!=\!\!\frac{q_0^3\,q_0'^3}{(q_0^2\!\!+\!\epsilon_{i0})(q_0'^2\!\!+\!\epsilon_{i0})}
\Bigg\{\!\epsilon^{-3}_{i0}\frac{\mathcal{E}'^2_0}{2}
\!-\!\epsilon^{-2}_{i0}\!\!\left[1\!\!+\!q'^2_0\right]\!\ln(q'^2_0)
\!-\!\epsilon^{-1}_{i0}2\Bigl[\ln^2(q'_0)\!-\!\mathcal{E}'_0\Bigr]\!\!+\!\!\Bigl[\epsilon^{-2}_{i0}\mathcal{E}'_0
\!\!-\!\epsilon^{-1}_{i0}\!\ln(q'^2_0)\Bigr]\nonumber\\
&&\hspace{-0.2cm}\times2\Bigl[\ln(\alpha_{i0})\!-\!\!2\!+\!\!\gamma\Bigr]
\!\!-\!\!\Bigg[\epsilon^{-2}_{i0}\!
\!\left[2\!-\!q^2_0\!\!-\!q'^2_0\right]
\!\!-\!\epsilon^{-1}_{i0}\!\Bigl[\ln(q^2_0)\!\!+\!\ln(q'^2_0)\Bigr]\!\Bigg]\!\Bigl[{\rm ci}(\alpha_{i0})\!-\!\frac{\sin(\alpha_{i0})}{\alpha_{i0}}\Bigr]\!\!-\!\!\Bigg[\epsilon_{i0}^{-2}\!\Bigl[q^2_0\!\!-\!q'^2_0\Bigr]\nonumber\\
&&\hspace{0.3cm}+\epsilon_{i0}^{-1}\!\Bigl[\ln(q^2_0)\!-\!\ln(q'^2_0)\Bigr]\!\Bigg]\!\Bigl[{\rm ci}(\alpha'_0)\!-\!\frac{\sin(\alpha'_0)}{\alpha'_0}\Bigr]\!\!-\!\!2\ln(q'^2_0)\!-\!q'^{-2}_0\!\!\!+\!q'^2_0
\!\!+\!\mathbf{A}(q'_0,\Delta x)\!\!+\!\mathcal{O}(\epsilon_{i0})\!\Bigg\}.
\label{totalproptophiekare}
\ee
For comoving separation~(\ref{comovsep}) and $K\!\!=\!\!1/2$, plots of the function $f_{1\lambda\bar\varphi^2_0\delta\bar{\varphi}^2_0}(t, t'\!, K)$ versus $a_0(t')$ are given in Fig.~\ref{fig:fluctoneloopquad}. The values of $\epsilon_{i0}$ are chosen as in Figs.~\ref{fig:flucttreecorr} and \ref{fig:flucttreelambdacorr}.
The plots imply that, for a given $\epsilon_{i0}$, the one loop $\mathcal{O}(\lambda)$ correction $\mathcal{V}_{\bar\varphi^2_0\delta\bar{\varphi}^2_0}$ grows---positively or negatively, depending on the sign choice in the potential---and asymptotes to a constant at late times during inflation. The growth is suppressed as the $\epsilon_{i0}$ increases, though not as sensitively as in $\mathcal{O}(\lambda)$ correction (\ref{fzerolambda}) depicted in Fig.~\ref{fig:flucttreelambdacorr}.

To complete the computation of one-loop correlator~(\ref{1loopcorr}), we need to evaluate the remaining VEVs that are quartic in the fluctuation field. That is the task of the next section.

\subsubsection{The VEVs quartic in the fluctuation field}
\label{subsubsect:flctqrt}

\begin{figure}
\includegraphics[width=11.5cm,height=6.5cm]{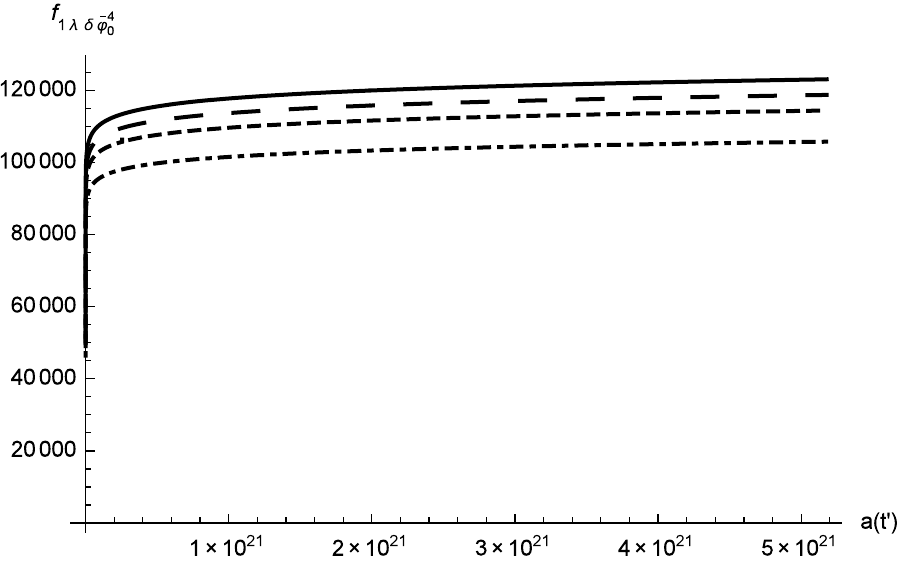}
\caption{Plots of the function $f_{1\lambda\delta\bar{\varphi}^4_0}(t, t'\!, K)$, defined in Eq.~(\ref{1loopremaintotal}). The ranges of parameters $a_0(t)$, $a_0(t')$, $K$ and $\epsilon_{i0}$ are chosen same as
in Fig.~\ref{fig:flucttreecorr}.}
\label{fig:fluctoneloopquart}
\end{figure}

The VEVs that are quartic in the fluctuation field in one-loop correlator~(\ref{1loopcorr}), for $D\!=\!4$, are
\be
&&\hspace{-0.5cm}\mathcal{V}_{\delta\bar{\varphi}^4_0}(q_0, q'_0, \Delta x)\!\equiv\!\mp\frac{\lambda}{18}\Bigg\{\!\frac{H'_0}{1\!\!+\!\epsilon'_0}\langle\Omega|
\delta\bar{\varphi}_0(x)\!\!\int_0^{t'_0}\!\!d\tilde{t}_0\Bigl[\!\frac{1\!\!+\!\tilde\epsilon_0}{\tilde{H}_0}\Bigr]^2
\!\delta\bar{\varphi}^3_0(\tilde{t}_0,\vec{x}\,')
|\Omega\rangle\nonumber\\
&&\hspace{1.45cm}+\frac{H_0}{1\!\!+\!\epsilon_0}\langle\Omega|\!\!
\int_0^{t_0}\!\!dt''_0\Bigl[\!\frac{1\!\!+\!\epsilon''_0}{H''_0}\Bigr]^2\!\delta\bar{\varphi}^3_0(t''_0,\vec{x})\,
\delta\bar{\varphi}_0(x')|\Omega\rangle\!\Bigg\} .\label{corrpropphi4}
\ee
Evaluation of the integrals in Eq.~(\ref{corrpropphi4}) are outlined in Appendix~\ref{app:oneloop-quart}. Combining Eqs.~(\ref{remaining1})-(\ref{intfor2ndsum}) in Eq.~(\ref{remainingtrms}) gives the first integral in Eq.~(\ref{corrpropphi4}) as
\be
&&\hspace{3.8cm}\frac{H'_0}{1\!\!+\!\epsilon'_0}\!\int_0^{t'_0}\!\!\!d\tilde{t}_0
\Bigl[\frac{1\!\!+\!\tilde\epsilon_0}{\tilde{H}_0}\Bigr]^2
\!\langle\Omega|\delta\bar{\varphi}_0(t_0,\vec{x})\,\delta\bar{\varphi}^3_0(\tilde{t}_0,\vec{x}\,')
|\Omega\rangle\nonumber\\
&&\hspace{-0.4cm}=\!\frac{3}{32\pi^4}\frac{H^2_{i0}q_0^3\,q_0'^3}{(q_0^2\!\!+\!\epsilon_{i0})(q_0'^2\!\!+\!\epsilon_{i0})}
\Bigg\{\!\epsilon^{-3}_{i0}\frac{\mathcal{E}'^3_0}{12}
\!+\!\epsilon^{-2}_{i0}\frac{\mathcal{E}'^2_0}{4}\Bigl[\ln(q'^2_0)\!-\!\!(1\!-\!\gamma)\Bigr]
\!\!+\!\epsilon^{-1}_{i0}\!\lefteqn{\left[\frac{\mathcal{E}'^2_0}{2}\!+\mathcal{E}'_0\Bigl[\ln^2(q'_0)\!+\!\gamma\ln(q'_0)
\!+\!1\Bigr]\right.}\nonumber\\
&&\hspace{-0.25cm}\left.+\frac{q'^{-2}_0\!-\!q'^2_0}{4}\!+\!\Bigl[1\!\!+\!\frac{q'^2_0}{2}\Bigr]\!\ln(q'^2_0)\right]
\!\!-\!\frac{1}{3}
\!+\!\frac{1\!\!-\!q'^{-4}_0}{8}\!-\!\!2\Bigl[1\!\!-\!q'^{-2}_0\Bigr]\!\Bigl[1\!\!-\!\frac{q'^2_0}{2}\Bigr]
\!\!-\!\frac{\mathcal{E}'^2_0}{4}
\!+\!\mathcal{E}'_0\Bigl[1\!\!+\!\frac{q'^{-2}_0}{2}\Bigr]\!\Bigl[\ln(q'^2_0)\!+\!\gamma\Bigr]
\nonumber\\
&&\hspace{-0.35cm}
+\!\gamma\Bigl[\ln^2(q'_0)\!\!+\!\ln(q'^3_0)\Bigr]\!\!\!+\!\frac{\ln^3(eq'^2_0)}{12}\!+\!\frac{\ln^2(eq'^2_0)}{4}
\!-\!\!\Bigg[\epsilon^{-2}_{i0}\frac{\mathcal{E}'^2_0}{4}\!+\!\epsilon^{-1}_{i0}\mathcal{E}'_0\ln(q'_0)
\!\!+\!\mathcal{E}'_0\Bigl[\!1\!\!+\!\frac{q'^{-2}_0}{2}\Bigr]\!\!\!+\!\ln^2(q'_0)\!\!+\!\ln(q'^3_0)\Bigg]\nonumber\\
&&\hspace{3.6cm}\times\Bigl[{\rm ci}(\alpha_{i0})\!-\!\frac{\sin(\alpha_{i0})}{\alpha_{i0}}\!-\!\ln(\alpha_{i0})\Bigr]\!\!+\!\frac{\mathbf{B}(q'_0,\Delta x)}{2}\!+\!\mathcal{O}(\epsilon_{i0})\!\Bigg\},\label{1loopremain1stint}
\ee
where we define
\be
&&\hspace{1.5cm}\mathbf{B}(q'_0,\Delta x)\!\equiv\!\!\sum_{n=1}^\infty\!\!\frac{(-1)^n(H_{i0}\Delta x)^{2n}}{2n(2n\!\!+\!\!1)!}e^{\frac{n}{\epsilon_{i0}}}\Bigg\{\!\epsilon^{-2}_{i0}
\Bigg[E_{-n-1}\!\Bigl(\!\frac{n}{\epsilon_{i0}}\!\Bigr)\!\!-\!q'^{2n+4}_0E_{-n-1}
\!\Bigl(\!\frac{nq'^2_0}{\epsilon_{i0}}\!\Bigr)
\nonumber\\
&&\hspace{1.5cm}-\Bigg\{\!E_{-n}\!\Bigl(\!\frac{n}{\epsilon_{i0}}\!\Bigr)
\!\!-\!q'^{2n+2}_0\!E_{-n}\!\Bigl(\!\frac{nq'^2_0}{\epsilon_{i0}}\!\Bigr)\!\!\Bigg\}\Bigg]\!\!-\!\epsilon^{-1}_{i0}\!
\Bigg[\!(2n\!+\!1)\Bigg\{\!E_{-n}\!\Bigl(\!\frac{n}{\epsilon_{i0}}\!\Bigr)\!\!-\!q'^{2n+2}_0
\!E_{-n}\!\Bigl(\!\frac{nq'^2_0}{\epsilon_{i0}}\!
\Bigr)\nonumber\\
&&\hspace{-0.45cm}-\!\Bigg[\!E_{1-n}\!\Bigl(\!\frac{n}{\epsilon_{i0}}\!\Bigr)\!\!-\!q'^{2n}_0
\!E_{1-n}\!\Bigl(\!\frac{nq'^2_0}{\epsilon_{i0}}\!\Bigr)
\!\Bigg]\!\Bigg\}\!\!-\!q'^{2n+2}_0\!E_{-n}\!\Bigl(\!\frac{nq'^2_0}{\epsilon_{i0}}\!\Bigr)\!\ln(q'^2_0)
\!\!+\!\frac{1}{(n\!\!+\!\!1)^2}\Bigg\{\!{}_2\mathcal{F}_2\Big(\!n\!\!+\!\!1,\! n\!\!+\!\!1;\! n\!\!+\!\!2,\! n\!\!+\!\!2;\!-\frac{n}{\epsilon_{i0}}\!\Big)
\nonumber\\
&&\hspace{0.4cm}-q'^{2n+2}_0{}_2\mathcal{F}_2\Big(\!n\!\!+\!\!1,\! n\!\!+\!\!1;\! n\!\!+\!\!2,\! n\!\!+\!\!2;\!-\frac{nq'^2_0}{\epsilon_{i0}}\!\Big)\!\!\Bigg\}\!\Bigg]
\!\!\!-\!\!2\Bigg[\!E_{-n}\!\Bigl(\!\frac{n}{\epsilon_{i0}}\!\Bigr)-q'^{2n+2}_0
E_{-n}\!\Bigl(\!\frac{nq'^2_0}{\epsilon_{i0}}\!
\Bigr)\!\!-\!\!\Bigg\{\!E_{1-n}\!\Bigl(\!\frac{n}{\epsilon_{i0}}\!\Bigr)\nonumber\\
&&\hspace{-0.45cm}-q'^{2n}_0\!
E_{1-n}\!\Bigl(\!\frac{nq'^2_0}{\epsilon_{i0}}\Bigr)\!\!\Bigg\}\!\Bigg]
\!\!\!+\!(2n\!\!+\!\!1)\!\Bigg[\!(n\!\!+\!1)\!\Bigg\{\!\!E_{1-n}\!\Bigl(\!\frac{n}{\epsilon_{i0}}\!\Bigr)
\!\!-\!q'^{2n}_0\!E_{1-n}\!\Bigl(\!\frac{nq'^2_0}{\epsilon_{i0}}\!\Bigr)\!\!-\!\Bigg\{\!\!E_{2-n}\!\Bigl(\!\frac{n}{\epsilon_{i0}}\!\Bigr)
\!\!-\!q'^{2n-2}_0E_{2-n}\!\Bigl(\!\frac{nq'^2_0}{\epsilon_{i0}}\!\Bigr)\!\!\Bigg\}\!\!\Bigg\}
\nonumber\\
&&\hspace{-0.2cm}-q'^{2n}_0\!E_{1-n}\!\Bigl(\!\frac{nq'^2_0}{\epsilon_{i0}}\!\Bigr)\!\ln(q'^2_0)
\!+\!\frac{1}{n^2}\Bigg\{\!{}_2
\mathcal{F}_2\Big(\!n, n; n\!\!+\!\!1, n\!\!+\!\!1;-\frac{n}{\epsilon_{i0}}\!\Big)\!\!-\!q'^{2n}_0{}_2\mathcal{F}_2\Big(\!n, n; n\!\!+\!\!1, n\!\!+\!\!1;-\frac{nq'^2_0}{\epsilon_{i0}}\!\Big)\!\!\Bigg\}\!\Bigg]\!\Bigg\} .\nonumber\\\label{Gnq}
\ee
Combining Eqs.~(\ref{frstprtintscnd})-(\ref{2ndint2ndprtint}) in Eq.~(\ref{barphi4scndint})
yields the second integral in Eq.~(\ref{corrpropphi4}) as
\be
&&\hspace{3.5cm}\frac{H_0}{1\!\!+\!\epsilon_0}
\!\int_0^{t_0}\!\!dt''_0\Bigl[\frac{1\!\!+\!\epsilon''_0}{H''_0}\Bigr]^2\!\langle\Omega|\delta\bar{\varphi}^3_0(t''_0,\vec{x})
\,\delta\bar{\varphi}_0(t'_0,\vec{x}\,')|\Omega\rangle\nonumber\\
&&\hspace{-0.25cm}=\!\frac{3}{32\pi^4}\frac{H^2_{i0}q_0^3\,q_0'^3}{(q_0^2\!\!+\!\epsilon_{i0})(q_0'^2\!\!+\!\epsilon_{i0})}
\Bigg\{\!\epsilon^{-3}_{i0}\frac{\mathcal{E}'^3_0}{12}
\!+\!\epsilon^{-2}_{i0}\frac{\mathcal{E}'^2_0}{4}\Bigl[\ln(q'^2_0)\!-\!(1\!\!-\!\!\gamma\!)\Bigr]
\!\!+\!\epsilon^{-1}_{i0}\!\lefteqn{\left[\frac{\mathcal{E}'^2_0}{2}\!+\!\mathcal{E}'_0\Bigl[\ln^2\!(q'_0)
\!+\!\!\gamma\ln(q'_0)
\!+\!1\Bigr]\right.}\nonumber\\
&&\hspace{-0.25cm}\left.-\frac{q'^2_0\!\!-\!q'^{-2}_0}{4}\!+\!\Bigl[1\!\!+\!\frac{q'^2_0}{2}\Bigr]\!\ln(q'^2_0)\right]
\!\!-\!\frac{1}{3}
\!+\!\frac{1\!\!-\!q'^{-4}_0}{8}\!-\!\!2\Bigl[1\!\!-\!q'^{-2}_0\Bigr]\!\Bigl[1\!\!-\!\frac{q'^2_0}{2}\Bigr]
\!-\!\frac{\mathcal{E}'^2_0}{4}\!+\!\mathcal{E}'_0\Bigl[1\!\!+\!\frac{q'^{-2}_0}{2}\Bigr]
\!\Bigl[\ln(q'^2_0)\!+\!\gamma\Bigr]\nonumber\\
&&\hspace{-0.2cm}
+\gamma\Bigl[\ln^2(q'_0)\!\!+\!\ln(q'^3_0)\Bigr]\!\!+\!\frac{\ln^3(eq'^2_0)}{12}\!+\!\frac{\ln^2(eq'^2_0)}{4}
\!-\!\!\left\{\!\epsilon^{-2}_{i0}\frac{\mathcal{E}^2_0}{4}
\!\!+\!\epsilon^{-1}_{i0}\mathcal{E}_0\!\ln(q_0)\!\!+\!\mathcal{E}_0
\!\Bigl[\!1\!\!+\!\frac{q^{-2}_0}{2}\Bigr]\!\!\!+\!\ln^2(q_0)
\!\!+\!\ln(q^3_0)\!\right\}
\nonumber\\
&&\hspace{0.7cm}\times\!\!\left[{\rm ci}(\alpha_{i0})\!-\!\frac{\sin(\alpha_{i0})}{\alpha_{i0}}\right]
\!\!+\!\left\{\!\epsilon^{-2}_{i0}\frac{\mathcal{E}'^2_0}{4}+\epsilon^{-1}_{i0}\mathcal{E}'_0\ln(q'_0)
\!+\!\mathcal{E}'_0\Bigl[1\!\!+\!\frac{q'^{-2}_0}{2}\Bigr]\!\!+\!\ln^2(q'_0)
\!+\!\ln(q'^3_0)\!\right\}\!\ln(\alpha_{i0})\nonumber\\
&&\hspace{0.6cm}+\Bigg\{\!\frac{\epsilon^{-2}_{i0}}{4}
\Bigl[\mathcal{E}^2_0\!-\!\mathcal{E}'^2_0\Bigr]\!\!+\!\epsilon^{-1}_{i0}\!\Bigl[\mathcal{E}_0\ln(q_0)
\!-\!\mathcal{E}'_0\ln(q'_0)\Bigr]
\!\!+\!\mathcal{E}_0\Bigl[\!1\!\!+\!\frac{q^{-2}_0}{2}\Bigr]\!\!-\!\mathcal{E}'_0
\Bigl[\!1\!\!+\!\frac{q'^{-2}_0}{2}\Bigr]
\!\!+\!\ln^2(q_0)\!-\!\ln^2(q'_0) \nonumber\\
&&\hspace{3.4cm}
\!+\!\ln(q^3_0)\!-\!\ln(q'^3_0)\!\Bigg\}\!\Bigl[{\rm ci}(\alpha'_0)\!-\!\frac{\sin(\alpha'_0)}{\alpha'_0}\Bigr]\!\!+\!\frac{\mathbf{B}(q'_0,\Delta x)}{2}\!+\!\mathcal{O}(\epsilon_{i0})\!\Bigg\}\; .
\label{1loopremain2ndint}
\ee
Adding up Eqs.~(\ref{1loopremain1stint}) and (\ref{1loopremain2ndint}) yields the contribution due to the VEVs that are quartic in the fluctuation field given in Eq.~(\ref{corrpropphi4}) as
\beeq
\mathcal{V}_{\delta\bar{\varphi}^4_0}\!\!\equiv\!\!\mp\lambda\frac{H^2_{i0}}{192\pi^4}
f_{1\lambda\delta\bar{\varphi}^4_0}(t, t'\!, \Delta x)\; ,\label{Vphi4}
\eneq
where
\be
&&\hspace{-0.3cm}f_{1\lambda\delta\bar{\varphi}^4_0}\!=\! \frac{q_0^3\,q_0'^3}{(q_0^2\!\!+\!\epsilon_{i0})(q_0'^2\!\!+\!\epsilon_{i0})}
\Bigg\{\!\epsilon^{-3}_{i0}\frac{\mathcal{E}'^3_0}{6}
\!+\!\epsilon^{-2}_{i0}\mathcal{E}'^2_0\!\Bigl[\ln(q'_0)\!-\!\frac{1\!\!-\!\!\gamma}{2}\Bigr]
\!\!\!+\!\lefteqn{\epsilon^{-1}_{i0}\!\Bigl[\mathcal{E}'^2_0\!\!+\!2\mathcal{E}'_0\Bigl[\ln^2(q'_0)
\!+\!\gamma\!\ln(q'_0)\!+\!1\Bigr]\Bigr.}\nonumber\\
&&\hspace{-0.25cm}\Bigl.+\frac{q'^{-2}_0}{2}\!-\!\frac{q'^2_0}{2}\!+\!\Bigl[\!2\!+\!q'^2_0\Bigr]
\!\ln(q'^2_0)\Bigr]
\!-\!\!\frac{2}{3}
\!+\!\frac{\!1\!\!-\!q'^{-4}_0}{4}\!-\!4\Bigl[\!1\!\!-\!q'^{-2}_0\Bigr]\!\!-\!\!2\mathcal{E}'_0
\!\!-\!\!\frac{\mathcal{E}'^2_0}{2}
\!+\!\Bigl[\!1\!\!+\!q'^{-2}_0\!\!-\!\!2 q'^2_0\Bigr]\!\Bigl[\ln(q'^2_0)\!+\!\gamma\Bigr]
\nonumber\\
&&\hspace{-0.25cm}
+2\gamma\Bigl[\ln^2(q'_0)\!+\!\ln(q'^3_0)\Bigr]\!\!+\!\frac{\ln^3(eq'^2_0)}{6}\!+\!\frac{\ln^2(eq'^2_0)}{2}
\!+\!\Bigg\{\!\epsilon^{-2}_{i0}\frac{\mathcal{E}'^2_0}{4}\!+\!\epsilon^{-1}_{i0}\mathcal{E}'_0\ln(q'_0)\!-\!
\frac{1\!\!-\!q'^{-2}_0}{2}\!+\!\mathcal{E}'_0\!+\!\ln^2(q'_0)
\nonumber\\
&&\hspace{-0.4cm}
+\ln(q'^3_0)\!\Bigg\}\!\ln(\alpha^2_{i0})\!-\!\Bigg\{\!\frac{\epsilon^{-2}_{i0}}{4}\Bigl[
\mathcal{E}^2_0
\!\!+\!\mathcal{E}'^2_0\Bigr]\!\!+\!\epsilon^{-1}_{i0}\Bigl[\mathcal{E}_0\!\ln(q_0)\!+\!\mathcal{E}'_0\!\ln(q'_0)\Bigr]
\!\!+\!\frac{q^{-2}_0}{2}\!+\!\frac{q'^{-2}_0}{2}\!-\!\!\left[q^2_0\!\!+\!q'^2_0\right]
\!\!+\!\!1\!\!+\!\ln^2(q_0)\nonumber\\
&&\hspace{-0.45cm}+\ln^2(q'_0)\!
\!+\!\ln(q^3_0)\!+\!\ln(q'^3_0)\!\Bigg\}\!\!\left[{\rm ci}(\alpha_{i0})\!-\!\!\frac{\sin(\alpha_{i0})}{\alpha_{i0}}\right]
\!\!+\!\!\Bigg\{\!\frac{\epsilon^{-2}_{i0}}{4}\Bigl[\mathcal{E}^2_0
\!-\!\mathcal{E}'^2_0\Bigr]
\!\!+\!\epsilon^{-1}_{i0}\Bigl[\mathcal{E}_0\!\ln(q_0)\!-\!\mathcal{E}'_0\!\ln(q'_0)\Bigr]\!\!+\!
\frac{q^{-2}_0}{2}\nonumber\\
&&\hspace{-0.55cm}-
\frac{q'^{-2}_0}{2}\!-\!\!\left[q^2_0\!\!-\!q'^2_0\right]\!\!\!+\!\ln^2(q_0)\!-\!\!\ln^2(q'_0)\!\!+\!\ln(q^3_0)\!-\!\ln(q'^3_0)
\!\Bigg\}\!\!\!\left[{\rm ci}(\alpha'_0)\!-\!\frac{\sin(\alpha'_0)}{\alpha'_0}\right]\!\!\!+\!\!\mathbf{B}(q'_0,\!\Delta x)\!\!+\!\mathcal{O}(\epsilon_{i0})\!\Bigg\}.
\label{1loopremaintotal}
\ee
Plots of the function $f_{1\lambda\delta\bar{\varphi}^4_0}(t, t'\!, K)$ versus $a_0(t')$,
for comoving separation~(\ref{comovsep}), $K\!\!=\!\!1/2$ and the four values of $\epsilon_{i0}$ that are used in the preceding figures, are given in Fig.~\ref{fig:fluctoneloopquart}.
They imply that, for a chosen $\epsilon_{i0}$, the one loop $\mathcal{O}(\lambda)$ correction $\mathcal{V}_{\delta\bar{\varphi}^4_0}$ grows---positively or negatively, depending on the sign choice in the potential---and asymptotes to an almost constant value at late times during inflation. The growth is only marginally suppressed as the $\epsilon_{i0}$ increases.
\subsubsection{The result}
The one-loop correlator at $\mathcal{O}(\lambda)$ is obtained mingling contributions~(\ref{Vphisqr}) and (\ref{Vphi4}) in Eq~(\ref{1loopcorr}), setting $D\!=\!4$, as
\beeq
\langle\Omega|
\delta\bar{\varphi}(t,\vec{x})\,\delta\bar{\varphi}(t'\!,\vec{x}\,')|\Omega\rangle_{\rm 1-loop}\!\simeq\!\mp\lambda\frac
{H^2_{i0}}{192\pi^4}\,\emph{f}_{1\lambda}(t,t'\!,\Delta x)\;,\label{1loopmassivefinal}
\eneq
where \be
&&\hspace{-0.3cm}\emph{f}_{1\lambda}\!\!=\!\!\frac{q^3_0q'^3_0}{(q^2_0\!\!+\!\!\epsilon_{i0})(q'^2_0\!\!+\!\!\epsilon_{i0})}
\Bigg\{\!\frac{\pi}{GH^2_{i0}}\!\Bigg\{\!\epsilon^{-3}_{i0}\frac{\mathcal{E}'^2_0}{2}
\!-\!\epsilon^{-2}_{i0}\!\!\left[1\!\!\!+\!q'^2_0\right]\!\ln(q'^2_0)
\!-\!\epsilon^{-1}_{i0}\!2\Bigl[\!\ln^2(q'_0)\!-\!\mathcal{E}'_0\Bigr]
\!\!\!+\!\!\Bigl[\epsilon^{-2}_{i0}\mathcal{E}'_0\!-\!\epsilon^{-1}_{i0}\!\ln(q'^2_0)\Bigr]\nonumber\\
&&\hspace{0.1cm}\times\!\Bigl[\ln(\alpha^2_{i0})\!-\!\!4\!\!+\!\!2\gamma\Bigr]\!\!-\!\!\Bigg[\epsilon^{-2}_{i0}
\!\!\left[2\!-\!q^2_0\!\!-\!q'^2_0\right]
\!\!-\!\epsilon^{-1}_{i0}\!\Bigl[\ln(q^2_0)\!+\!\ln(q'^2_0)\Bigr]\!\Bigg]\!\Bigl[{\rm ci}(\alpha_{i0})\!-\!\!\frac{\sin(\alpha_{i0})}{\alpha_{i0}}\Bigr]
\!\!-\!\!\Bigg[\epsilon_{i0}^{-2}\!\Bigl[q^2_0\!\!-\!q'^2_0\Bigr]\nonumber\\
&&\hspace{-0.36cm}
+\epsilon_{i0}^{-1}\!\Bigl[\ln(q^2_0)\!-\!\ln(q'^2_0)\Bigr]\!\Bigg]\!\Bigl[{\rm ci}(\alpha'_0)\!-\!\!\frac{\sin(\alpha'_0)}{\alpha'_0}\Bigr]\!\!-\!\!2\ln(q'^2_0)\!-\!q'^{-2}_0
\!\!+\!q'^2_0\!\!+\!\!\mathbf{A}\!(q'_0,\!\Delta x)\!\Bigg\}\!\!\!+\!\epsilon^{-3}_{i0}\frac{\mathcal{E}'^3_0}{6}
\!\!+\!\epsilon^{-2}_{i0}\mathcal{E}'^2_0\!\Bigl[\ln(q'_0)\nonumber\\
&&\hspace{-0.1cm}-\frac{1\!\!-\!\!\gamma}{2}\Bigr]
\!\!+\!\epsilon^{-1}_{i0}\!\!\left[\mathcal{E}'^2_0\!\!\!+\!\!2\mathcal{E}'_0\Bigl[\ln^2(q'_0)\!\!+\!\!\gamma\!\ln(q'_0)
\!+\!\!1\Bigr]\!\!+\!\!\frac{q'^{-2}_0\!\!\!-\!q'^2_0}{2}\!+\!\!\Bigl[\!2\!\!+\!q'^2_0\Bigr]\!\!\ln(q'^2_0)\right]
\!\!-\!\frac{2}{3}\!\!+\!\!\frac{1\!\!-\!q'^{-4}_0}{4}\!-\!4\!\Bigl[\!1\!\!-\!q'^{-2}_0\Bigr]
\nonumber\\
&&\hspace{-0.15cm}
-\!2\mathcal{E}'_0\!\!-\!\frac{\mathcal{E}'^2_0}{2}\!+\!\!\Bigl[1\!\!+\!q'^{-2}_0\!\!-\!\!2q'^2_0\Bigr]\!\Bigl[\ln(q'^2_0)\!+\!\gamma\Bigr]
\!\!+\!\!2\gamma\Bigl[\ln^2(q'_0)\!+\!\ln(q'^3_0)\Bigr]\!\!+\!\frac{\ln^3(eq'^2_0)}{6}\!+\!\frac{\ln^2(eq'^2_0)}{2}
\!+\!\Bigg\{\!\epsilon^{-2}_{i0}\frac{\mathcal{E}'^2_0}{4}
\nonumber
\ee
\be
&&\hspace{-0.3cm}
+\epsilon^{-1}_{i0}\mathcal{E}'_0\!\ln(q'_0)\!-\!
\frac{1\!\!-\!q'^{-2}_0}{2}\!+\!\mathcal{E}'_0\!\!+\!\ln^2(q'_0)
\!+\!\ln(q'^3_0)\!\Bigg\}\!\ln(\alpha^2_{i0})\!-\!\!\Bigg\{\!\frac{\epsilon^{-2}_{i0}}{4}\!
\Bigl[\mathcal{E}^2_0
\!\!+\!\mathcal{E}'^2_0\Bigr]\!\!+\!\epsilon^{-1}_{i0}\!\Bigl[\mathcal{E}_0\!\ln(q_0)\!\!+\!\mathcal{E}'_0\!\ln(q'_0)\Bigr]\nonumber\\
&&\hspace{-0.3cm}+\frac{q^{-2}_0\!\!+\!q'^{-2}_0}{2}
\!-\!\!\left[q^2_0\!\!+\!q'^2_0\right]\!\!+\!\!1\!\!+\!\ln^2(q_0)\!\!+\!\ln^2(q'_0)
\!\!+\!\ln(q^3_0)\!+\!\ln(q'^3_0)\!\Bigg\}\!\!\left[{\rm ci}(\alpha_{i0})\!-\!\frac{\sin(\alpha_{i0})}{\alpha_{i0}}\right]\!\!+\!\!\Bigg\{\!\frac{\epsilon^{-2}_{i0}}{4}\Bigl[\mathcal{E}^2_0
\!\!-\!\mathcal{E}'^2_0\Bigr]\nonumber\\
&&\hspace{0.5cm}+\epsilon^{-1}_{i0}\!\Bigl[\mathcal{E}_0\ln(q_0)\!-\!\!\mathcal{E}'_0\ln(q'_0)\Bigr]\!\!+\!
\frac{q^{-2}_0\!\!-\!q'^{-2}_0}{2}\!-\!\!\left[q^2_0\!\!-\!q'^2_0\right]\!\!+\!\ln^2(q_0)\!-\!\ln^2(q'_0)
\!+\!\ln(q^3_0)\!-\!\ln(q'^3_0)
\!\Bigg\}
\nonumber\\
&&\hspace{4.5cm}\times\!\!\left[{\rm ci}(\alpha'_0)\!-\!\frac{\sin(\alpha'_0)}{\alpha'_0}\right]\!\!\!+\!\mathbf{B}(q'_0,\Delta x)\!+\!\mathcal{O}(\epsilon_{i0})\!\Bigg\}
\; .\label{f1loop}
\ee
In the next section, we obtain the coincident one-loop correlator.

\subsubsection{Coincident one-loop correlator}
\label{subsubsect:coin1loopcorr}
Using the equal spacetime limit of Eq.~(\ref{f1loop}) in Eq.~(\ref{1loopmassivefinal}) yields the coincident one-loop correlator at $\mathcal{O}(\lambda)$ as
\be
&&\hspace{0.05cm}\langle\Omega|
\delta\bar{\varphi}^2(t,\vec{x})|\Omega\rangle_{\rm 1-loop}\!=\!\mp\lambda\frac
{H^2_{i0}}{192\pi^4}\frac{q^6_0}{\left[q^2_0\!\!+\!\!\epsilon_{i0}\right]^2}
\Bigg\{\!\frac{\pi}{GH^2_{i0}}\!\Bigg[\epsilon^{-3}_{i0}\frac{\mathcal{E}^2_0}{2}
\!-\!\epsilon^{-2}_{i0}2\!\left[\ln(q^2_0)\!+\!\mathcal{E}_0\Bigl[1\!\!-\!\frac{\ln(q^2_0)}{2}\Bigr]\right]\nonumber\\
&&\hspace{0.25cm}+\epsilon^{-1}_{i0}2\Bigl[\ln(q^2_0)\!-\!\frac{\ln^2(q^2_0)}{4}\!+\!\mathcal{E}_0\Bigr]\!\!-\!2\ln(q^2_0)
\!-\!q^{-2}_0\Bigl[2\mathcal{E}_0\!-\!\mathcal{E}_0^2\Bigr]
\!\!+\!\mathcal{O}(\epsilon_{i0})\Bigg]\!\!\!+\!\epsilon^{-3}_{i0}\frac{\mathcal{E}^3_0}{6}
\!+\!\epsilon^{-2}_{i0}\mathcal{E}_0^2\frac{\ln(q^2_0)}{2}\nonumber\\
&&\hspace{-0.3cm}+\epsilon^{-1}_{i0}3\!\left[\ln(q^2_0)
\!+\!\mathcal{E}_0q^{-2}_0\Bigl[1\!\!+\!\frac{\ln^2(q^2_0)}{6}\!-\!\frac{\mathcal{E}_0}{2}
\Bigl[1\!\!+\!\frac{\ln^2(q^2_0)}{3}\Bigr]
\!\!-\!\frac{\mathcal{E}^2_0}{3}\Bigr]\right]
\!\!+\!\frac{9}{2}q^{-2}_0\ln(q^2_0)\!+\!\frac{3}{2}\ln^2(q^2_0)\!+\!\frac{\ln^3(q^2_0)}{6}
\nonumber\\
&&\hspace{0cm}+\frac{9}{2}\mathcal{E}_0q^{-4}_0\!\!\left[1\!\!-\!\frac{\ln(q^2_0)}{3}
\!-\!\frac{7}{6}\mathcal{E}_0\Bigl[1\!\!+\!\frac{2}{21}\ln(q^2_0)\Bigr]
\!\!+\!\frac{2}{9}\mathcal{E}_0^2\Bigl[1\!\!+\!2\ln(q^2_0)\Bigr]\!\!-\!\frac{\mathcal{E}_0^3}{9}\right]
\!\!+\!\mathcal{O}(\epsilon_{i0})\Bigg\}\!\!+\!\mathcal{O}(\lambda^2)\;.
\label{1loopeqlspctm}
\ee
Adding up the tree-order coincident correlator at $\mathcal{O}(\lambda^0)$ and $\mathcal{O}(\lambda)$, and the one-loop coincident correlator at $\mathcal{O}(\lambda)$ we obtain the coincident correlator up to $\mathcal{O}(\lambda^2)$. The result is given in the next section and is used to compute the quantum corrected power spectrum up to $\mathcal{O}(\lambda^2)$ in Sec.~\ref{sec:power}.

\section{Coincident correlation function of the fluctuations}
\label{sec:coincorr}
Coincident correlation function of the inflaton fluctuations is obtained by adding up Eqs.~(\ref{0tree0}), (\ref{coincorrtree}) and (\ref{1loopeqlspctm}) as
\be
&&\hspace{0cm}\langle\Omega|\delta{\bar\varphi}^2(t,\vec{x})|\Omega\rangle_{\rm stoch}\!=\!
\frac{H_{i0}^2}{4\pi^2}\frac{q^4_0}{\left[q^2_0\!\!+\!\epsilon_{i0}\right]^2}
\Bigg\{\!\Bigl[1\!\!-\!\mathcal{E}_0\Bigr]\!\Bigg\{\!\ln(a_0)\!-\!
\ln\Bigl(\!\frac{q_0^2\!\!+\!\epsilon_{i0}}{q_0^3\!\left(1\!\!+\!\epsilon_{i0}\right)}\Bigr)\!\Bigg\}\nonumber
\ee
\be
&&\hspace{-0.4cm}\mp\frac{\lambda}{144\pi^2}\Bigg\{\!\frac{\pi}{GH^2_{i0}}\Bigg\{\!\epsilon^{-3}_{i0}\Bigg[\!1\!\!-\!q_0
\!\!-\!\!\frac{7}{2}\mathcal{E}_0\Bigl[1\!\!-\!\frac{5}{7}q_0\Bigr]
\!\!+\!\frac{37}{8}\mathcal{E}^2_0\Bigl[1\!\!-\!\frac{4}{37}q_0\Bigr]\!\!-\!\frac{9}{4}\mathcal{E}^3_0\Bigg]
\!\!-\!\epsilon^{-2}_{i0}q_0^{-2}\Bigg[\!1\!\!-\!q_0
\!+\!\frac{15}{2}\Bigl[1\!\!-\!\frac{2}{15}q_0\Bigr]\nonumber\\
&&\hspace{-0.3cm}\times\!\ln(q^2_0)\!\!+\!\frac{9}{2}\mathcal{E}_0\!\!\left[\!1\!\!-\!\frac{4}{9}q_0\!-\!\frac{37}{9}
\Bigl[1\!\!-\!\frac{2}{37}q_0\Bigr]\!\ln(q^2_0)\right]
\!\!-\!\frac{79}{8}\mathcal{E}^2_0\Bigl[\!1\!\!-\!\frac{12}{79}q_0\!\!-\!\frac{117}{79}\!\ln(q^2_0)\Bigr]
\!\!+\!\frac{37}{8}\mathcal{E}^3_0
\!\left[\!1\!\!-\!\frac{29}{37}\ln(q^2_0)\right]\!\!\Bigg]\nonumber\\
&&\hspace{0cm}+\epsilon^{-1}_{i0}q_0^{-4}3\!\Bigg[\!1\!\!-\!q_0\!\!-\!\frac{5}{3}\!\Bigl[\!1\!\!+
\!\frac{3}{5}q_0\Bigr]\!\ln(q^2_0)
\!+\!\frac{\ln^2(q^2_0)}{2}\!-\!\frac{17}{3}\mathcal{E}_0\!
\left[\!1\!\!-\!\frac{19}{34}q_0\!-\!\Bigl[\!1\!\!+\!\frac{9}{34}q_0\Bigr]\!\ln(q^2_0)
\!+\!\frac{9}{34}\ln^2(q^2_0)\right]\nonumber\\
&&\hspace{-0.25cm}+\frac{121}{12}\mathcal{E}^2_0\!\!\left[\!1\!\!-\!\frac{20}{121}q_0\!-\!\frac{75}{121}\!
\Bigl[\!1\!\!+\!\!\frac{2}{25}q_0\Bigr]\!\ln(q^2_0)
\!+\!\frac{18}{121}\!\ln^2(q^2_0)\right]\!\!-\!\frac{95}{12}\mathcal{E}^3_0
\Bigl[\!1\!\!-\!\!\frac{27}{95}\ln(q^2_0)\!\!+\!\frac{6}{95}\!\ln^2(q^2_0)\Bigr]
\!\!\!+\!\frac{29}{12}\mathcal{E}^4_0\Bigg]\nonumber\\
&&\hspace{-0.1cm}+q_0^{-6}3\Bigg[\!1\!\!-\!q_0\!\!-\!\frac{5}{3}\Bigl[\!1\!\!+\!\!\frac{3}{5}q_0\Bigr]\!\ln(q^2_0)
\!-\!\frac{17}{3}{\mathcal{E}_0}\!\left[\!1\!\!-\!\frac{7}{34}q_0\!-\!\frac{23}{17}\!
\Bigl[\!1\!+\!\!\frac{9}{46}q_0\Bigr]\!\ln(q^2_0)\right]\!\!+\!\frac{155}{12}\mathcal{E}^2_0
\!\left[\!1\!\!+\!\!\frac{12}{155}q_0\!-\!\frac{147}{155}\right.\nonumber\\
&&\hspace{0.1cm}\times\!\!\left.\Bigl[\!1\!\!+\!\!\frac{2}{49}q_0\Bigr]\!\ln(q^2_0)\!\right]
\!\!-\!\!\frac{175}{12}\mathcal{E}^3_0\!\!\left[\!1\!\!+\!\frac{8}{175}q_0\!\!-\!\frac{99}{175}\!\ln(q^2_0)\right]
\!\!+\!\!\frac{179}{24}\mathcal{E}^4_0\!\!\left[\!1\!\!-\!\frac{48}{179}\ln(q^2_0)\right]
\!\!-\!\frac{29}{24}\mathcal{E}^5_0\Bigg]\!\!\!+\!\mathcal{O}(\epsilon_{i0})\!\Bigg\}
\nonumber\\
&&\hspace{-0.2cm}+\Bigl[1\!\!-\!\mathcal{E}_0\Bigr]\!\!\left[\epsilon^{-3}_{i0}\frac{\mathcal{E}^3_0}{2}
\!+\!\epsilon^{-2}_{i0}\frac{3}{2}\mathcal{E}_0^2\!\ln(q^2_0)\right]\!\!+\!\epsilon^{-1}_{i0}
9\Bigg[\!\!\ln(q^2_0)\!+\!\mathcal{E}_0\!\!
\left[\!1\!\!-\!\ln(q^2_0)\!+\!\frac{\ln^2(q^2_0)}{6}\right]
\!\!-\!\!\frac{\mathcal{E}_0^2}{2}\!\!\left[\!1\!\!+\!\frac{\ln^2(q^2_0)}{3}\!\right]\!\!-\!\frac{\mathcal{E}_0^3}{3}
\!\Bigg]\nonumber\\
&&\hspace{0.8cm}+\frac{27}{2}\ln(q^2_0)
\!+\!\frac{9}{2}\Bigl[\!1\!\!-\!\mathcal{E}_0\Bigr]\!\!\left[\ln^2(q^2_0)
\!+\!\frac{\ln^3(q^2_0)}{9}\right]\!\!+\!\frac{27}{2}\mathcal{E}_0q^{-2}_0\!\!
\left[\!1\!\!-\!\frac{\ln(q^2_0)}{3}\!-\!\frac{7}{6}\mathcal{E}_0\!\!
\left[\!1\!\!+\!\frac{2}{21}\!\ln(q^2_0)\right]\right.\nonumber\\
&&\hspace{4.5cm}\left.+\frac{2}{9}\mathcal{E}_0^2\Bigl[\!1\!\!+\!\!2\ln(q^2_0)\Bigr]\!\!-\!\frac{\mathcal{E}_0^3}{9}\right]
\!\!+\!\mathcal{O}(\epsilon_{i0})\!\Bigg\}\!\!+\!\mathcal{O}(\lambda^2) \;.\label{coincorrQ}
\ee

Correlation function and the power spectrum are related by the Wiener-Khinchin theorem which states that the spatial correlation between simultaneous values of the field measured at two spatial points and power spectrum form a Fourier transform pair of each other. In the next section, we use coincident correlator (\ref{coincorrQ}) to compute the quantum corrected power spectrum of inflaton fluctuations up to $\mathcal{O}(\lambda^2)$.

\section{Power Spectrum of the fluctuations}
\label{sec:power}
The spatial two-point correlation function for an untruncated  {\it free} fluctuation field is obtained using free field mode expansion~(\ref{phizero}) as
\beeq
\langle\Omega|\delta\varphi_0(t,\vec{x})\delta\varphi_0(t\!,\vec{x}\,')|\Omega\rangle\!=\!\!
\int\!\!\frac{d^{D-1}k}{(2\pi)^{D-1}}|u_0(t, k)|^2 e^{i\vec{k}\cdot(\vec{x}-\vec{x}\,')}\; .\label{corrpow}
\eneq
The norm square of amplitude function $|u_0(t, k)|^2$ is a measure of power for each mode $k$ of the fluctuation field at time $t$. It is, however, convenient to evaluate the integral in Eq.~(\ref{corrpow}) and define a measure of power $\Delta^2_{\delta\varphi_0}(t, k)$ as
\beeq
\hspace{0.3cm}\langle\Omega|\delta\varphi_0(t,\vec{x})\delta\varphi_0(t\!,\vec{x}\,')|\Omega\rangle\!=\!\!
\int_0^\infty\!\frac{dk}{k}\frac{\sin(k\Delta x)}{k\Delta x}\Delta^2_{{\delta\varphi}_0}(t, k)\; ,\label{treecorrpow}
\eneq
where the time dependent tree-order power spectrum
\beeq\hspace{0.5cm}\Delta^2_{{\delta\varphi}_0}(t, k)\!=\!\frac{k^{D-1}}{\pi^\frac{D}{2}}\frac{\Gamma\!\left(\frac{D}{2}\right)}{\Gamma\!\left(D\!-\!\!1\right)}|u_0(t, k)|^2\; . \label{treemodepow}\eneq
The quantum corrected fluctuation field $\delta\varphi(t,\vec{x})$ [Eq.~(\ref{fullfieldexp})] is a {\it full} field hence, unlike the free field, it does not have a mode expansion. The quantum corrected power spectrum $\Delta^2_{\delta\varphi}(t, k)$ can be defined \cite{MP} in two different ways: (i) using the quantum corrected two point correlation function via Eq.~(\ref{treecorrpow}) or (ii) using the quantum corrected mode function in Eq.~(\ref{treemodepow}). The two definitions yield slightly different results. We use definition (i) to compute quantum corrected power spectrum in this paper.

In the coincidence limit, Eq.~(\ref{treecorrpow}) implies \beeq
\hspace{0cm}
\langle\Omega|\delta{\varphi}^2(t,\vec{x})|\Omega\rangle\!\equiv\!\!
\int_0^\infty\!\frac{dk}{k}\Delta^2_{\delta\varphi}(t, k)\; .
\eneq
The stochastic contribution to coincident correlator, on the other hand, ought to involve the usual \cite{Dod} time-independent power spectrum
\be\hspace{0.5cm}\Delta^2_{\delta\varphi}(k)\!\equiv\!\lim_{t\gg t_k}\Delta^2_{\delta\varphi}(t, k)\; ,\ee
where $t_k$ is the time of first horizon crossing defined by Eq.~(\ref{TK}). Thus, for the IR truncated full fluctuation field
\beeq
\hspace{2.3cm}\langle\Omega|\delta{\bar\varphi}^2(t,\vec{x})|\Omega\rangle_{\rm stoch}\!\equiv\!\!
\int_0^\infty\!\frac{dk}{k}\Theta\Bigl(\!\frac{Ha}{1\!\!+\!\epsilon}\!-\!k\Bigr)\Delta^2_{\delta\bar\varphi}(k)
\!=\!\!\int_{k_i=\frac{H_i}{1\!+\!\epsilon_i}}^{k=\frac{Ha}{1\!+\!\epsilon}}\frac{dk}{k}\Delta^2_{\delta\bar\varphi}(k)\; .
\eneq
Taking the time derivative of both sides yields
\beeq
\frac{d}{dt}\langle\Omega|\delta{\bar\varphi}^2(t,\vec{x})|\Omega\rangle_{\rm stoch}\!=\!\frac{\dot{k}}{k}\Delta^2_{\delta\bar\varphi}(k)\Big|_{k=\frac{Ha}{1\!+\!\epsilon}}\; ,
\eneq
where
\beeq
\frac{\dot{k}}{k}\,\Big|_{k=\frac{Ha}{1\!+\!\epsilon}}\!=\!\left[H[1\!\!-\!\epsilon]
\!-\!\frac{\dot{\epsilon}}{1\!\!+\!\epsilon}\right]\; .\label{dotk}
\eneq
Thus, the power for the horizon sized mode $k$ at time $t$ (i.e. $t\!=\!t_k$) is
\be
\hspace{1.5cm}\Delta^2_{\delta\bar\varphi}\Bigl(\!k\!=\!\frac{Ha}{1\!\!+\!\epsilon}\Bigr)\!\!=\!\!
\left[H[1\!\!-\!\epsilon]\!-\!\frac{\dot{\epsilon}}{1\!\!+\!\epsilon}\right]^{-1}\!\!\frac{d}{dt}
\langle\Omega|\delta{\bar\varphi}^2(t,\vec{x})|\Omega\rangle_{\rm stoch}\;.\label{stochpow}
\ee
One can, however, look at the power at {\it any} time $t$ and that time corresponds to a mode $k$ through the relation $a\!=\!k\frac{1+\epsilon}{H}$. So by considering Eq.~(\ref{stochpow}) at different times one can access the power for different $k$ values. Employing Eq.~(\ref{chnvarder}), converting the derivative with respect to $q$ to the derivative with respect to $q_0$ via the chain rule
\beeq
\hspace{0.3cm}\frac{d}{dq}\!=\![\frac{dq}{dq_0}]^{-1}\frac{d}{dq_0}\!=\![\frac{dq}{dq_0}]^{-1}\Bigl[\frac{\partial}{\partial q_0}\!-\!2q_0\frac{\partial}{\partial \mathcal{E}_0}\Bigr]\; ,\label{DQ}
\eneq
and using Eq.~(\ref{coincorrQ}) in Eq.~(\ref{stochpow}), in $D\!=\!4$ dimensions, yield
\be
&&\hspace{1cm}\Delta^2_{\delta\bar\varphi}\Bigl(\!k\!=\!\frac{H(t)\,a(t)}{1\!\!+\!\epsilon(t)}\Bigr)\!\!=\!-\!\!
\left[1\!\!-\!\epsilon(t)\!-\!\frac{\dot{\epsilon}(t)}{H(t)\left[1\!\!+\!\epsilon(t)\right]}\right]^{-1}
\!\frac{\epsilon_{i0}}{q}\frac{d}{dq}\langle\Omega|\delta{\bar\varphi}^2(t,\vec{x})|\Omega\rangle_{\rm stoch}\nonumber\\
&&\hspace{-0.3cm}=\!\!\frac{H^2_{i0}}{4\pi^2}\Bigg\{\!\frac{q^6_0}{\left[q^2_0\!\!+\!\!\epsilon_{i0}\right]^2}\!\!
\left[\!1\!\!+\!\frac{q^2_0\!\!+\!\!3\epsilon_{i0}}{q^4_0\!\!-\!\!3\epsilon^2_{i0}}
\ln\Bigl(\!\frac{q_0^2\!\!+\!\epsilon_{i0}}{a_0q_0^3\!\left(\!1\!\!+\!\!\epsilon_{i0}\right)}\!\Bigr)^{2\epsilon_{i0}}\right]
\!\!\mp\!\frac{\lambda}{144\pi^2}\Bigg\{\!\!\frac{\pi}{GH^2_{i0}}
\Bigg[\!\epsilon^{-2}_{i0}q_0^{-1}4\lefteqn{\!\left[\!1\!\!-\!\frac{5}{4}q_0
\!\!-\!\mathcal{E}_0\Bigl(\!1\!\!-\!\frac{7}{2}q_0\!\Bigr)\right.}\nonumber\\
&&\hspace{-0.4cm}\left.-\frac{\mathcal{E}^2_0}{8}\Bigl(\!1\!\!+\!25q_0\!\Bigr)\!\right]
\!\!+\!\!\epsilon^{-1}_{i0}q^{-4}_04\!\!\left[\!1\!\!+\!\!\frac{q_0}{2}\!+\!5\Bigl(\!1\!\!+\!\frac{q_0}{20}\!\Bigr)\!
\ln(q^2_0)
\!-\!\mathcal{E}_0\Bigl(\!1\!\!+\!q_0
\!\!+\!\!\frac{27}{2}\!\Bigl[\!1\!\!+\!\frac{5}{108}q_0\Bigr]\!\ln(q^2_0)\!\Bigr)
\!\!-\!\!\frac{9}{4}\mathcal{E}^2_0\Bigl(\!1\!\!-\!\!\frac{5}{18}q_0\Bigr.\right.\nonumber\\
&&\hspace{-0.3cm}\left.\Bigl.-\frac{16}{3}\!\Bigl[\!1\!\!+\!\!\frac{q_0}{32}
\Bigr]\!\ln(q^2_0)\!\Bigr)\!\!+\!\frac{35}{16}\mathcal{E}^3_0\Bigl(\!1\!\!-\!\!\frac{8}{5}\!\ln(q^2_0)\!\Bigr)\!\right]\!
\!\!-\!q^{-6}_05\!\!\left[\!1\!\!-\!\!4q_0\!\!-\!\!\frac{3}{5}\Bigl(\!1\!\!-\!\frac{2}{3}q_0
\!\Bigr)\!\ln(q^2_0)\!\!+\!\!\frac{3}{5}\!\ln^2(q^2_0)
\!\!+\!\frac{2}{5}\mathcal{E}_0\Bigl(\!1\!\!+\!\!16q_0\right.\nonumber\\
&&\hspace{-0.3cm}+\frac{13}{2}\!\Bigl[\!1\!\!-\!\!\frac{7}{26}q_0\Bigr]\!\!\ln(q^2_0)
\!-\!\frac{9}{2}\!\ln^2(q^2_0)\!\Bigr)\!\!+\!\frac{21}{10}\mathcal{E}^2_0\Bigl(\!1\!\!-\!\!\frac{17}{21}q_0
\!+\!\!\frac{23}{14}\!\Bigl[\!1\!\!-\!\!\frac{2}{23}q_0\Bigr]\!\!\ln(q^2_0)\!+\!\!\frac{6}{7}\!\ln^2(q^2_0)\!\Bigr)
\!\!-\!\frac{91}{10}\mathcal{E}^3_0\Bigl(\!1\!\!+\!\!\frac{6}{91}q_0
\nonumber\\
&&\hspace{-0.25cm}\left.
-\frac{29}{182}\!\ln(q^2_0)\!\!+\!\!\frac{6}{91}\!\ln^2(q^2_0)\!\Bigr)\!\!+\!\frac{28}{5}\mathcal{E}^4_0\!\right]
\!\!+\!\!\mathcal{O}(\epsilon_{i0})\!\Bigg]
\!\!\!+\!\!\epsilon^{-2}_{i0}3\!\!\left[\mathcal{E}_0^2\!\!-\!\frac{4}{3}\mathcal{E}_0^3\right]
\!\!\!+\!\!\epsilon^{-1}_{i0}6q^{-4}_0\!\!\left[\mathcal{E}_0\!\ln(q^2_0)
\!\!-\!\mathcal{E}_0^2\Bigl(\!1\!\!+\!\!\frac{7}{2}\!\ln(q^2_0)\!\Bigr)\right.
\nonumber\\
&&\hspace{0cm}
\left.+\frac{11}{6}\mathcal{E}_0^3\Bigl(\!1\!\!+\!\frac{24}{11}\!\ln(q^2_0)\!\Bigr)
\!\!-\!\!\frac{5}{6}\mathcal{E}_0^4\Bigl(\!1\!\!+\!\frac{9}{5}\!\ln(q^2_0)\!\Bigr)\!\right]\!\!\!-\!\!18q^{-6}_0
\!\!\left[\ln(q^2_0)\!\!-\!\frac{\ln^2(q^2_0)}{6}
\!+\!\mathcal{E}_0\Bigl(\!1\!\!-\!\!\frac{7}{3}\!\ln(q^2_0)\!\!+\!\frac{5}{6}\!\ln^2(q^2_0)\!\Bigr)
\right.
\nonumber\\
&&\hspace{-0.55cm}\left.-\frac{17}{6}\mathcal{E}^2_0
\Bigl(\!1\!\!-\!\frac{7}{17}\!\ln(q^2_0)\!\!+\!\!\frac{9}{17}\!\ln^2(q^2_0)\!\Bigr)
\!\!+\!\frac{5}{3}\mathcal{E}_0^3\Bigl(\!1\!\!+\!\!\frac{2}{5}\!\ln(q^2_0)\!\!+\!\!\frac{7}{10}\!\ln^2(q^2_0)\!\Bigr)
\!\!+\!\frac{7}{6}\mathcal{E}_0^4\Bigl(\!1\!\!-\!\frac{3}{7}\!\ln(q^2_0)
\!\!-\!\!\frac{2}{7}\!\ln^2(q^2_0)\!\Bigr)\!\!-\!\mathcal{E}_0^5\!\right]\nonumber\\
&&\hspace{6.5cm}+\mathcal{O}(\epsilon_{i0})\!\Bigg\} .
\label{powertreelamb}
\ee
Through the relations\beeq
q_0^2(t_k)\!=\!1\!\!-\!\mathcal{E}_0(t_k)\!=\!1\!-\!2\epsilon_{i0}
\ln\Bigl(\!a_0(t_k)\Bigr)\;,
\eneq
and
\beeq
a_0(t_k)\!=\!\frac{k}{H_0(t_k)}\left[1\!\!+\!\epsilon_0(t_k)\right]\;,\label{asub0tk}
\eneq
one obtains the power for any mode
\beeq
\frac{k}{H_{i0}}\!=\!\frac{a_0(t_k)\, q_0^3(t_k)}{q_0^2(t_k)\!+\!\epsilon_{i0}} \;,\label{kkk}
\eneq
which is a monotonically increasing function of $a_0$.
Note that Eqs.~(\ref{asub0tk}) and (\ref{kkk}) reduce~\cite{GKVO} to $\frac{k}{H_{i0}}\!=\!a_0(t_k)$ in de Sitter limit.

In the noninteracting limit, \beeq\Delta^2_{{\delta\bar\varphi}_0}(k)\!=\!\frac{H^2_{i0}}{4\pi^2}\mathfrak{P}_0(k)\;,\eneq
where the dimensionless function $\mathfrak{P}_0(k)$, whose exact form can be read off from Eq.~(\ref{powertreelamb}), is expanded in $\epsilon_{i0}$ as\be
\hspace{-0.5cm}\mathfrak{P}_0(k)
\!\!=\!2q^2_0\!\!-\!\!1\!\!-\!\epsilon_{i0}\!
\Bigl[\!1\!\!+\!q^{-2}_0\!\!+\!\ln(q^2_0)\Bigr]\!\!+\!\mathcal{O}(\epsilon^2_{i0})\; .
\ee
The $\mathfrak{P}_0(k)$ is positive definite during inflation and decreases monotonically as $a_0$, i.e., $k/H_{i0}$ increases. Thus, as the comoving wave number $k$ decreases, i.e., as the scale increases, the power $\Delta^2_{\delta{\bar\varphi}_0}$ increases. The amplitude of fluctuations grow toward the larger scales and the spectrum is said to be red-tilted.
The $\mathcal{O}(\lambda)$ correction to the power spectrum
\be\Delta^2_{\delta{\bar\varphi}_{\lambda}}\!(k)
\!\!=\!\mp\lambda\frac{H^2_{i0}}{576\pi^4}\mathfrak{P}_\lambda(k)\; ,\nonumber\ee
where the dimensionless function $\mathfrak{P}_\lambda(k)$, which can be read off from Eq.~(\ref{powertreelamb}), is also positive definite and decreases monotonically as $a_0$ or $k/H_{i0}$ increases. As the scale increases, so does the $\mathfrak{P}_\lambda$. Thus, for the $+$ $(-)$ sign choice in potential~(\ref{Vpert}), the $\mathcal{O}(\lambda)$ correction $\Delta^2_{\delta{\bar\varphi}_{\lambda}}(k)$ reduces (increases) the power $\Delta^2_{\delta{\bar\varphi}_0}(k)$ in the noninteracting limit.

The tilt is quantified by the spectral index. Taylor expanding the $\ln(\Delta^2(k))$ around the $\ln(k_P)$, $k_P$ being a pivotal wavenumber, yields
\be
\ln(\Delta^2(k))\!\!=\!\ln(\Delta^2(k_p))\!+\!n(k_P)\ln\Bigl(\!\frac{k}{k_P}\Bigr)
\!\!+\!\frac{1}{2!}\alpha(k_P)\ln^2\Bigl(\!\frac{k}{k_P}\Bigr)\!+\cdots\; ,
\ee
where the expansion coefficients,
\beeq
\hspace{-0.5cm} n(k_P)\!\!\equiv\!\frac{d\ln(\Delta^2(k))}{d\ln(k)}\Big|_{k_P}\; ,\label{indx}
\eneq
and
\beeq
\alpha(k_P)\!\!\equiv\!\frac{d^2\ln(\Delta^2(k))}{d(\ln(k))^2}\Big|_{k_P}\!\!=\!\frac{dn(k)}{d\ln(k)}\Big|_{k_P}\; .\label{rnn}
\eneq
Conventionally, the scale independent Harrison-Zeldovich spectrum is defined, for inflaton fluctuations, so that the corresponding expansion coefficient $n_{\delta\bar\varphi}$, the spectral index, equals $1$ rather then more logical $0$, as in the case of graviton fluctuations. Therefore, the tilt for the inflaton fluctuations, is defined as deviation of $n_{\delta\bar\varphi}$ from $1$, \beeq n_{\delta\bar\varphi}\!\left(\!\!k\!=\!\frac{Ha}{1\!\!+\!\epsilon}\!\right)\!\!-\!1
\!\equiv\!\frac{d\ln\!\left(\Delta^2_{\delta\bar\varphi}\!\left(k\!=\!\frac{Ha}{1+\epsilon}\right)\right)}{d\ln(k)}\; ,\label{spctr}\eneq
which measures the variation of the power spectrum with scale.
The logarithmic derivative can be written as\beeq
\frac{d}{d\ln(k)}\!=\!k\frac{d}{dk}\!=\!k\frac{d t}{d k}\frac{d}{d t}\!=\!-\!\!\left[1\!\!-\!\epsilon\!-\!\frac{\dot{\epsilon}}
{H\left[1\!\!+\!\epsilon\right]}\right]^{-1}
\!\!\frac{\epsilon_{i0}}{q}\frac{d}{dq}\;.\label{dlnk}
\eneq
where we used Eqs.~(\ref{chnvarder}) and (\ref{dotk}) in the last equality. Hence the tilt
\be
&&\hspace{0.5cm}n_{\delta\bar\varphi}\!\left(\!\!k\!=\!\frac{H(t)\,a(t)}{1\!\!+\!\epsilon(t)}\right)\!\!-\!1
\!=\!-\!\!\left[1\!\!-\!\epsilon(t)\!-\!\frac{\dot{\epsilon}(t)}
{H(t)\left[1\!\!+\!\epsilon(t)\right]}\right]^{-1}
\!\!\frac{\epsilon_{i0}}{q}\frac{d}{dq}\ln\!\left(\!\Delta^2_{\delta\bar\varphi}
\!\Bigl(\!k\!=\!\frac{H(t)\,a(t)}{1\!\!+\!\epsilon(t)}\Bigr)\!\!\right)\nonumber\\
&&\hspace{0.cm}\!=\!-\frac{4\epsilon_{i0}}{\mathbf{N}}\Bigg\{\!q_0^{10}
\!\!+\!3\epsilon_{i0}q_0^6\Bigl[\!1\!\!-\!\frac{5}{6}\mathcal{E}_0\Bigr]
\!\!-\!6\epsilon^2_{i0}q_0^4\Bigl[\!1\!\!-\!\frac{5}{4}\mathcal{E}_0\Bigr]
\!\!-\!\!18\epsilon^3_{i0}q_0^2\Bigl[\!1\!\!-\!\frac{19}{12}\mathcal{E}_0\Bigr]\!\!+\!9\epsilon^4_{i0}
\Bigl[\!1\!\!+\!\frac{\mathcal{E}_0}{2}\Bigr]
\!\!+\!27\epsilon^5_{i0}\!\!-\!\mathbf{M}\!\Bigg\}\nonumber\\
&&\hspace{-0.15cm}\mp\frac{\lambda}{36\pi^2}\frac{q^{-3}_0}{\left[1\!\!-\!\!2\mathcal{E}_0\right]^2}
\Bigg\{\!\!\frac{\pi}{GH^2_{i0}}5\!
\Bigg[\!\epsilon^{-1}_{i0}\!\!\left[\!1\!-\!\frac{13}{5}\mathcal{E}_0\Bigl(\!1\!\!+\!\frac{8}{13}q_0\!\Bigr)
\!\!+\!\frac{91}{40}\mathcal{E}^2_0\Bigl(\!1\!\!+\!\frac{12}{7}q_0\!\Bigr)
\!\!-\!\frac{13}{20}\mathcal{E}_0^3\Bigl(\!1\!\!+\!\frac{46}{13}q_0\!\Bigr)
\right]\!\!-\!\frac{3q^{-3}_0}{1\!\!-\!\!2\mathcal{E}_0}\nonumber\\
&&\hspace{-0.35cm}\times\!\!\left[\!1\!\!-\!\frac{9}{5}q_0\!\!-\!\frac{8}{15}
\Bigl(\!1\!\!+\!\frac{19}{16}q_0\!\Bigr)\!\ln(q_0^2)
\!\!-\!\frac{25}{6}\mathcal{E}_0\Bigl[\!1\!\!-\!\frac{363}{250}q_0\!\!-\!\frac{88}{125}
\Bigl(\!1\!\!+\!\frac{261}{352}q_0\!\Bigr)\!\ln(q_0^2)\Bigr]
\!\!+\!\!\frac{49}{8}\mathcal{E}_0^2\Bigl[\!1\!\!-\!\frac{296}{245}q_0\!\!-\!\frac{664}{735}\Bigr.\right.\nonumber\\
&&\hspace{-0.55cm}\left.\Bigl.\times\!\Bigl(\!1\!\!+\!\!\frac{41}{83}q_0\!\Bigr)\!\ln(q_0^2)\!\Bigr]
\!\!-\!\frac{119}{40}\mathcal{E}_0^3\!\Bigl[\!1\!\!-\!\frac{67}{51}q_0\!\!-\!\frac{520}{357}
\Bigl(\!1\!\!+\!\!\frac{177}{520}q_0\!\Bigr)\!\ln(q_0^2)\!\Bigr]
\!\!-\!\frac{2}{3}\mathcal{E}_0^4\!\Bigl[\!1\!\!+\!\!\frac{49}{40}q_0
\!\!+\!\!\frac{17}{10}\Bigl(\!1\!\!+\!\!\frac{q_0}{4}\!\Bigr)\!\ln(q_0^2)\!\!+\!\!\frac{7}{10}\mathcal{E}_0^5\right.\nonumber\\
&&\hspace{-0.5cm}\left.\times\!\Bigl(\!1\!\!-\!\frac{2}{21}\!\ln(q^2_0)\!\Bigr)\!\Bigr]\!\right]
\!\!\!+\!\!\mathcal{O}(\epsilon_{i0})\!\Bigg]\!\!\!+\!\!\epsilon^{-1}_{i0}\!3q_0
\!\Bigl[\mathcal{E}_0\!\!-\!4\mathcal{E}_0^2\!\!+\!\frac{17}{3}\mathcal{E}_0^3
\!\!-\!\!\frac{8}{3}\mathcal{E}_0^4\Bigr]
\!\!+\!\!\frac{3q^{-3}_0}{1\!\!-\!\!2\mathcal{E}_0}\!\!\left[\!\Bigl[\!1\!\!-\!\!7\mathcal{E}_0\Bigr]\!\!\ln(q^2_0)
\!\!+\!\!3 \mathcal{E}_0^2\Bigl[\!1\!\!+\!\!\frac{22}{3}\!\ln(q^2_0)\Bigr]\right.\nonumber\\
&&\hspace{-0.55cm}\left.-\frac{32}{3}\mathcal{E}_0^3\Bigl[\!1\!\!+\!\!\frac{29}{8}\ln(q^2_0)\Bigr]
\!\!+\!\!\frac{41}{3}\mathcal{E}_0^4\Bigl[\!1\!\!+\!\!\frac{117}{41}\!\ln(q^2_0)\Bigr]
\!\!-\!\!\frac{22}{3}\mathcal{E}_0^5\Bigl[\!1\!\!+\!\!\frac{63}{22}\!\ln(q^2_0)\Bigr]
\!\!+\!\!\frac{4}{3}\mathcal{E}_0^6\Bigl[\!1\!\!+\!\!\frac{7}{2}\!\ln(q^2_0)\Bigr]\right]
\!\!\!+\!\!\mathcal{O}(\epsilon_{i0})\!\Bigg\} ,\label{tilt}
\ee
where\be
&&\hspace{-0.5cm}\mathbf{N}(q_0, \epsilon_{i0})\!=\!\Bigl[q^4_0\!\!-\!3\epsilon^2_{i0}\Bigr]^3\Bigg\{\!1\!\!+\!\frac{q^2_0\!\!+\!\!3\epsilon_{i0}}
{q^4_0\!\!-\!\!3\epsilon^2_{i0}}
\ln\Bigl(\!\frac{q_0^2\!\!+\!\epsilon_{i0}}{a_0q_0^3\!\left(\!1\!\!+\!\!\epsilon_{i0}\right)}
\!\Bigr)^{2\epsilon_{i0}}\!\Bigg\}\;,\\
&&\hspace{-0.5cm}\mathbf{M}(q_0, \epsilon_{i0})\!=\!\epsilon^2_{i0}\Bigl[q^6_0\!\!+\!3\epsilon_{i0}q^4_0
\!\!+\!21\epsilon^2_{i0}q^2_0\!\!+\!27\epsilon^3_{i0}\Bigr]\!
\ln\Bigl(\!\frac{q_0^2\!\!+\!\epsilon_{i0}}{q_0^3\!\left(\!1\!\!+\!\!\epsilon_{i0}\right)}\!\Bigr)\;.
\ee
Recall that $q_0(t_k)$ and $k$ are related via Eq.~(\ref{kkk}). In the noninteracting limit, the tilt
\beeq
n_{\delta\bar{\varphi}_0}\!(k)\!-\!\!1
\!\!\equiv\!\mathfrak{N}_0(k)\!=\!-\frac{4\epsilon_{i0}}{1\!\!-\!\!2\mathcal{E}_0}\!+\!\mathcal{O}(\epsilon^2_{i0})\; ,
\eneq
whose exact form is given in Eq.~(\ref{tilt}), is nonsingular for $\mathcal{E}_0\!\!<\!1/2$ which yields \beeq \epsilon_{i0}\!<\!\frac{1}{4\ln(a_0)}\;.\label{ei060}
\eneq
During an inflation which lasts $60$ e-foldings, for example, Eq.~(\ref{ei060}) implies $\epsilon_{i0}\!<\!0.0041$. The dimensionless function $\mathfrak{N}_0(k)$, for an $\epsilon_{i0}$ satisfying Eq.~(\ref{ei060}), is negative definite. Thus, the tilt is red. The $\mathfrak{N}_0(k)$ decreases monotonically (becomes more negative) as $a_0$, i.e., $k/H_{i0}$, increases.

The $\mathcal{O}(\lambda)$ correction to the spectral index
\be n_{{\bar\varphi}_{\lambda}}\!(k)
\!\!\equiv\!\mp\frac{\lambda}{36\pi^2}\mathfrak{N}_\lambda(k)\; ,\nonumber\ee
where the dimensionless function $\mathfrak{N}_\lambda(k)$, which can be read off from Eq.~(\ref{tilt}), is nonsingular for $\mathcal{E}_0\!\!<\!1/2$. Thus, the dimensionless function $\mathfrak{N}_\lambda(k)$, for an $\epsilon_{i0}$ satisfying Eq.~(\ref{ei060}), is positive definite and increases monotonically as $a_0$, hence $k/H_{i0}$, increases. Thus, for the $+$ $(-)$ sign choice in potential~(\ref{Vpert}), the $\mathcal{O}(\lambda)$ correction $n_{{\bar\varphi}_{\lambda}}\!(k)$ enhances (reduces) the red~tilt.

The measured tilt \cite{plnck1} in primordial power spectrum of the inflaton fluctuations implied by the {\it Planck} TT$+$lowP$+$BAO data, is $-0.032\pm 0.0045$ at $68\%$ confidence level. For $a_0(t_k)=e^{50}$, choosing $\epsilon_{i0}\!=\!0.00305$, which corresponds to the physical wave number $k_{\rm phys}\!=\!k/a_0(t_k)\!=\!0.83 H_{i0}$ [Eq.~(\ref{kkk})], for example, the tilt we get from Eq.~(\ref{tilt})~is \beeq
n_{\delta{\bar\varphi}}\!-\!\!1\!\!=\!-0.032\!\mp\!\lambda\Bigl(12.996\frac{\pi}{GH^2_{i0}}\!+\!1.866\Bigr)\;.\nonumber\eneq

The measured tilt \cite{plnck1} implied by the {\it Planck} TT,TE,EE$+$lowP data, on the other hand, is $-0.0348\pm 0.0047$ at $68\%$ confidence level. For $a_0(t_k)=e^{50}$ and $\epsilon_{i0}\!=\!0.00315$, the tilt that Eq.~(\ref{tilt}) yields is \beeq n_{\delta{\bar\varphi}}\!-\!\!1\!\!=\!-0.0348\!\mp\!\lambda\Bigl(13.657\frac{\pi}{GH^2_{i0}}\!+\!1.993\Bigr)\; ,\nonumber\eneq
in agreement with observation within the range provided by the Planck Collaboration.

Another physical quantity that can be computed and compared with measurements is the running [Eq.~(\ref{rnn})] of the spectral index,
\beeq \alpha_{\delta\bar\varphi}\!\left(\!\!k\!=\!\frac{Ha}{1\!\!+\!\epsilon}\!\right)
\!\equiv\!\frac{dn_{\delta\bar\varphi}\!\left(k\!=\!\frac{Ha}{1+\epsilon}\right)}{d\ln(k)}\; .\label{running}\eneq
Employing Eq.~(\ref{dlnk}) in Eq.~(\ref{running}) we find the running in our model, up to $\mathcal{O}(\lambda^2)$, as
\be
&&\hspace{0.9cm}\alpha_{\delta\bar\varphi}\!\left(\!\!k\!=\!\frac{H(t)\,a(t)}{1\!\!+\!\epsilon(t)}\right)
\!=\!-\!\!\left[1\!\!-\!\epsilon(t)\!-\!\frac{\dot{\epsilon}(t)}
{H(t)\left[1\!\!+\!\epsilon(t)\right]}\right]^{-1}
\!\!\frac{\epsilon_{i0}}{q}\frac{d}{dq}
n_{\delta\bar\varphi}\!\left(\!\!k\!=\!\frac{H(t)\,a(t)}{1\!\!+\!\epsilon(t)}\right)
\nonumber\\
&&\hspace{0.1cm}=\!-\frac{4\epsilon^2_{i0}}{
\left[q^2_0\!\!+\!3\epsilon_{i0}\right]^2}\Bigg\{\! \frac{4}{\mathbf{N}^2}\!\left[q^4_0\!\!-\!3\epsilon^2_{i0}\right]^2\!\Bigl[q^8_0\!+\!\frac{13}{2}\epsilon_{i0}q^6_0
\!+\!\frac{15}{2}\epsilon^2_{i0}q^4_0\!-\!\frac{15}{2}\epsilon^3_{i0}q^2_0\!-\!\frac{27}{2}\epsilon^4_{i0}\Bigr]^2
\!\!\!+\!\frac{\epsilon_{i0}q^2_0\left[q^2_0\!\!+\!\epsilon_{i0}\right]}{q^4_0\!\!-\!3\epsilon^2_{i0}}\nonumber\\
&&\hspace{-0.6cm}\times\!\Bigg[\!\frac{1}{\mathbf{N}}\Bigl[q^{10}_0\!\!+\!\!3\epsilon_{i0}q^8_0
\!\!+\!\!66\epsilon^2_{i0}q^6_0\!\!+\!\!198\epsilon^3_{i0}q^4_0\!\!+\!\!153\epsilon^4_{i0}q^2_0\!\!+\!\!27\epsilon^5_{i0}\Bigr]
\!\!-\!\frac{2}{\left[q^4_0\!\!-\!3\epsilon^2_{i0}\right]^3}\Bigl[q^{10}_0\!\!+\!\!6\epsilon_{i0}q^8_0
\!\!+\!54\epsilon^2_{i0}q^6_0\!\!+\!\!180\epsilon^3_{i0}q^4_0\nonumber\\
&&\hspace{-0.5cm}+\!189\epsilon^4_{i0}q^2_0\!\!+\!\!54\epsilon^5_{i0}\Bigr]\!
\Bigg]
\!\Bigg\}\!\!\mp\!\frac{\lambda}{6\pi^2}\frac{1}{\left[1\!\!-\!\!2\mathcal{E}_0\right]^3}\Bigg\{\!\frac{\pi}{G H^2_{i0}}q^{-5}_0\frac{37}{6}\!\Bigg[\!1\!\!-\!\frac{24}{37}q_0
\!\!-\!\!\frac{255}{74}\mathcal{E}_0\!\Bigl[\!1\!\!-\!\frac{32}{51}q_0\Bigr]\!\!+\!\!\frac{1305}{296}\mathcal{E}^2_0\!
\Bigl[\!1\!\!-\!\frac{736}{1305}q_0\Bigr]\nonumber\\
&&\hspace{1cm}-\frac{183}{74}\mathcal{E}^3_0\!\Bigl[\!1\!\!-\!\frac{80}{183}q_0\Bigr]
\!\!+\!\frac{\mathcal{E}^4_0}{2}\!\Bigl[\!1\!\!-\!\frac{8}{37}q_0\Bigr]\!\!+\!\mathcal{O}(\epsilon_{i0})\Bigg]
\!\!+\!\!1\!\!-\!4\mathcal{E}_0\!+\!8\mathcal{E}_0^2
\!\!-\!\frac{16}{3}\mathcal{E}_0^3
\!+\!\mathcal{O}(\epsilon_{i0})\!\Bigg\}
\; .\label{alphamassive}
\ee
In the noninteracting limit, the running
$\alpha_{\delta\bar{\varphi}_0}\!(k)
\!\equiv\!\mathfrak{A}_0(k)$,
whose exact form is given in Eq.~(\ref{alphamassive}), can be expanded in powers of $\epsilon_{i0}$ as
\beeq
\mathfrak{A}_0(k)\!=\!-\frac{16\epsilon^2_{i0}}{(2q_0^2\!-\!\!1)^2}\!+\!\mathcal{O}(\epsilon^3_{i0})\; .
\eneq
The dimensionless function $\mathfrak{A}_0(k)$ is negative definite and decreases monotonically as $a_0$, i.e.,  $k/H_{i0}$, increases. The $\mathcal{O}(\lambda)$ correction to the running
\be \alpha_{\delta{\bar\varphi}_{\lambda}}\!(k)
\!\!\equiv\!\mp\frac{\lambda}{6\pi^2}\mathfrak{A}_\lambda(k)\; ,\nonumber\ee
where the dimensionless function $\mathfrak{A}_\lambda(k)$, which can be read off from Eq.~(\ref{tilt}), is positive definite and  increases monotonically as $a_0$, hence $k/H_{i0}$, increases. Thus, for the $+$ $(-)$ sign choice in potential~(\ref{Vpert}), the $\mathcal{O}(\lambda)$ correction $\alpha_{\delta{\bar\varphi}_{\lambda}}\!(k)$ enhances (reduces) the $\alpha_{\delta\bar{\varphi}_0}\!(k)$, the negative running in the noninteracting limit.

Observations \cite{plnck1} also constrain the running of the spectral index. The $\alpha_{\delta{\bar\varphi}}$ implied by the {\it Planck} TT$+$lowP$+$BAO data, is $-0.0125\pm 0.0091$ at $68\%$ confidence level. For $a_0(t_k)=e^{50}$, choosing $\epsilon_{i0}\!=\!0.00305$, as we did in the estimation of the tilt, the running we get from Eq.~(\ref{alphamassive}) is \beeq \alpha_{\delta{\bar\varphi}}\!=\!-0.001\!\mp\!\lambda\Bigl(0.585\frac{\pi}{GH^2_{i0}}\!+\!0.106\Bigr)\; .\eneq

The $\alpha_{\delta{\bar\varphi}}$ implied by the {\it Planck} TT,TE,EE$+$lowP data, on the other hand, is $-0.0085\pm 0.0076$ at $68\%$ confidence level. For $a_0(t_k)=e^{50}$, choosing $\epsilon_{i0}\!=\!0.00315$, as we did in the estimation of the tilt, the running that Eq.~(\ref{alphamassive}) yields is
\beeq
\alpha_{\delta{\bar\varphi}}\!=\!-0.0012\!\mp\!\lambda\Bigl(0.685\frac{\pi}{GH^2_{i0}}\!+\!0.122\Bigr)\; ,\eneq
in agreement with observation within the range provided by the Planck Collaboration.

\section{Conclusions}
\label{sec:conclusions}

We extended our method \cite{vacuum,GKVO} to compute two-point correlation function of infrared truncated fields, the usual power spectrum $\Delta^2(k)\!\equiv\!\lim_{t\gg t_k}\Delta^2(t, k)$, spectral index $n(k)$ and running of the spectral index $\alpha(k)$ with quantum corrections to more general spacetimes where the expansion rate $H$ is not constant. We applied the method to study the quantum fluctuations of a self-interacting inflaton in the simplest model of inflationary spacetime.

In Sec.~\ref{sec:model}, we introduced the background metric of a spatially flat Friedman-Robinson-Walker spacetime and presented the Lagrangian of the model where a minimally coupled massive scalar with a quartic self-interaction $\pm\frac{\lambda}{4!}\varphi^4$, which we treat perturbatively, drives the inflation. We considered the inflaton field as a sum of the averaged background field $\bar\varphi(x)$, which we treat classically, plus the fluctuation field $\delta\varphi(x)$ which we treat quantum mechanically. We solved the Einstein's equations and the equations of motion for the inflaton to obtain the expansion rate $H(t)$, scale factor $a(t)$, slow-roll parameter $\epsilon(t)$ and the background field $\bar\varphi(t)$, with $\mathcal{O}(\lambda)$ corrections, in terms of the initial values of the expansion rate $H_{i0}$ and the slow-roll parameter $\epsilon_{i0}$.

Section~\ref{sec:InfFluct} is devoted to the fluctuations of the inflaton field. The mode expansion for the fluctuation field $\delta\varphi_0(x)$ of the free inflaton is given in Sec.~\ref{subsect:freethry}. The fluctuation field $\delta\varphi(x)$ of the full inflaton is expressed in terms of the free inflaton fluctuations in Sec.~\ref{subsect:interactingtheory}.

Following the Starobinsky's approach, in Sec.~\ref{sec:twopointcorrelator}, we computed the quantum corrected two point correlation function for the fluctuations of the infrared truncated inflaton field in leading order. The correlation function is evaluated at the full solution of the effective field equation. Therefore the tree-order and one-loop correlators respectively get $\mathcal{O}(\lambda)$ and $\mathcal{O}(\lambda^2)$ corrections, due to the backreaction of the interactions on the spacetime geometry. We obtained the tree-order correlator with the $\mathcal{O}(\lambda)$ correction in Sec.~\ref{subsect:treecorr} and the one-loop correlator at $\mathcal{O}(\lambda)$ in Sec.~\ref{subsect:1loopcorr}. The tree-order correlator in the noninteracting limit grows in time and asymptotes to a constant at late times for a physical distance chosen as a constant fraction of the Hubble length. The growth is reduced as the $\epsilon_{i0}\!=\!\frac{m^2}{3H^2_{i0}}$ increases. For the $+$ ($-$) sign choice in the interaction potential, the $\mathcal{O}(\lambda)$ correction at tree-order enhances (reduces) the growth whereas the $\mathcal{O}(\lambda)$ correction at one-loop order reduces (enhances) the growth.

In Sec.~\ref{sec:coincorr}, we obtained the coincidence limit of the quantum corrected two-point correlation function. In Sec.~\ref{sec:power}, we introduced a method to compute the power spectrum $\Delta^2_{\delta\bar\varphi}(k)$ [Eq.~(\ref{powertreelamb})] of the inflaton fluctuations as a time derivative of the coincident correlator. In the noninteracting limit, the power spectrum $\Delta^2_{\delta{\bar\varphi}_0}(k)$ is positive definite but decreases as $k$ increases. The $\mathcal{O}(\lambda)$ correction to the power spectrum $\Delta^2_{\delta{\bar\varphi}_\lambda}(k)$, in the interacting theory, reduces (increases) the power for the $+$ ($-$) sign choice in the interaction potential. The power spectrum is red tilted. In the non-interacting limit, the tilt $1\!-\!n_{\delta{\bar\varphi}_0}(k)$ is negative definite and decreases as $k$ increases. The $\mathcal{O}(\lambda)$ correction to the tilt $n_{\delta{\bar\varphi}_\lambda}(k)$, in the interacting theory, enhances (reduces) the red-tilt for the $+$ ($-$) sign choice in the interaction potential. The spectral index [Eq.~(\ref{tilt})] and the running of the spectral index [Eq.~(\ref{alphamassive})] are in accordance with the observations within reasonable ranges of values for the  $\epsilon_{i0}$ and the number of e-foldings.
\begin{center}
\begin{acknowledgments}
We thank Richard P. Woodard for stimulating discussions.
\end{acknowledgments}
\end{center}

\begin{appendix}
\section{The comoving time $t$ in terms of $q(t)$, $H_{i0}$ and $\epsilon_{i0}$}
\label{App:comovt}
In this Appendix, we express the comoving time $t$ in the interacting theory in terms of
$q(t)$ defined in Eq.~(\ref{Q}) and the initial values $H_{i0}$ and $\epsilon_{i0}$.

We integrate Eq.~(\ref{Hwhole}), neglecting the term linear in $\epsilon_{i0}$ among the $\mathcal{O}(\lambda)$-terms
as a zeroth order approximation in our iteration, and obtain
\beeq
z^3\!\!+\!Az^2\!\!+\!Bz\!\!+\!C\!\cong\! 0\; ,\label{polycube}
\eneq
where we define
\be
z\!\equiv\! H_{i0}t,\;\varrho\!\equiv\!\pm\frac{\lambda\xi}{12}\epsilon^{-2}_{i0},\; A\!\equiv\!-\!\frac{\epsilon^{-1}_{i0}}{2\varrho}
\Bigl[1\!\!+\!3\varrho\Bigr],\; B\!\equiv\!\frac{\epsilon^{-2}_{i0}}{\varrho}
\Bigl[1\!\!+\!\frac{\varrho}{2}\Bigr],\; C\!\equiv\!-\!\frac{\epsilon^{-3}_{i0}}{2\varrho}\Bigl[1\!\!-\!q^2\Bigr]\;.
\ee
Let
\be
Q\!\equiv\!\frac{3B\!-\!A^2}{9},\;R\!\equiv\!\frac{9AB\!-\!27C\!-\!2A^3}{54},
\;S\!\equiv\!\sqrt[3]{\!R\!\!+\!\!\sqrt{D}},\;T\!\equiv\!\sqrt[3]{\!R\!\!-\!\!\sqrt{D}}\, ,
\ee
where the discriminant $D\!\equiv\!Q^3\!\!+\!R^2$. The roots of the cubic polynomial are
\be
z_1\!\!=\!S\!+\!T\!\!-\!\frac{A}{3},\;z_2\!\!=\!-\frac{S\!+\!T}{2}\!-\!\frac{A}{3}\!+\!i\frac{\sqrt{3}}{2}(S\!-\!T)
,\;z_3\!\!=\!-\frac{S\!+\!T}{2}\!-\!\frac{A}{3}\!-\!i\frac{\sqrt{3}}{2}(S\!-\!T)\; .\label{z123}
\ee
If the discriminant $D\!>\!0$, one root is real and two are complex conjugate of each other. If $D\!=\!0$, all roots are real and at least two are equal. If $D\!<\!0$, as is the case for cubic polynomial~(\ref{polycube}), all roots are real and unequal. The physically relevant root,
\beeq
z_2\!\!=\!H_{i0}t\!\!=\!\epsilon^{-1}_{i0}\Bigl[1\!\!-\!\!q\Bigr]\!\!
\left[1\!\!+\!\frac{\varrho}{2}\Bigl[1\!\!-\!\!2q\Bigr]
\!\!+\!\mathcal{O}(\lambda^2)\right]\; ,\label{comovtimecube}
\eneq
yields the comoving time given in Eq.~(\ref{time}).

\section{Special functions}
\label{App:expintincopletegamma}

In this Appendix, we define various special functions we use in the manuscript.

\subsection{Exponential integral function $E_\beta(z)$}

The exponential integral function   \beeq
E_\beta(z)\!\equiv\!\!\int_1^\infty\!\! dt\, t^{-\beta}\, e^{-tz}\; ,\label{expintdefn}
\eneq
where $\beta$ and $z$ are $c$-numbers.

\subsection{Incomplete gamma function $\Gamma(\beta, z)$}
The incomplete gamma function
\beeq
\Gamma(\beta, z)\!\equiv\!\!\int_z^\infty\!\! dt\, t^{\beta-1}\, e^{-t}\; ,\label{incompgammadefn}
\eneq
satisfies the recurrence relation
\beeq
\Gamma(\beta\!+\!1, z)\!=\!\beta\Gamma(\beta, z)\!+\!z^{\beta} e^{-z}\; .\label{incompgammarecurr}
\eneq
Incomplete gamma function can be expressed in terms of the ordinary gamma function plus an alternating power series as
\beeq
\Gamma(\beta, z)\!=\!\Gamma(\beta)\!-\!\!\sum_{n=0}^\infty\frac{(-1)^nz^{\beta+n}}{n!(\beta\!+\!n)}\; .\label{gammaaspowerseries}
\eneq
The right side of Eq.~(\ref{gammaaspowerseries}) is replaced by its limiting value if $\beta$ is a negative integer or zero.

\subsection{Cosine integral function $ci(z)$}

The cosine integral function \be
{\rm ci}(z) \!\equiv\! -\!\int_z^{\infty} \!\!dt {\cos(t)
\over t} \!=\! \gamma \!+\! \ln(z) \!+\!\!\int_0^z \!\!dt {\cos(t)
\!-\!1 \over t} \; ,\label{cosintdefn}
\ee where $\gamma \approx 0.577$ is the Euler-Mascheroni number. The following identity involving
${\rm ci}(z)$,\beeq
{\rm ci}(z)\!-\!\frac{\sin(z)}{z}\!=\!\gamma\!-\!1\!+\!\ln(z)\!+\!\!\sum_{n=1}^\infty
                 \frac{(-1)^{n}z^{2n}}{2n (2n\!\!+\!\!1)!}\; ,\label{cieksisin}
\eneq
is useful.
\subsection{Generalized hypergeometric function ${}_2\mathcal{F}_2(\alpha_1,\alpha_2; \beta_1,\beta_2; z)$}

Generalized hypergeometric function,
\beeq
{}_2\mathcal{F}_2(\alpha_1,\alpha_2; \beta_1,\beta_2; z)\!=\!\frac{\Gamma(\beta_1)\,\Gamma(\beta_2)}
{\Gamma(\alpha_1)\,\Gamma(\alpha_2)}
\sum_{n=0}^\infty\frac{\Gamma(\alpha_1\!\!+\!n)\,\Gamma(\alpha_2\!\!+\!n)}
{\Gamma(\beta_1\!\!+\!n)\,\Gamma(\beta_2\!\!+\!n)}\left(\frac{z^n}{n!}\right)\; ,\label{hypergpq}
\eneq
where $\alpha_i$, $\beta_j$ and $z$ are $c$-numbers.

\section{Tree-order correlator in D-dimensions}
\label{App:dcor}
Three-order correlator in an arbitrary spacetime dimensions is given in Eqs.~(\ref{corrtreeord}) and (\ref{corrtreeordincomp}). Using Eq.~(\ref{nu}), which implies $\nu\!\!=\!(D\!-\!1)/2\!+\!\mathcal{O}(\epsilon)$, in these equations we find
\beeq
\hspace{0cm}\langle\Omega|\delta\bar\varphi_0(t,\vec{x}) \delta\bar\varphi_0(t'\!\!,\vec{x}\,'\!)|\Omega\rangle\!\simeq\!\!\frac{\Gamma(D\!-\!\!1)}{2^{D-1}\pi^{\frac{D}{2}}
\Gamma(\frac{D}{2})}
\!\Bigl[\!\frac{HH'}{(1\!\!+\!\epsilon\!)(1\!\!+\!\epsilon'\!)}\Bigr]^{\frac{D}{2}-1}\!\Bigl[{\rm ci}(\alpha')\!-\!\!\frac{\sin(\alpha')}{\alpha'}\!-\!{\rm ci}(\alpha_i)\!+\!\frac{\sin(\alpha_i)}{\alpha_i}\Bigr] ,\label{Dcor}
\eneq
Power series expansions of the functions inside the second square brackets in Eq.~(\ref{Dcor}),
\beeq
{\rm ci}(\alpha')\!-\!\!\frac{\sin(\alpha')}{\alpha'}\!-\!{\rm ci}(\alpha_i)\!+\!\frac{\sin(\alpha_i)}{\alpha_i}=
\ln\Bigl(\!\frac{H'a'}{H_i}\Bigr)\!-\!\ln\Bigl(\!\frac{1\!\!+\!\epsilon'}{1\!\!+\!\epsilon_i}\Bigr)
\!\!+\!\!\sum_{n=1}^\infty\frac{(-1)^n\!\!\left[{\alpha'}^{2n}\!-\!\alpha_i^{2n}\right]}{2n(2n\!\!+\!1)!}\;,\label{functseries}
\eneq
provides a useful representation for the tree-order correlator.
Using equal space and equal spacetime limits of Eq.~(\ref{functseries}) in Eq.~(\ref{Dcor}), respectively, yields\be
&&\hspace{-1cm}\langle\Omega|\delta\bar\varphi_0(t,\vec{x}) \delta\bar\varphi_0(t'\!,\vec{x})|\Omega\rangle\!\simeq\!\frac{\Gamma(D\!-\!\!1)}{2^{D-1}\pi^{\frac{D}{2}}\Gamma(\frac{D}{2})}
\Bigl[\!\frac{HH'}{(1\!\!+\!\epsilon)(1\!\!+\!\epsilon')}\Bigr]^{\frac{D}{2}-1}\Bigl[
\ln\Bigl(\!\frac{H'a'}{H_i}\Bigr)\!-\!\ln\Bigl(\!\frac{1\!\!+\!\epsilon'}{1\!\!+\!\epsilon_i}\Bigr)\Bigr]\; ,\label{Dcoreqspace}
\ee
and
\be
&&\hspace{-1cm}\langle\Omega|\delta\bar\varphi^2_0(t,\vec{x}) |\Omega\rangle\!\simeq\!\frac{\Gamma(D\!-\!\!1)}{2^{D-1}\pi^{\frac{D}{2}}\Gamma(\frac{D}{2})}
\Bigl(\!\frac{H}{1\!\!+\!\epsilon}\Bigr)^{D-2}\Bigl[
\ln\Bigl(\!\frac{Ha}{H_i}\Bigr)\!-\!\ln\Bigl(\!\frac{1\!\!+\!\epsilon}{1\!\!+\!\epsilon_i}\Bigr)\Bigr]\; .\label{Dcoreqsptime}
\ee
The $D\!\rightarrow\!4$ limits of Eqs.~(\ref{Dcor}), (\ref{Dcoreqspace}) and (\ref{Dcoreqsptime}) are used in Sec.~\ref{sec:twopointcorrelator}.
\section{Integrals of the one-loop correlator}
\label{App:onecorr}
One-loop correlator is presented in Sec.~\ref{subsect:1loopcorr}. Integrals of the one-loop correlator that involve VEVs that are quadratic in the background and fluctuation fields are computed in Sec.~\ref{subsubsect:backgrndsqr}. Integrals that involve VEVs that are quartic in the fluctuation field, on the other hand, are computed in Sec.~\ref{subsubsect:flctqrt}. In Secs. \ref{app:oneloop-quad} and \ref{app:oneloop-quart} of this appendix we, respectively, outline the details of those integrals.

\subsection{Integrals of Sec.~\ref{subsubsect:backgrndsqr}}
\label{app:oneloop-quad}

The first integral that we need to evaluate in Eq.~(\ref{corrpropphi2}) is
\be
&&\hspace{1.5cm}\frac{H'_0}{1\!\!+\!\epsilon'_0}\!\int_0^{t_0'}\!\!\!d\tilde{t}_0
\Bigl[\frac{1\!\!+\!\tilde\epsilon_0}{\tilde{H}_0}\Bigr]^2
\!\!\bar\varphi_0^2(\tilde{t}_0)\langle\Omega|
\delta\bar{\varphi}_0(t_0,\vec{x})\,\delta\bar{\varphi}_0(\tilde{t}_0,\vec{x}\,')
|\Omega\rangle_{\bf 0}\nonumber\\
&&\hspace{-1cm}=\!\frac{1}{4\pi^2}\frac{H_0}{1\!\!+\!\epsilon_0}\frac{H'_0}{1\!\!+\!\epsilon'_0}\!\int_0^{t'_0}
\!\!\!d\tilde{t}_0
\,\frac{1\!\!+\!\tilde\epsilon_0}{\tilde{H}_0}
\bar\varphi_0^2(\tilde{t}_0)\Bigl[\ln\Bigl(\!\frac{\tilde{H}_0\tilde{a}_0}{H_{i0}}\!\Bigr)
\!-\!\ln\Bigl(\!\frac{1\!\!+\!\tilde\epsilon_0}
{1\!\!+\!\epsilon_{i0}}\!\Bigr)
\!\!+\!\!\sum_{n=1}^\infty\!\frac{(-1)^n\!\left[\tilde\alpha_0^{2n}
\!\!-\!\alpha_{i0}^{2n}\right]}{2n(2n\!\!+\!\!1)!}\Bigr]\; ,\label{1stvevintg}
\ee
where we used the fact that $\bar\varphi_0(t)$ is just a $c$-number as far as the perturbations are concerned and Eq.~(\ref{treecorrzeroorder0}). To evaluate the integral we make change of variable~(\ref{time}) which implies
\beeq
d\tilde{t}_0\!=\!-\frac{\epsilon^{-1}_{i0}}{H_{i0}}d\tilde{q}_0\; .\label{dterotree}
\eneq
The first integral in Eq.~(\ref{1stvevintg}) is evaluated employing Eqs.~(\ref{barbarvar}), (\ref{Hep2}), (\ref{lnaHlnQ}) and (\ref{dterotree}) as
\be
&&\hspace{-1.3cm}\int_0^{t'_0}\!\!\!d\tilde{t}_0
\,\frac{1\!\!+\!\tilde\epsilon_0}{\tilde{H}_0}\bar\varphi_0^2(\tilde{t}_0)
\ln\Bigl(\!\frac{\tilde{H}_0\tilde{a}_0}{H_{i0}}\!\Bigr)\!=\!\frac{\epsilon^{-2}_{i0}}{4\pi GH^2_{i0}}\!\int_{q'_0}^1\!\! d\tilde{q}_0\frac{\tilde{q}^2_0\!+\!\epsilon_{i0}}{\tilde{q}_0}\Bigl[\ln(\tilde{q}_0)
\!+\!\frac{\epsilon_{i0}^{-1}}{2}\!\left[1\!-\!\tilde{q}^2_0\right]\Bigr]\nonumber\\
&&\hspace{-0.2cm}=\!\frac{1}{8\pi GH^2_{i0}}\Bigg\{\!\epsilon^{-3}_{i0}\frac{\mathcal{E}'^2_0}{4}\!-\!\epsilon^{-2}_{i0}
\Bigl[\left[1\!\!+\!q_0'^2\right]\!\ln(q_0')\!+\mathcal{E}'_0\Bigr]\!-\!\epsilon^{-1}_{i0}\!\ln^2(q_0')\Bigg\}\; .\label{1stint}
\ee
The second integral in Eq.~(\ref{1stvevintg}) is evaluated employing Eqs.~(\ref{barbarvar}), (\ref{Hep2}), (\ref{lnratioeps}) and (\ref{dterotree}) as
\be
&&\hspace{-1.1cm}\int_0^{t'_0}\!\!\!d\tilde{t}_0
\,\frac{1\!\!+\!\tilde\epsilon_0}{\tilde{H}_0}\bar\varphi_0^2(\tilde{t}_0)\ln\Bigl(\!\frac{1\!\!+\!\tilde\epsilon_0}
{1\!\!+\!\epsilon_{i0}}\!\Bigr)\!=\!\frac{\epsilon^{-2}_{i0}}{4\pi GH^2_{i0}}\!\int_{q'_0}^1\!\! d\tilde{q}_0\frac{\tilde{q}^2_0\!+\!\epsilon_{i0}}{\tilde{q}_0}
\ln\Bigl(\frac{\tilde{q}_0^2\!\!+\!\epsilon_{i0}}{\tilde{q}_0^2\!\left(1\!\!+\!\epsilon_{i0}\right)}\Bigr)\nonumber\\
&&\hspace{-0.7cm}=\!\frac{-1}{8\pi GH^2_{i0}}\Bigg\{\!\epsilon^{-1}_{i0}\Bigl[\ln(q'^2_0)\!+\!\mathcal{E}'_0\Bigr]
\!\!-\!\ln(q'^2_0)\!-\!\frac{\mathcal{E}'_0}{2}\!+\!\frac{1\!-\!q'^{-2}_0}{2}\!+\!\mathcal{O}(\epsilon_{i0})\!\Bigg\}\;.\label{2ndint}
\ee
The third integral in Eq.~(\ref{1stvevintg}) is
\be
&&\hspace{-0.6cm}\int_0^{t'_0}\!\!\!d\tilde{t}_0
\frac{1\!\!+\!\tilde\epsilon_0}{\tilde{H}_0}\bar\varphi_0^2(\tilde{t_0})\!
\sum_{n=1}^\infty\!\frac{(-1)^n{\tilde\alpha_0}^{2n}}{2n(2n\!\!+\!\!1)!}\!=\!\!\sum_{n=1}^\infty\!\frac{(-1)^n(\Delta x)^{2n}}{2n(2n\!\!+\!\!1)!}\!\!\int_0^{t'_0}\!\!\!d\tilde{t}_0
\frac{1\!\!+\!\tilde\epsilon_0}{\tilde{H}_0}\bar\varphi_0^2(\tilde{t}_0)\Bigl[\frac{\tilde{H}_0\tilde{a}_0}
{1\!\!+\!\tilde\epsilon_0}\Bigr]^{2n}\; .\label{thirdint1loop}
\ee
Employing Eqs.~(\ref{barbarvar}), (\ref{Hep2}), (\ref{alpzrprm}) and (\ref{dterotree}) in the above integral brings it to the form
\be
\hspace{0.2cm}\int_0^{t'_0}\!\!\!d\tilde{t}_0
\frac{1\!\!+\!\tilde\epsilon_0}{\tilde{H}_0}\bar\varphi_0^2(\tilde{t}_0)\Bigl[\frac{\tilde{H}_0\tilde{a}_0}
{1\!\!+\!\tilde\epsilon_0}\Bigr]^{2n}\!\!=\!\frac{\epsilon^{-2}_{i0}}{4\pi GH^2_{i0}}H^{2n}_{i0}e^{\frac{n}{\epsilon_{i0}}}\!\!\int_{q'_0}^1\!\!d\tilde{q}_0\,\tilde{q}^2_0
\Bigl[\frac{\tilde{q}_0^3}{\tilde{q}_0^2\!\!+\!\epsilon_{i0}}\Bigr]^{2n-1}\!\!e^{-\frac{n}{\epsilon_{i0}}\tilde{q}^2_0}\; .\label{D6}
\ee
Evaluating the final form of the integral in Eq.~(\ref{D6}) and using the result in Eq.~(\ref{thirdint1loop}) yields
\be
&&\hspace{-1.45cm}\int_0^{t'_0}\!\!\!d\tilde{t}_0
\frac{1\!\!+\!\tilde\epsilon_0}{\tilde{H}_0}\bar\varphi_0^2(\tilde{t_0})\!
\sum_{n=1}^\infty\!\frac{(-1)^n{\tilde\alpha_0}^{2n}}{2n(2n\!\!+\!\!1)!}\!=\!\frac{1}{8\pi GH^2_{i0}}\!\!\sum_{n=1}^\infty\!\frac{(-1)^n(H_{i0}\Delta x)^{2n}}{2n(2n\!\!+\!\!1)!}e^{\frac{n}{\epsilon_{i0}}}\!\Bigg\{
\!\epsilon^{-2}_{i0}\!\lefteqn{\Bigl[q'^{2n+2}_0E_{-n}\Bigl(\!\frac{nq'^2_0}{\epsilon_{i0}}\!\Bigr)\Bigr.}\nonumber\\
&&\hspace{-1.37cm}\Bigl. -E_{-n}\Bigl(\!\frac{n}{\epsilon_{i0}}\!\Bigr)\!\Bigr]\!\!-\!\!(2n\!\!-\!\!1)\!
\Bigg[\epsilon^{-1}_{i0}\!\Bigl[q'^{2n}_0 \! E_{1-n}\Bigl(\!\frac{nq'^2_0}{\epsilon_{i0}}\!\Bigr)
\!\!-\!\!E_{1-n}\Bigl(\!\frac{n}{\epsilon_{i0}}\!\Bigr)\!\Bigr]
\!\!-\!n\!\Bigl[q'^{2n-2}_0 \! E_{2-n}\Bigl(\!\frac{nq'^2_0}{\epsilon_{i0}}\!\Bigr)
\!\!-\!\!E_{2-n}\Bigl(\!\frac{n}{\epsilon_{i0}}\!\Bigr)\!\Bigr]\!\Bigg]\!
\Bigg\}.
\label{3rdint}
\ee
The fourth integral in Eq.~(\ref{1stvevintg})
\be
&&\hspace{-0.2cm}\int_0^{t'_0}\!\!\!d\tilde{t}_0
\,\frac{1\!\!+\!\tilde\epsilon_0}{\tilde{H}_0}
\bar\varphi_0^2(\tilde{t}_0)\!\sum_{n=1}^\infty\!\frac{(-1)^n\alpha_{i0}^{2n}}{2n(2n\!\!+\!\!1)!}
\!=\!\frac{\epsilon^{-2}_{i0}}{4\pi GH^2_{i0}}\!\sum_{n=1}^\infty\!\frac{(-1)^n \alpha_{i0}^{2n}}{2n(2n\!\!+\!\!1)!}\!\int_{q'_0}^1\!\!d\tilde{q}_0
\frac{\tilde{q}_0^2\!\!+\!\epsilon_{i0}}{\tilde{q}_0}\nonumber\\
&&\hspace{0.2cm}=\!\frac{1}{8\pi GH^2_{i0}}\Bigl[\epsilon^{-2}_{i0}\mathcal{E}'_0\!-\!\epsilon^{-1}_{i0}\!\ln(q'^2_0)\Bigr]
\Bigl[1\!\!-\!\!\gamma\!+\!{\rm ci}(\alpha_{i0})\!-\!\frac{\sin(\alpha_{i0})}{\alpha_{i0}}\!-\!\ln(\alpha_{i0})\Bigr]\; .\label{4thint}
\ee
Thus, the first integral in Eq.~(\ref{corrpropphi2}) is obtained using Eqs.~(\ref{1stint}), (\ref{2ndint}), (\ref{3rdint}) and (\ref{4thint}) in Eq.~(\ref{1stvevintg}). The result is given in Eq.~(\ref{1loop1stint}).

The second integral that we need to compute in Eq.~(\ref{corrpropphi2}) is
\be
\hspace{0.5cm}\frac{H_0}{1\!\!+\!\epsilon_0}\!
\int_0^{t_0}\!\!dt''_0\Bigl[\frac{1\!\!+\!\epsilon''_0}{H''_0}\Bigr]^2\!\bar\varphi^2_0(t''_0)
\langle\Omega|\delta\bar{\varphi}_0(t''_0,\vec{x})\,
\delta\bar{\varphi}_0(t'_0,\vec{x}\,')|\Omega\rangle\;. \label{barphi2scndint}
\ee
To evaluate the remaining VEV in the integrand we need to break up the integral into two as $\int_0^{t_0}\!dt''_0\!=\!\int_0^{t'_0}\!dt''_0\!+\!\int_{t'_0}^{t_0}\!dt''_0$. In the first integral on the right side $t''_0\!\leq\!t'_0$, whereas $t'_0\!\leq\!t''_0$ in the second. The first part \be
\frac{H_0}{1\!\!+\!\epsilon_0}\!
\int_0^{t'_0}\!\!dt''_0\Bigl[\frac{1\!\!+\!\epsilon''_0}{H''_0}\Bigr]^2\!\!\bar\varphi^2_0(t''_0)
\langle\Omega|\delta\bar{\varphi}_0(t''_0,\vec{x})\,
\delta\bar{\varphi}_0(t'_0,\vec{x}\,')|\Omega\rangle\;, \label{1l1p}
\ee
yields exactly the same result obtained in Eq.~(\ref{1loop1stint}). The second part, on the other hand,
\be
&&\hspace{1.4cm}\frac{H_0}{1\!\!+\!\epsilon_0}\!
\int_{t'_0}^{t_0}\!\!dt''_0\Bigl[\frac{1\!\!+\!\epsilon''_0}{H''_0}\Bigr]^2\!\!\bar\varphi^2_0(t''_0)
\langle\Omega|\delta\bar{\varphi}_0(t''_0,\vec{x})\,
\delta\bar{\varphi}_0(t'_0,\vec{x}\,')|\Omega\rangle\nonumber\\
&&\hspace{-1.3cm}=\frac{1}{4\pi^2}\frac{H_0}{1\!\!+\!\epsilon_0}\frac{H'_0}{1\!\!+\!\epsilon'_0}
\!\Bigl[\ln\Bigl(\!\frac{H'_0a'_0}{H_{i0}}\!\Bigr)
\!-\!\ln\Bigl(\!\frac{1\!\!+\!\epsilon'_0}
{1\!\!+\!\epsilon_{i0}}\!\Bigr)
\!\!+\!\!\sum_{n=1}^\infty\!\frac{(-1)^n\!\!\left[{\alpha'_0}^{2n}\!\!-\!\alpha_{i0}^{2n}\right]}{2n(2n\!\!+\!\!1)!}\Bigr]
\!\int_{t'_0}^{t_0}\!\!dt''_0
\,\frac{1\!\!+\!\epsilon''_0}{H''_0}
\bar\varphi_0^2(t''_0)
\; ,\label{2ndpart}
\ee
where the integral
\be
&&\hspace{-1.1cm}\int_{t'_0}^{t_0}\!\!dt''_0
\frac{1\!\!+\!\epsilon''_0}{H''_0}
\bar\varphi_0^2(t''_0)\!=\!\frac{\epsilon^{-2}_{i0}}{4\pi GH^2_{i0}}\!\int_{q_0}^{q'_0}\!\!dq''_0
\frac{q''^2_0\!\!+\!\epsilon_{i0}}{q''_0}\nonumber\\
&&\hspace{-1.5cm}=\!-\frac{1}{8\pi GH^2_{i0}}\Bigg\{\!\epsilon_{i0}^{-2}\Bigl[q^2_0\!\!-\!q'^2_0\Bigr]
\!\!+\!\epsilon_{i0}^{-1}\Bigl[\ln(q^2_0)\!-\!\ln(q'^2_0)\Bigr]\!\Bigg\}\; .\label{oneloopint2prt2}
\ee
Hence, employing integral~(\ref{oneloopint2prt2})---summing up the infinite series multiplying it---in Eq.~(\ref{2ndpart}) yields
\be
&&\hspace{-0.1cm}\frac{H_0}{1\!\!+\!\epsilon_0}\!
\int_{t'_0}^{t_0}\!\!dt''_0\Bigl[\frac{1\!\!+\!\epsilon''_0}{H''_0}\Bigr]^2\!\!\bar\varphi^2_0(t''_0)
\langle\Omega|\delta\bar{\varphi}_0(t''_0,\vec{x})\,
\delta\bar{\varphi}_0(t'_0,\vec{x}\,')|\Omega\rangle\nonumber\\
&&\hspace{-1cm}=\!\frac{1}{32\pi^3G}\frac{q_0^3\,q_0'^3}{(q_0^2\!\!+\!\epsilon_{i0})(q_0'^2\!\!+\!\epsilon_{i0})}
\Bigg\{\!\epsilon_{i0}^{-2}\Bigl[q^2_0\!\!-\!q'^2_0\Bigr]
\!\!+\!\epsilon_{i0}^{-1}\Bigl[\ln(q^2_0)\!-\!\ln(q'^2_0)\Bigr]\!\Bigg\}\nonumber\\
&&\hspace{0.9cm}\times\!\Bigg[{\rm ci}(\alpha_{i0})\!-\!\frac{\sin(\alpha_{i0})}{\alpha_{i0}}
\!-\!\Bigl[{\rm ci}(\alpha'_0)\!-\!\frac{\sin(\alpha'_0)}{\alpha'_0}\Bigr]\Bigg]
\; .\label{1l2p}
\ee
The second integral in Eq.~(\ref{corrpropphi2}) is, therefore, obtained combining  Eq.~(\ref{1loop1stint}) and
Eqs.~(\ref{1l1p})-(\ref{1l2p}). The result is given in Eq.~(\ref{1loop2ndint}).

In the next section, we outline the evaluation of the integrals involving VEVs that are quartic in the fluctuation field in one-loop correlator~(\ref{1loopcorr}).

\subsection{Integrals of Sec.~\ref{subsubsect:flctqrt}}
\label{app:oneloop-quart}

The first integral in Eq.~(\ref{corrpropphi4})
\be
&&\hspace{3.75cm}\frac{H'_0}{1\!\!+\!\epsilon'_0}\!\int_0^{t'_0}\!\!d\tilde{t}_0
\Bigl[\frac{1\!\!+\!\tilde\epsilon_0}{\tilde{H}_0}\Bigr]^2
\!\langle\Omega|\delta\bar{\varphi}_0(t_0,\vec{x})\,\delta\bar{\varphi}^3_0(\tilde{t}_0,\vec{x}\,')
|\Omega\rangle\nonumber\\
&&\hspace{1.6cm}=\!\!\frac{H'_0}{1\!\!+\!\epsilon'_0}
\!\int_0^{t'_0}\!\!d\tilde{t}_0
\Bigl[\frac{1\!\!+\!\tilde\epsilon_0}{\tilde{H}_0}\Bigr]^2
3\!\cdot\!1\langle\Omega|\delta\bar{\varphi}_0(t_0,\vec{x})\,\delta\bar{\varphi}_0(\tilde{t}_0,\vec{x}\,')|\Omega\rangle
\langle\Omega|\delta\bar{\varphi}^2_0(\tilde{t}_0,\vec{x}\,')|\Omega\rangle\nonumber\\
&&\hspace{-0.5cm}=\!\!\frac{3}{16\pi^4}\frac{H_0}{1\!\!+\!\epsilon_0}\frac{H'_0}{1\!\!+\!\epsilon'_0}
\!\!\int_0^{t'_0}\!\!\!\!d\tilde{t}_0\!\frac{\tilde{H}_0}{1\!\!+\!\tilde\epsilon_0}
\Bigg\{\!\!\Bigl[\ln\Bigl(\!\!\frac{\tilde{H}_0\tilde{a}_0}{H_{i0}}\!\Bigr)
\!\!-\!\!\ln\Bigl(\!\frac{1\!\!+\!\tilde\epsilon_0}
{1\!\!+\!\epsilon_{i0}}\!\Bigr)\!\Bigr]^2\!\!\!\!+\!\!\!
\sum_{n=1}^\infty\!\!\frac{(-1)^n\!\!\left[{\tilde\alpha}^{2n}_0\!\!-\!\alpha_{i0}^{2n}\right]}{2n(2n\!\!+\!\!1)!}
\!\Bigl[\ln\Bigl(\!\!\frac{\tilde{H}_0\tilde{a}_0}{H_{i0}}\!\Bigr)
\!\!-\!\!\ln\Bigl(\!\frac{1\!\!+\!\tilde\epsilon_0}
{1\!\!+\!\epsilon_{i0}}\!\Bigr)\!\Bigr]\!\!\Bigg\}\label{remainingtrms}\nonumber\\
\ee
where we used correlator~(\ref{treecorrzeroorder0}) and its coincident limit~(\ref{cointreebit}) at $\mathcal{O}(\lambda^0)$. The integral of the square bracketed terms can be written, by expanding the integrand, as
\be
\hspace{-0.6cm}\int_0^{t'_0}\!\!\!\!\!d\tilde{t}_0\!\frac{\tilde{H}_0}{1\!\!+\!\tilde\epsilon_0}\!
\Bigl[\ln\Bigl(\!\!\frac{\tilde{H}_0\tilde{a}_0}{H_{i0}}\!\Bigr)
\!\!-\!\!\ln\Bigl(\!\frac{1\!\!+\!\tilde\epsilon_0}
{1\!\!+\!\epsilon_{i0}}\!\Bigr)\!\Bigr]^2\!\!\!\!\!
=\!\!\!\int_0^{t'_0}\!\!\!\!\!d\tilde{t}_0\!\frac{\tilde{H}_0}{1\!\!+\!\tilde\epsilon_0}\!
\Bigl[\ln^2\!\Bigl(\!\!\frac{\tilde{H}_0\tilde{a}_0}{H_{i0}}\!\Bigr)
\!\!-\!\!2\ln\Bigl(\!\!\frac{\tilde{H}_0\tilde{a}_0}{H_{i0}}\!\Bigr)\!\ln\Bigl(\!\frac{1\!\!+\!\tilde\epsilon_0}
{1\!\!+\!\epsilon_{i0}}\!\Bigr)\!\!\!+\!\ln^2\!\Bigl(\!\frac{1\!\!+\!\tilde\epsilon_0}
{1\!\!+\!\epsilon_{i0}}\!\Bigr)\!\Bigr].\label{remaining1}
\ee
The first and second integrals in Eq.~(\ref{remaining1}) are
\be
&&\hspace{-0.65cm}\int_0^{t'_0}\!\!\!d\tilde{t}_0\frac{\tilde{H}_0}{1\!\!+\!\tilde\epsilon_0}
\ln^2\Bigl(\!\frac{\tilde{H}_0\tilde{a}_0}{H_{i0}}\!\Bigr)\!\!=\!\epsilon^{-1}_{i0}\!\!\!\int_{q'_0}^1\!\!\! d\tilde{q}_0\frac{\tilde{q}^3_0}{\tilde{q}^2_0\!\!+\!\epsilon_{i0}}\Bigl[\ln(\tilde{q}_0)
\!+\!\frac{\epsilon^{-1}_{i0}}{2}[1\!-\!\tilde{q}^2_0]\Bigr]^2
\!\!=\epsilon^{-3}_{i0}\frac{\mathcal{E}'^3_0}{24}
\!+\!\epsilon^{-2}_{i0}\frac{\mathcal{E}'^2_0}{4}\ln(q'_0)\nonumber\\
&&\hspace{-0.4cm}
-\frac{\epsilon^{-1}_{i0}}{2}\!\!\left[\frac{q'^2_0}{4}
\!-\!\frac{q'^{-2}_0}{4}\!-\!\mathcal{E}'_0\ln^2(q'_0)\!-\!\ln(q'_0)\right]
\!\!+\!\frac{\ln^3(eq'^2_0)}{24}\!+\!\frac{q'^{-2}_0}{2}\ln(eq'_0)\!-\!\frac{q'^{-4}_0}{16}
\!-\!\frac{23}{48}\!+\!\mathcal{O}(\epsilon_{i0})\; ,
\ee
and
\be
&&\hspace{-0.6cm}\int_0^{t'_0}\!\!\!d\tilde{t}_0\frac{\tilde{H}_0}{1\!\!+\!\tilde\epsilon_0}
\ln\Bigl(\!\frac{\tilde{H}_0\tilde{a}_0}{H_{i0}}\!\Bigr)\!\ln\Bigl(\!\frac{1\!\!+\!\tilde\epsilon_0}
{1\!\!+\!\epsilon_{i0}}\!\Bigr)\!\!=\!\epsilon^{-1}_{i0}\!\!\!\int_{q'_0}^1\!\!\! d\tilde{q}_0\frac{\tilde{q}^3_0}{\tilde{q}^2_0\!\!+\!\epsilon_{i0}}\Bigl[\ln(\tilde{q}_0)
\!+\!\frac{\epsilon^{-1}_{i0}}{2}[1\!-\!\tilde{q}^2_0]\Bigr]\!
\ln\Bigl(\!\frac{\tilde{q}_0^2\!\!+\!\epsilon_{i0}}{\tilde{q}_0^2\!\left(1\!\!+\!\epsilon_{i0}\right)}\Bigr)\nonumber\\
&&\hspace{-0.4cm}=\!-\frac{\epsilon^{-1}_{i0}}{2}\!\!\left[\frac{\mathcal{E}'^2_0}{4}
\!+\!\frac{\mathcal{E}'_0}{2}\!+\!\ln(q'_0)\right]
\!+\!\frac{\mathcal{E}'^2_0}{16}\!+\!\frac{3}{4}\Bigl[\!\frac{1\!\!-\!q'^{-2}_0}{2}\Bigr]
\!\!-\!\frac{\ln^2(q'_0)}{2}\!+\!\Bigl[\frac{q'^{2}_0}{2}\!-\!\frac{5}{4}\Bigr]\!\ln(q'_0)
\!+\!\mathcal{O}(\epsilon_{i0})\;,
\ee
respectively. The third integral in Eq.~(\ref{remaining1}), on the other hand, is of $\mathcal{O}(\epsilon_{i0})$
\be
\hspace{-0.6cm}\int_0^{t'_0}\!\!\!\!d\tilde{t}_0\frac{\tilde{H}_0}{1\!\!+\!\tilde\epsilon_0}
\ln^2\!\Bigl(\!\frac{1\!\!+\!\tilde\epsilon_0}
{1\!\!+\!\epsilon_{i0}}\!\Bigr)\!\!=\!\epsilon^{-1}_{i0}\!\!\!\int_{q'_0}^1\!\!\! d\tilde{q}_0\frac{\tilde{q}^3_0}{\tilde{q}^2_0\!\!+\!\epsilon_{i0}}\!
\ln^2\Bigl(\!\frac{\tilde{q}_0^2\!\!+\!\epsilon_{i0}}{\tilde{q}_0^2\!\left(1\!\!+\!\epsilon_{i0}\right)}\Bigr)
\!\!=\!\epsilon_{i0}\Bigl[\frac{q'^{-2}_0}{2}\!-\!\frac{q'^2_0}{2}\!+\!\ln(q'^2_0)\Bigr]\!\!+\!\mathcal{O}(\epsilon^2_{i0})
\;,
\ee
hence we neglect it. The remaining three terms in Eq.~(\ref{remainingtrms}) involve series expansions of which the first is
\be
&&\hspace{0.3cm}\sum_{n=1}^\infty\!\frac{(-1)^n(\Delta x)^{2n}}{2n(2n\!\!+\!\!1)!}
\!\!\int_0^{t'_0}\!\!\!d\tilde{t}_0\frac{\tilde{H}_0}{1\!\!+\!\tilde\epsilon_0}\Bigl[\!\frac{\tilde{H}_0\tilde{a}_0}
{1\!\!+\!\tilde\epsilon_{0}}\!\Bigr]^{2n}\!
\ln\Bigl(\!\frac{\tilde{H}_0\tilde{a}_0}{H_{i0}}\!\Bigr)\; ,
\ee
where the integral
\be
&&\hspace{-0.8cm}\int_0^{t'_0}\!\!\!d\tilde{t}_0\frac{\tilde{H}_0}{1\!\!+\!\tilde\epsilon_0}
\Bigl[\!\frac{\tilde{H}_0\tilde{a}_0}
{1\!\!+\!\tilde\epsilon_{0}}\!\Bigr]^{2n}\!
\ln\Bigl(\!\frac{\tilde{H}_0\tilde{a}_0}{H_{i0}}\!\Bigr)
\!=\!\epsilon^{-1}_{i0}H^{2n}_{i0}e^{\frac{n}{\epsilon_{i0}}}\!\!\int_{q'_0}^1\!\! d\tilde{q}_0\Bigl[\frac{\tilde{q}^3_0}{\tilde{q}^2_0\!\!+\!\epsilon_{i0}}\Bigr]^{2n+1}e^{-\frac{n}{\epsilon_{i0}}\tilde{q}^2_0}
\Bigl[\ln(\tilde{q}_0)
\!+\!\frac{\epsilon_{i0}^{-1}}{2}\!\left[\!1\!\!-\!\tilde{q}^2_0\right]\!\Bigr]
\nonumber\\
&&\hspace{-0.2cm}\!=\!\frac{H^{2n}_{i0}}{4}e^{\frac{n}{\epsilon_{i0}}}\Bigg\{\!\epsilon^{-2}_{i0}\!
\Bigg[E_{-n-1}\!\Bigl(\!\frac{n}{\epsilon_{i0}}\Bigr)\!-\!q'^{2n+4}_0E_{-n-1}\!\Bigl(\frac{nq'^2_0}{\epsilon_{i0}}\Bigr)
\!\!-\!\!\left\{\!E_{-n}\!\Bigl(\!\frac{n}{\epsilon_{i0}}\Bigr)
\!-\!q'^{2n+2}_0E_{-n}\!\Bigl(\!\frac{nq'^2_0}{\epsilon_{i0}}\Bigr)\!\right\}\!\!\Bigg]\nonumber\\
&&\hspace{0.2cm}-\epsilon^{-1}_{i0}\!
\Bigg[\!(2n\!\!+\!\!1)\!\left\{\!E_{-n}\!\Bigl(\!\frac{n}{\epsilon_{i0}}\Bigr)\!-\!q'^{2n+2}_0E_{-n}
\!\Bigl(\!\frac{nq'^2_0}{\epsilon_{i0}}\Bigr)
\!\!-\!\!\left\{\!E_{1-n}\!\Bigl(\!\frac{n}{\epsilon_{i0}}\Bigr)
\!-\!q'^{2n}_0E_{1-n}\!\Bigl(\!\frac{nq'^2_0}{\epsilon_{i0}}\Bigr)\!\right\}\!\right\}\nonumber\\
&&\hspace{-0.3cm}+\frac{1}{(n\!\!+\!\!1)^2}\!\left\{\! {}_2\mathcal{F}_2\Big(\!n\!\!+\!\!1, n\!\!+\!\!1; n\!\!+\!\!2, n\!\!+\!\!2;-\frac{n}{\epsilon_{i0}}\!\Big)\!-\!q'^{2n+2}_0{}_2\mathcal{F}_2\Big(\!n\!\!+\!\!1, n\!\!+\!\!1; n\!\!+\!\!2, n\!\!+\!\!2;-\frac{nq'^2_0}{\epsilon_{i0}}\!\Big)\!\right\}\!\!\Bigg]\nonumber\\
&&\hspace{0cm}+(2n\!\!+\!\!1)\Bigg[\!(n\!\!+\!\!1)\!\left\{\!E_{1-n}\!\Bigl(\!\frac{n}{\epsilon_{i0}}\Bigr)
\!-\!q'^{2n}_0E_{1-n}\!\Bigl(\!\frac{nq'^2_0}{\epsilon_{i0}}\Bigr)
\!\!-\!\!\left\{\!E_{2-n}\!\Bigl(\!\frac{n}{\epsilon_{i0}}\Bigr)
\!-\!q'^{2n-2}_0E_{2-n}\!\Bigl(\!\frac{nq'^2_0}{\epsilon_{i0}}\Bigr)\!\right\}\!\right\}\nonumber\\
&&\hspace{0.3cm}+\frac{1}{n^2}\!\left\{\! {}_2\mathcal{F}_2\Big(\!n, n; n\!\!+\!\!1, n\!\!+\!\!1;-\frac{n}{\epsilon_{i0}}\!\Big)\!-\!q'^{2n}_0{}_2\mathcal{F}_2\Big(\!n, n; n\!\!+\!\!1, n\!\!+\!\!1;-\frac{nq'^2_0}{\epsilon_{i0}}\!\Big)\!\right\}\!\!\Bigg]\!\!\!+\!\mathcal{O}(\epsilon_{i0})\Bigg\}.
\ee
We define the generalized hypergeometric function ${}_2\mathcal{F}_2$ in Eq.~(\ref{hypergpq}). The second remaining term which involves a series expansion in Eq.~(\ref{remainingtrms}) is
\be
&&\hspace{-0.5cm}
\sum_{n=1}^\infty\!\!\frac{(-1)^n(\Delta x)^{2n}}{2n(2n\!\!+\!\!1)!}\!\!
\int_0^{t'_0}\!\!d\tilde{t}_0\frac{\tilde{H}_0}{1\!\!+\!\tilde\epsilon_0}\Bigl[\!\frac{\tilde{H}_0\tilde{a}_0}
{1\!\!+\!\tilde\epsilon_{0}}\!\Bigr]^{2n}\!\!\ln\Bigl(\!\frac{1\!\!+\!\tilde\epsilon_0}
{1\!\!+\!\epsilon_{i0}}\!\Bigr)\; ,\label{secndsum}
\ee
where the integral in Eq.~(\ref{secndsum}) is\be
&&\hspace{-1cm}
\int_0^{t'_0}\!\!d\tilde{t}_0\frac{\tilde{H}_0}{1\!\!+\!\tilde\epsilon_0}\Bigl[\!\frac{\tilde{H}_0\tilde{a}_0}
{1\!\!+\!\tilde\epsilon_{0}}\!\Bigr]^{2n}\!\!\ln\Bigl(\!\frac{1\!\!+\!\tilde\epsilon_0}
{1\!\!+\!\epsilon_{i0}}\!\Bigr)\!=\!\epsilon^{-1}_{i0}H^{2n}_{i0}e^{\frac{n}{\epsilon_{i0}}}\!\!\int_{q'_0}^1\!\! d\tilde{q}_0\Bigl[\frac{\tilde{q}^3_0}{\tilde{q}^2_0\!\!+\!\epsilon_{i0}}\Bigr]^{2n+1}\!
e^{-\frac{n}{\epsilon_{i0}}\tilde{q}^2_0}
\ln\Bigl(\!\frac{\tilde{q}_0^2\!\!+\!\epsilon_{i0}}{\tilde{q}_0^2\!\left(1\!\!+\!\epsilon_{i0}\right)}\Bigr)\nonumber\\
&&\hspace{-0.9cm}\!=\!-\frac{H^{2n}_{i0}}{2}e^{\frac{n}{\epsilon_{i0}}}\Bigg\{\!E_{1-n}\!\Bigl(\!\frac{n}{\epsilon_{i0}}\Bigr)
\!-\!q'^{2n}_0E_{1-n}\!\Bigl(\!\frac{nq'^2_0}{\epsilon_{i0}}\Bigr)
\!\!-\!\!\left[E_{-n}\!\Bigl(\!\frac{n}{\epsilon_{i0}}\Bigr)
\!-\!q'^{2n+2}_0E_{-n}\!\Bigl(\!\frac{nq'^2_0}{\epsilon_{i0}}\Bigr)\!\right]\!\!+\!\mathcal{O}(\epsilon_{i0})\!\Bigg\}\; .
\ee
Final term we evaluate in Eq.~(\ref{remainingtrms}) is
\be
\sum_{n=1}^\infty\!\frac{(-1)^n(\Delta x)^{2n}}{2n(2n\!\!+\!\!1)!}\Bigl[\frac{H_{i0}}
{1\!\!+\!\epsilon_{i0}}\Bigr]^{2n}\!\!\!
\int_0^{t'_0}\!\!d\tilde{t}_0\frac{\tilde{H}_0}{1\!\!+\!\tilde\epsilon_0}
\Bigl[\ln\Bigl(\!\frac{\tilde{H}_0\tilde{a}_0}{H_{i0}}\!\Bigr)
\!\!-\!\!\ln\Bigl(\!\frac{1\!\!+\!\tilde\epsilon_0}
{1\!\!+\!\epsilon_{i0}}\!\Bigr)\!\Bigr]\;,\label{summm}
\ee
where the sum in Eq.~(\ref{summm})
\be
\sum_{n=1}^\infty\!\frac{(-1)^n(\Delta x)^{2n}}{2n(2n\!\!+\!\!1)!}\Bigl[\frac{H_{i0}}
{1\!\!+\!\epsilon_{i0}}\Bigr]^{2n}\!\!\!=\!1\!\!-\!\!\gamma\!+\!{\rm ci}(\alpha_{i0})\!-\!\frac{\sin(\alpha_{i0})}{\alpha_{i0}}\!-\!\ln(\alpha_{i0})\; ,
\ee
and the integral in Eq.~(\ref{summm})
\be
&&\hspace{-1.3cm}\int_0^{t'_0\!}\!\!d\tilde{t}_0\frac{\tilde{H}_0}{1\!\!+\!\tilde\epsilon_0}
\Bigl[\ln\Bigl(\!\frac{\tilde{H}_0\tilde{a}_0}{H_{i0}}\!\Bigr)
\!\!-\!\!\ln\Bigl(\!\frac{1\!\!+\!\tilde\epsilon_0}
{1\!\!+\!\epsilon_{i0}}\!\Bigr)\!\Bigr]\!\!=\!\epsilon^{-1}_{i0}\!\!\!\int_{q'_0}^1\!\!\! d\tilde{q}_0\frac{\tilde{q}^3_0}{\tilde{q}^2_0\!\!+\!\epsilon_{i0}}\!\left[
\ln(\tilde{q}_0)
\!+\!\frac{\epsilon^{-1}_{i0}}{2}[1\!-\!\tilde{q}^2_0]\!-\!\ln\Bigl(\!\frac{\tilde{q}_0^2
\!\!+\!\epsilon_{i0}}{\tilde{q}_0^2\!\left(1\!\!+\!\epsilon_{i0}\right)}\Bigr)\!\right]\label{intfor2ndsumline1}
\\
&&\hspace{1.1cm}=\!\epsilon^{-2}_{i0}\frac{\mathcal{E}'^2_0}{8}
\!+\!\epsilon^{-1}_{i0}\frac{\mathcal{E}'_0}{2}\ln(q'_0)\!+\!\frac{\mathcal{E}'_0}{2}
\Bigl[1\!\!+\!\frac{q'^{-2}_0}{2}\Bigr]\!\!+\!\frac{\ln^2(q'_0)}{2}\!+\!\frac{\ln(q'^{3}_0)}{2}
\!+\!\mathcal{O}(\epsilon_{i0})\; .\label{intfor2ndsum}
\ee
Thus, the first integral in Eq.~(\ref{corrpropphi4}) is obtained using Eqs.~(\ref{remaining1})-(\ref{intfor2ndsum}) in Eq.~(\ref{remainingtrms}). The result is given in Eq.~(\ref{1loopremain1stint}).

To evaluate the second integral in Eq.~(\ref{corrpropphi4}),
\be
\frac{H_0}{1\!\!+\!\epsilon_0}
\!\int_0^{t_0}\!\!\!dt''_0\Bigl[\frac{1\!\!+\!\epsilon''_0}{H''_0}\Bigr]^2
\!\langle\Omega|\delta\bar{\varphi}^3_0(t''_0,\vec{x})\,
\delta\bar{\varphi}_0(t'_0,\vec{x}\,')|\Omega\rangle\; ,\label{barphi4scndint}
\ee
we break up the integral into two as $\int_0^{t}\!dt''\!=\!\int_0^{t'}\!dt''\!+\!\int_{t'}^t\!dt''$. In the first integral $t''\!\leq\!t'$, whereas $t'\!\leq\!t''$ in the second. The first part of integral~(\ref{barphi4scndint})\be
&&\hspace{-0.4cm}\frac{H_0}{1\!\!+\!\epsilon_0}\!
\int_0^{t'_0}\!\!dt''_0\Bigl[\frac{1\!\!+\!\epsilon''_0}{H''_0}\Bigr]^2\!\langle\Omega|\delta\bar{\varphi}^3_0(t''_0,\vec{x})
\,\delta\bar{\varphi}_0(t'_0,\vec{x}\,')|\Omega\rangle\; ,\label{frstprtintscnd}
\ee
yields---following the same steps through Eqs.~(\ref{remaining1})-(\ref{intfor2ndsum})---exactly the same result obtained in Eq.~(\ref{1loopremain1stint}). The remaining part of integral~(\ref{barphi4scndint}) is
\be
&&\hspace{1.1cm}\frac{H_0}{1\!\!+\!\epsilon_0}\!
\int_{t'_0}^{t_0}\!\!dt''_0\Bigl[\frac{1\!\!+\!\epsilon''_0}{H''_0}\Bigr]^2
\!\langle\Omega|\delta\bar{\varphi}^3_0(t''_0,\vec{x})
\,\delta\bar{\varphi}_0(t'_0,\vec{x}\,')|\Omega\rangle\nonumber\\
&&\hspace{-1.3cm}=\!\frac{H_0}{1\!\!+\!\epsilon_0}\!
\int_{t'_0}^{t_0}\!\!dt''_0\Bigl[\frac{1\!\!+\!\epsilon''_0}{H''_0}\Bigr]^2
\!3\!\cdot\!1\langle\Omega|\delta\bar{\varphi}_0(t''_0,\vec{x})\,
\delta\bar{\varphi}_0(t'_0,\vec{x}\,')|\Omega\rangle\langle\Omega|\delta\bar{\varphi}^2_0(t''_0,\vec{x})|\Omega\rangle\nonumber\\
&&\hspace{-0.7cm}=\!\frac{3}{16\pi^4}\frac{H_0}{1\!\!+\!\epsilon_0}\frac{H'_0}{1\!\!+\!\epsilon'_0}
\Bigg\{\!\!\ln\Bigl(\!\frac{H'_0 a'_0}{H_{i0}}\!\Bigr)
\!\!-\!\!\ln\Bigl(\!\frac{1\!\!+\!\epsilon'_0}
{1\!\!+\!\epsilon_{i0}}\!\Bigr)\!\!+\!\!
\sum_{n=1}^\infty\!\!\frac{(-1)^n\!\left[\alpha'^{2n}_0\!\!-\!\alpha_{i0}^{2n}\right]}{2n(2n\!\!+\!\!1)!}\Bigg\}\nonumber\\
&&\hspace{1.8cm}\times\!\!\int_{t'_0}^{t_0}\!\!dt''_0\frac{H''_0}{1\!\!+\!\epsilon''_0}\Bigl[\ln\Bigl(\!\frac{H''_0 a''_0}{H_{i0}}\!\Bigr)
\!\!-\!\!\ln\Bigl(\!\frac{1\!\!+\!\epsilon''_0}
{1\!\!+\!\epsilon_{i0}}\!\Bigr)\Bigr]
\label{}\; .\label{intscndprt}
\ee
The terms inside the curly brackets in Eq.~(\ref{intscndprt}) can be written in terms of simple analytic functions using Eq.~(\ref{functseries}). The integral in Eq.~(\ref{intscndprt}) is the same as integral~(\ref{intfor2ndsumline1})---except for the limits of integration. It is evaluated as
\be
&&\hspace{2cm}\int_{t'_0}^{t_0}\!\!dt''_0\frac{H''_0}{1\!\!+\!\epsilon''_0}\Bigl[\ln\Bigl(\!\frac{H''_0 a''_0}{H_{i0}}\!\Bigr)
\!\!-\!\!\ln\Bigl(\!\frac{1\!\!+\!\epsilon''_0}
{1\!\!+\!\epsilon_{i0}}\!\Bigr)\Bigr]\nonumber\\
&&\hspace{-0.3cm}=\frac{\epsilon^{-2}_{i0}}{8}\!\left[\mathcal{E}^2_0
\!\!-\!\mathcal{E}'^2_0\right]\!\!-\!\frac{\epsilon^{-1}_{i0}}{2}
\!\Bigl[q^2_0\ln(q_0)\!-\!q'^2_0\ln(q'_0)\!-\!\left[\,\ln(q_0)\!-\!\ln(q'_0)\right]\Bigr]\nonumber\\
&&\hspace{-1.1cm}+\frac{1}{2}\!\left[\frac{q^{-2}_0\!\!-\!\!q'^{-2}_0}{2}
\!-\!\!\left[\,q^2_0\!\!-\!q'^2_0\right]\!\!+\!\ln^2(q_0)
\!-\!\ln^2(q'_0)\!+\!\ln(q^3_0)\!-\!\ln(q'^3_0)\right]\!\!+\!\mathcal{O}(\epsilon_{i0})\; .\label{2ndint2ndprtint}
\ee
The second integral in Eq.~(\ref{corrpropphi4}) is, therefore, obtained combining  Eqs.~(\ref{frstprtintscnd})-(\ref{2ndint2ndprtint}) in Eq.~(\ref{barphi4scndint}). The result is given in Eq.~(\ref{1loopremain2ndint}).
\end{appendix}


\begin{thebibliography}{99}

\bibitem{vacuum}  V. K. Onemli, Phys. Rev. D {\bf 91}, 103537 (2015).

\bibitem{GKVO} G. Karakaya and V. K. Onemli, Phys. Rev. D {\bf 97}, 123531 (2018).


\bibitem{Star} A.~A.~Starobinsky, in {\it Field Theory, Quantum
Gravity and Strings}, edited by H.~J.~de~Vega and N.~Sanchez
(Springer-Verlag, Berlin, 1986), p. 107.

\bibitem{StarYok} A.~A.~Starobinsky and J.~Yokoyama, Phys. Rev.~D~{\bf 50},
6357 (1994).

\bibitem{OW1} V. K. Onemli and R. P. Woodard, Classical Quantum Gravity {\bf 19}, 4607 (2002).

\bibitem{OW2} V. K. Onemli and R. P. Woodard, Phys. Rev. D {\bf70}, 107301 (2004).

\bibitem{BOW} T. Brunier, V. K. Onemli, and R. P. Woodard, Classical Quantum Gravity {\bf 22}, 59 (2005).

\bibitem{KO} E. O. Kahya and V. K. Onemli, Phys. Rev. D {\bf 76}, 043512 (2007).

\bibitem{KOW1} E. O. Kahya, V. K. Onemli, and R. P. Woodard, Phys. Rev. D {\bf 81}, 023508 (2010).

\bibitem{O1}  V. K. Onemli, Phys. Rev. D {\bf 89}, 083537 (2014).

\bibitem{sfl} References on quantum theory of scalar field fluctuations include:
S. Weinberg, Phys. Rev. D {\bf 72}, 043514 (2005); {\bf74}, 023508 (2006);
K. Chaicherdsakul, Phys. Rev. D {\bf75}, 063522 (2007); P. Adshead, R. Easther, and E. A. Lim, Phys. Rev. D {\bf79},
063504 (2009); D. Boyanovsky, H. J. de Vega, and N. G. Sanchez, Nucl.
Phys. {\bf B747}, 25 (2006); Phys. Rev. D {\bf72}, 103006 (2005); M. Sloth, Nucl. Phys. {\bf B748}, 149 (2006); Nucl. Phys. {\bf B775}, 78 (2007); D. Seery, J. E. Lidsey, and M. S. Sloth, J.~Cosmol. Astropart. Phys.~01 (2007) 027; M. van der Meulen and J. Smit, J.~Cosmol. Astropart. Phys.~11 (2007) 023; D. H. Lyth, J.~Cosmol. Astropart. Phys.~12 (2007) 016; D. Seery, J.~Cosmol. Astropart. Phys.~11 (2007) 025; 02 (2008) 006; 05 (2009) 021; Classical Quantum Gravity~{\bf 27},
124005 (2010); Y. Urakawa and K. I. Maeda, Phys. Rev. D {\bf78}, 064004 (2008); A. Riotto and M. Sloth,
J.~Cosmol. Astropart. Phys.~04 (2008) 030; 10 (2011) 003; P. Adshead, R. Easther, and E. A. Lim, Phys. Rev.
D {\bf 79}, 063504 (2009); Y. Urakawa and T. Tanaka, Prog. Theor. Phys. {\bf 122}, 779 (2009); {\bf 122}, 1207 (2009); {\bf 125}, 1067 (2011); Phys. Rev. D {\bf82}, 121301 (2010); J.~Cosmol. Astropart. Phys.~05 (2011)
014; Y. Urakawa, Prog. Theor. Phys. {\bf 126}, 961 (2011); S. B. Giddings and M. S. Sloth,
J.~Cosmol. Astropart. Phys.~07 (2010) 015; 01 (2011) 023; Phys. Rev.
D {\bf84}, 063528 (2011); {\bf 86}, 083538 (2012);  E. O. Kahya, V. K. Onemli, and R. P. Woodard, Phys. Lett. B {\bf 694}, 101 (2010); C. P. Burgess, R. Holman, L. Leblond, and S. Shandera, J.~Cosmol. Astropart. Phys.~03 (2010) 033;
D. Boyanovsky, Phys. Rev. D {\bf 85}, 123525 (2012); {\bf 86}, 023509 (2012); {\bf93}, 083507 (2016); {\bf98}, 023515 (2018); K. Feng, Y.-F. Cai, and Y.-S. Piao, Phys. Rev. D {\bf 86}, 103515 (2012);
K. Larjo and D. Lowe, Phys. Rev. D {\bf 87}, 083506 (2013); E. T. Akhmedov, F. K. Popov, and V. M. Slepukhin,
Phys. Rev. D {\bf 88}, 024021 (2013); J. Serreau and R. Parentani, Phys. Rev. D {\bf 87},
085012 (2013); J. Serreau, Phys. Lett. B {\bf 728}, 380 (2014); E. T. Akhmedov, Int. J. Mod. Phys. D {\bf 23}, 1430001 (2014); L. Lello, D. Boyanovsky, and
R. Holman, Phys. Rev. D {\bf 89}, 063533 (2014); M. Herranen, T. Markkanen, and A.
Tranberg, J. High Energy Phys. 05 (2014) 026; V. K. Onemli, arXiv:1510.02272; X. Chen, Y. Wang, and Z.-Z. Xianyu, J. High Energy Phys. 08 (2016) 051; C. Armendariz-Picon and G. \c{S}eng\"{o}r, J.~Cosmol. Astropart. Phys.~11 (2016) 016; T. Markkanen, J. High Energy Phys. 01 (2018) 116; P. Adshead, C. P. Burgess, R. Holman, and S. Shandera, J.~Cosmol. Astropart. Phys.~02 (2018) 016; E. T. Akhmedov and F. Bascone, Phys. Rev. D {\bf 97}, 045013 (2018); E.T. Akhmedov, K. V. Bazarov, D. V. Diakonov, U. Moschella, F. K. Popov, and C. Schubert, arXiv:1905.09344.

\bibitem{W3}
S.~P.~Miao and R.~P.~Woodard, Phys. Rev.~D~{\bf 74},
044019 (2006).

\bibitem{W4}
R.~P.~Woodard, J. Phys. Conf. Ser. {\bf 68}, 012032 (2007).

\bibitem{PTW1}
T.~Prokopec, N.~C.~Tsamis, and R.~P.~Woodard, Classical Quantum Gravity {\bf 24}, 201 (2007).

\bibitem{PTW2}
T.~Prokopec, N.~C.~Tsamis, and R.~P.~Woodard, Ann. Phys. (Amsterdam)~{\bf 323}, 1324 (2008).

\bibitem{W1}
R.~P.~Woodard, Nucl. Phys.~B, Proc. Suppl. {\bf 148}, 108 (2005).

\bibitem{TWstgrav}
N.~C.~Tsamis and R.~P.~Woodard, Nucl. Phys.~{\bf B724}, 295 (2005).

\bibitem{MW} S.~P.~Miao and R.~P.~Woodard, Classical Quantum Gravity {\bf 25}, 145009 (2008).

\bibitem{MRST} T.~Markkanen, A.~Rajantie, S.~Stopyra, and T.~Tenkanen, J.~Cosmol. Astropart. Phys.~08 (2019) 001.

\bibitem{TT} J.~Tokuda and T.~Tanaka, J.~Cosmol. Astropart. Phys.~02 (2018) 014; 11 (2018) 022.

\bibitem{AMPP1} E. T. Akhmedov, U. Moschella, K. E. Pavlenko, and F. K. Popov, Phys. Rev. D {\bf96}, 025002 (2017).

\bibitem{AMPP2} E. T. Akhmedov, U. Moschella, and F. K. Popov, Phys. Rev. D {\bf99}, 086009 (2019).

\bibitem{MR} I. Moss and G. Rigopoulos, J.~Cosmol. Astropart. Phys.~05 (2017) 009.

\bibitem{R1} G. Rigopoulos, J.~Cosmol. Astropart. Phys.~07 (2016) 035.

\bibitem{DB1} D. Boyanovsky, Phys. Rev. D {\bf93}, 043501 (2016); New J. Phys. {\bf17}, 063017 (2015).

\bibitem{CW1} C. Wetterich, J.~Cosmol. Astropart. Phys.~05 (2016) 041; Phys. Rev. D {\bf92}, 083507 (2015).

\bibitem{JSS} R. K. Jain, M. Sandora, and M. S. Sloth, J.~Cosmol. Astropart. Phys.~06 (2015) 016.

\bibitem{PVAW} H. Assadullahi, H. Firouzjahi, M. Noorbala, V. Vennin, and D. Wands, J.~Cosmol. Astropart. Phys.~06 (2016) 043.

\bibitem{AFNVW} C. Pattison, V. Vennin, H. Assadullahi, and D. Wands, J.~Cosmol. Astropart. Phys.~07 (2019) 031.

\bibitem{GS1} F. Gautier and J. Serreau, Phys. Rev. D {\bf92}, 105035 (2015).

\bibitem{DB2} D. Boyanovsky, Phys. Rev. D {\bf92}, 023527 (2015).

\bibitem{GS2} M. Guilleux and J. Serreau, Phys. Rev. D {\bf92}, 084010 (2015).

\bibitem{Dod} S. Dodelson, {\it Modern Cosmology} (Academic Press, San Diego, 2003).

\bibitem{FinMarVacVen} F.~Finelli, G.~Marozzi,~G. P. Vacca, and G. Venturi,  Phys. Rev.~D~{\bf 65},
103521 (2002).

\bibitem{MP} S. P. Miao and S. Park, Phys. Rev. D {\bf 89}, 064053 (2014).

\bibitem{plnck1} P. A. R. Ade {\it et al}. [Planck Collaboration], Astron. Astrophys. {\bf 594}, A20 (2016).

\end{thebibliography}
\end{document}